%% file: letter.tex
\documentclass[10pt]{IEEEtran}
\usepackage[noadjust]{cite}
\usepackage{amsmath}
\usepackage{amssymb}
\usepackage{amsthm}
\usepackage{mathtools}
\usepackage{bbm}
\usepackage{algorithm}
\usepackage[noend]{algpseudocode}
\usepackage{bm}
\usepackage{comment}
\usepackage{tabularx,booktabs}
\usepackage{subcaption}
\usepackage[capitalize]{cleveref}
\DeclareMathOperator*{\argmin}{arg\,min}
\DeclareMathOperator*{\argmax}{arg\,max}
\DeclareMathOperator{\tr}{tr}

\DeclareMathOperator{\E}{\mathbb{E}}
\usepackage{pgfplots}
\pgfplotsset{compat=1.15}
\usepgfplotslibrary{fillbetween}
\usetikzlibrary{patterns,arrows,plotmarks}
\usepgfplotslibrary{groupplots}
\pgfdeclarelayer{background}
\pgfsetlayers{background,main}
\usetikzlibrary{automata,positioning}
\usetikzlibrary{decorations}
\usetikzlibrary{shapes.arrows}
\usetikzlibrary{tikzmark}
\usetikzlibrary{calc}
\usetikzlibrary{decorations.markings}
\algrenewcommand\algorithmicindent{10pt}
\usepgfplotslibrary{colorbrewer}
\usepackage{glossaries}    
    
\newacronym[plural=MDPs,firstplural=Markov Decision Processes (MDPs)]{mdp}{MDP}{Markov Decision Process}
\newacronym{iot}{IoT}{Internet of Things}
\newacronym{fec}{FEC}{Forward Error Correction}
\newacronym{snr}{SNR}{Signal to Noise Ratio}
\newacronym{pmf}{pmf}{probability mass function}
\newacronym{iid}{IID}{Independent and Identically Distributed}
\newacronym{aoi}{AoI}{Age of Information}
\newacronym{voi}{VoI}{Value of Information}
\newacronym{paoi}{PAoI}{Peak Age of Information}
\newacronym{qaoi}{QAoI}{Age of Information at Query}
\newacronym{mse}{MSE}{Mean Square Error}
\newacronym{leo}{LEO}{Low Earth Orbit}
\newacronym{geo}{GEO}{Geosynchronous Earth Orbit}
\newacronym{pec}{PEC}{Packet Erasure Channel}
\newacronym{pdf}{PDF}{Probability Density Function}
\newacronym{cdf}{CDF}{Cumulative Distribution Function}
\newacronym{pq}{PQ}{Permanent Query}
\newacronym{qapa}{QAPA}{Query Arrival Process Aware}
\newacronym{harq}{HARQ}{Hybrid Automated Repeat Request}
\newacronym{bs}{BS}{Base Station}
\newacronym{mmse}{MMSE}{Minimum Mean Square Error}
\newacronym{maf}{MAF}{Maximum Age First}
\newacronym{tdma}{TDMA}{Time Division Multiple Access}

\errorcontextlines\maxdimen
\newcommand{\ALGtikzmarkcolor}{black}
\newcommand{\ALGtikzmarkextraindent}{0pt}
\newcommand{\ALGtikzmarkverticaloffsetstart}{-.5ex}
\newcommand{\ALGtikzmarkverticaloffsetend}{-.5ex}
\makeatletter
\newcounter{ALG@tikzmark@tempcnta}

\newcommand\ALG@tikzmark@start{%
    \global\let\ALG@tikzmark@last\ALG@tikzmark@starttext%
    \expandafter\edef\csname ALG@tikzmark@\theALG@nested\endcsname{\theALG@tikzmark@tempcnta}%
    \tikzmark{ALG@tikzmark@start@\csname ALG@tikzmark@\theALG@nested\endcsname}%
    \addtocounter{ALG@tikzmark@tempcnta}{1}%
}

\def\ALG@tikzmark@starttext{start}
\newcommand\ALG@tikzmark@end{%
    \ifx\ALG@tikzmark@last\ALG@tikzmark@starttext
    \else
        \tikzmark{ALG@tikzmark@end@\csname ALG@tikzmark@\theALG@nested\endcsname}%
        \tikz[overlay,remember picture] \draw[\ALGtikzmarkcolor] let \p{S}=($(pic cs:ALG@tikzmark@start@\csname ALG@tikzmark@\theALG@nested\endcsname)+(\ALGtikzmarkextraindent,\ALGtikzmarkverticaloffsetstart)$), \p{E}=($(pic cs:ALG@tikzmark@end@\csname ALG@tikzmark@\theALG@nested\endcsname)+(\ALGtikzmarkextraindent,\ALGtikzmarkverticaloffsetend)$) in (\x{S},\y{S})--(\x{S},\y{E});%
    \fi
    \gdef\ALG@tikzmark@last{end}%
}

\algrenewcommand\alglinenumber[1]{\tiny #1:}

\apptocmd{\ALG@beginblock}{\ALG@tikzmark@start}{}{\errmessage{failed to patch}}
\pretocmd{\ALG@endblock}{\ALG@tikzmark@end}{}{\errmessage{failed to patch}}
\makeatother

\definecolor{msecol}{HTML}{0011af}
\definecolor{avgcol}{HTML}{8819a0}
\definecolor{varcol}{HTML}{bf418d}
\definecolor{maxcol}{HTML}{e37076}
\definecolor{cntcol}{HTML}{f9a256}
\definecolor{mafcol}{HTML}{ffd700}

\def \fwidth{0.99\columnwidth}
\def \fheight {0.75\columnwidth}

\def \fswidth{0.9\linewidth}
\def \fsheight {0.165\linewidth}

\newcommand{\edit}[1]{{#1}}

\begin{document}
\title{Scheduling of Sensor Transmissions Based on Value of Information for Summary Statistics}

\author{Federico Chiariotti,~\IEEEmembership{Member,~IEEE,} Anders E. Kal\o{}r,~\IEEEmembership{Student Member,~IEEE,} Josefine Holm,~\IEEEmembership{Student Member,~IEEE,}
Beatriz Soret,~\IEEEmembership{Member,~IEEE,} and Petar Popovski,~\IEEEmembership{Fellow,~IEEE}%
\thanks{This work has been in part supported by the Danish Council for Independent Research, Grant Nr. 8022-00284B (SEMIOTIC). F. Chiariotti, A. E. Kal\o{}r, J. Holm and P. Popovski are with the Department of Electronic Systems, Aalborg University, Denmark (email: \{fchi, aek, jho, petarp\}@es.aau.dk). B. Soret is with the Telecommunication Research Institute (TELMA) , Universidad de Malaga, Spain  (email: bsoret@ic.uma.es).}\vspace{-0.8cm}
}

\maketitle

\begin{abstract}
The optimization of \gls{voi} in sensor networks integrates awareness of the measured process in the communication system. However, most existing scheduling algorithms do not consider the specific needs of monitoring applications, but define \gls{voi} as a generic \gls{mse} of the whole system state regardless of the relevance of individual components. In this work, we consider different summary statistics, i.e., different functions of the state, which can represent the useful information for a monitoring process, particularly in safety and industrial applications. We propose policies that minimize the estimation error for different summary statistics, showing significant gains by simulation.
\end{abstract}
 
\glsresetall

\section{Introduction}
Over the past few years, the unprecedented development of the \gls{iot} has made the remote estimation of stochastic processes a central problem in communications and automation~\cite{soderlund2019optimization}, where a set of sensors transmit observations to a central \gls{bs}. The possibility to process sensor data either at the \gls{bs} or in a distributed fashion through in-network processing~\cite{awad2017distributed} has led the research community to focus extensively on the scheduling of sensor updates in severely resource-constrained wireless network environments.

For a wide range of remote estimation problems, the \emph{freshness} of the observations at the \gls{bs} is a good proxy for the estimation quality. This promotes \gls{aoi}~\cite{kaul2012real} as a measure of the time that has passed since the last update from a given sensor. However, if the destination has a model of the observed processes, it is often better to directly minimize the uncertainty of the process estimates instead of the \gls{aoi}~\cite{ayan2019age}. The problem of scheduling \gls{iot} sensors with this goal has been considered for several different policies~\cite{gupta2006stochastic,hashemi2020randomized}, whose objective is to minimize the \gls{mse} of a Kalman filter, considering communication constraints. More recently, the problem of minimizing the \gls{mse} of the process estimates has been referred to as \gls{voi}~\cite{ayan2019age}. \edit{A recent work~\cite{li2021sensor} tries to maximize the accuracy of a more complex unscented filter, aiming at optimal sensor selection for maneuvering tasks, and \gls{voi} can also be used for data muling applications in underwater or drone networks~\cite{duan2020voi}. Another interesting twist to this is the application of \gls{voi} concepts not over time, but in space, placing sensors in the positions that will result in the highest overall accuracy for the estimation of a spatial process~\cite{hoseyni2019voi}.}

\begin{figure}
    \centering    
    \resizebox{0.6\columnwidth}{!}{\input{./fig/scenario.tex}}
    \caption{Representation of the scenario.\vspace{-0.3cm}}
 \label{fig:scenario}
\end{figure}
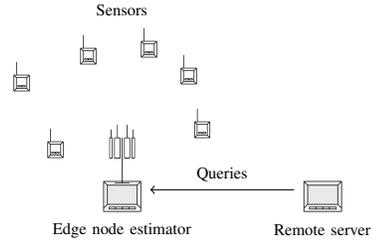

However, there are cases where minimizing the \gls{mse} is not be the best thing to do: for example, if the application needs to compute a non-linear function of the state, such as the maximum value among all sensors. While minimizing the \gls{mse} implicitly gives equal value to all sensors, some might have a larger weight in the non-linear function (e.g., sensors with a higher value for the maximum function). Examples in industrial settings include: (1) triggering a safety warning if the temperature of any of the components in a machine reaches a safety limit, (2) monitoring if the difference in the strain on different parts of a structure is outside the design parameters. Such a scenario is represented in Fig.~\ref{fig:scenario}: the remote server sends queries to the \gls{bs}, which correspond to the non-linear function, and the \gls{bs} needs to schedule transmission so as to maximize the accuracy. \edit{This setup was also used in our previous work~\cite{chiariotti2022query}. The scheduling in this scenario is driven by the \gls{bs}, which selects the sensor that it believes to have the most useful information at each time slot; the opposite scenario, in which sensors themselves decide whether to transmit or not, is an interesting but different problem, as it requires sensors to maintain an estimate of the system state and a decision algorithm, which consume energy, as well as to coordinate among themselves to avoid collisions. Our scenario is directly applicable to wake-up radio~\cite{froytlog2019ultra,shiraishi2020wake} and similar schemes with low-power sensors.}

We propose heuristic strategies to schedule sensor updates in a linear dynamic system, which explicitly aim to minimize the error of various summary statistics. We derive the one-step optimal strategies for some well-known function, and give a general Monte Carlo-based algorithm that can deal with different query functions. The simulations show that the proposed strategies can significantly reduce the error on a number of summary statistics, with more significant gains in case of highly non-linear summary statistics.

The rest of this letter is organized as follows. The system model is presented in Sec.~\ref{sec:system}, and one-step policies for various summary statistics are derived in Sec.~\ref{sec:policies}. Numerical results are presented in Sec.~\ref{sec:results}, and finally Sec.~\ref{sec:concl} concludes the paper and presents some possible avenues of future work.

\section{System Model}\label{sec:system}
We consider a system with $N$ sensors, which are connected through time-slotted wireless links to a \gls{bs} equipped with computing and storage resources. Without loss of generality, we assume that the time slots occur at $t=1,2,\ldots$ and the sensors are indexed by $n=1,\ldots,N$. We assume that each sensor observes a value in an $N$-dimensional process, whose state $\mathbf{x}(t)=[x_1,\ldots,x_N]^T$ evolves according to
\begin{equation}
    \mathbf{x}(t) = \mathbf{A}\mathbf{x}(t-1) + \mathbf{v}(t),\label{eq:process_model}
\end{equation}
where $\mathbf{A}\in\mathbb{R}^{N\times N}$ is the transition matrix, $\mathbf{v}(t)\sim\mathcal{N}(\mathbf{0},\mathbf{\Sigma}_{\mathbf{v}})$ is the process noise with covariance matrix $\mathbf{\Sigma}_{\mathbf{v}}\in\mathbb{R}^{N\times N}$, and $\mathbf{x}(0)=\mathbf{0}$. The sensors observe the processes with additive white Gaussian measurement noise $\mathbf{w}(t)\sim\mathcal{N}(\mathbf{0},\mathbf{\Sigma}_{\mathbf{w}})$, i.e., $\mathbf{y}(t) = \mathbf{x}(t)+\mathbf{w}(t)$. In general, the covariance matrices $\mathbf{\Sigma}_{\mathbf{v}}$ and $\mathbf{\Sigma}_{\mathbf{w}}$ are not diagonal.
Note that although we assume that the number of sensors is equal to the dimension of the process (to simplify the notation), the analysis can be easily extended to more general observable systems.

\edit{We consider a \gls{tdma} air interface, in which each time slot, $t$, contains a downlink phase and an uplink phase. The downlink is used by the \gls{bs} to schedule the sensor, $a(t)$, that transmits its observation $y_{a(t)}(t)$ in the uplink phase. The channel is modeled as a packet erasure channel with error probability $\varepsilon_{a(t)}$, which captures errors both in the transmission of the scheduling decision and the observation.} We also assume that the process dynamics are known to the \gls{bs}\edit{, a standard assumption in Kalman filtering, which is practical if the monitored system is well-understood, even if its instantaneous state is hard to measure directly. This condition is common for many \gls{iot} applications~\cite{huang2019epkf,wang2018differentially}, in which well-known processes are estimated by sensors over wide areas}. We also denote the row vector of length $N$ whose only non-zero value is the $n$-th, which is 1, as $\mathbf{1}_n$, and the $N\times N$ identity matrix as $\mathbf{I}_N$.

\subsection{Kalman Filter Estimation}

We assume that the \gls{bs} maintains a distribution over its belief of the state $p(\mathbf{x}(t))$ using a Kalman filter. The Kalman filter is the \gls{mmse} estimator for the model defined in \cref{eq:process_model}~\cite{kalman1960new}, in which case $p(\mathbf{x}(t))\sim\mathcal{N}(\hat{\mathbf{x}}(t),\bm{\psi}(t))$. The mean vector $\hat{\mathbf{x}}(t)$ and the covariance matrix $\bm{\psi}(t)$ are updated at each timestep $t$ based on the outcome of the scheduling process. The Kalman filter operates in two steps: a \emph{prior} update, which only depends on the system statistics, and a \emph{posterior} update, which integrates new observations. The prior update operation is given by:
\begin{align}
    \hat{\mathbf{x}}(t)&=\mathbf{A}\hat{\mathbf{x}}(t-1) \triangleq \hat{\mathbf{x}}_{\mathrm{F}}(t),\label{eq:blind}\\
    \bm{\psi}(t)&=\mathbf{A}\bm{\psi}(t-1)\mathbf{A}^T+\mathbf{\Sigma}_{\mathbf{v}} \triangleq \bm{\psi}_{\mathrm{F}}(t).\label{eq:var_blind}
\end{align}
If the transmission of the update fails, an event we denote as $\mathrm{F}$, the \gls{bs} can only rely on the prior update for its estimate. If the update is received, it can be used to improve the estimate. We then compute the Kalman filter gain $\mathbf{k}(t)$:
\begin{align}
\mathbf{k}(t)=&\bm{\psi}_{\mathrm{F}}(t)\mathbf{1}_{a(t)}^T\left[\mathbf{1}_{a(t)}\left(\bm{\psi}_{\mathrm{F}}(t)+\mathbf{\Sigma}_{\mathbf{w}}\right)\mathbf{1}_{a(t)}^T\right]^{-1}.
\end{align}
We finally get the updated estimate in case of a success event:
\begin{align}
\begin{split}
    \hat{\mathbf{x}}(t)&=\hat{\mathbf{x}}_F(t)+\mathbf{k}(t)\mathbf{1}_{a(t)}(\mathbf{y}(t)-\hat{\mathbf{x}}_F(t))\triangleq \hat{\mathbf{x}}_{\mathrm{S},a(t)}(t)
    \end{split}\\
    \bm{\psi}(t)&=(\mathbf{I}_N-\mathbf{k}(t)\mathbf{1}_{a(t)})\bm{\psi}_{\mathrm{F}}(t) \triangleq \bm{\psi}_{\mathrm{S},a(t)}(t),\label{eq:var_update}
\end{align}
Note that the recursive structure of the Kalman filter and the independence of the transmission errors imply that $\hat{\mathbf{x}}(t)$ and $\bm{\psi}(t)$ are sufficient statistics for the state estimate given the full history $\mathcal{H}(t)$ of past actions and observations.

\subsection{Summary Statistics}

Unlike the majority of \gls{voi} applications, in which the \gls{bs} aims to minimize the \gls{mse} of $\hat{\mathbf{x}}(t)$, we consider the case in which an external user requests \emph{summary statistics} about the state of the system: these correspond to a predefined, fixed function of the system state, e.g., the average value or the number of states with values within a given interval. Formally, we define a summary statistic as a function $z:\mathbb{R}^N\rightarrow\mathbb{R}$ of the true state $\mathbf{x}(t)$. However, because $\mathbf{x}(t)$ is unknown to the \gls{bs}, it can only provide an \emph{approximate} answer to the query based on its state belief $p(\mathbf{x}(t))$. We will consider estimators of the summary statistics on the form
\begin{equation}
  \hat{z}(t)=\E_{\mathbf{x}\sim \mathcal{N}(\hat{\mathbf{x}}(t),\bm{\psi}(t))}[z(\mathbf{x})],\label{eq:E_z}
\end{equation}
which corresponds to the minimum \gls{mse} estimator of $z(\mathbf{x}(t))$, given the current observation~\cite{humpherys2012fresh,kay93}. This is different from minimizing the \gls{mse} of $\hat{\mathbf{x}}(t)$, particularly when function $z(\cdot)$ is non-linear or the sensors have different weights. We denote the squared error as $\nu_z(t)$:
\begin{equation}
    \nu_z(t)=(z(\mathbf{x}(t))-\hat{z}(t))^2.
\end{equation}

\section{Scheduling Strategies}\label{sec:policies}

In our scenario, we seek a scheduling strategy, that is, a function $\pi_z$ from the current Kalman state (which is represented by vector $\hat{\mathbf{x}}(t)$ and matrix $\bm{\psi}(t)$) to an action $a(t)$, that minimizes the expected error for a given summary statistic $z$:
\begin{equation}
    \underset{\pi_z\in\Pi}{\text{minimize}}\E\left[\nu_z(t)\mid\pi_z\right],\label{eq:onestep_def}
\end{equation}
While we only consider the error in the next time step, the optimal solutions are expected to perform well with respect to the long-term error due to the linearity of the observed process (despite a non-linear summary statistic). Computing $\E\left[\nu_z(t)|a(t)\right]$ is not simple, but it can be expressed in terms of the two possible transmission outcomes:
\begin{equation}
\begin{aligned}
 \E\left[\nu_z(t)\mid a(t)\right]=&(1-\varepsilon_{a(t)})\E\left[\nu_z(t)\mid\hat{\mathbf{x}}_{\mathrm{S},a(t)}(t),\bm{\psi}_{\mathrm{S},a(t)}(t)\right]\\
 &+\varepsilon_{a(t)}\E\left[\nu_z(t)\mid\hat{\mathbf{x}}_{\mathrm{F}}(t),\bm{\psi}_{\mathrm{F}}(t)\right],\label{eq:split_sched}
 \end{aligned}
\end{equation}
where the expectation is over the state evolution.
Since $\bm{\psi}_{\mathrm{S},a(t)}(t)$ and $\bm{\psi}_{\mathrm{F}}(t)$ can be computed using~\eqref{eq:var_update} and~\eqref{eq:var_blind}, we can iterate over the possible actions and find the optimal scheduler, as long as we can estimate the \gls{mse} for a given observation. In the following, we derive the optimal schedulers for some well-known summary statistics, along with giving a Monte Carlo-based approximate scheduler that can deal with more complex statistics for which the \gls{mse} is hard to express in closed form. Using the result from~\eqref{eq:split_sched} we can obtain the optimal scheduling decision at time $t$ as:
\begin{equation}
    a^*_z(t)=\argmin_{a(t)\in\{1,\ldots,N\}} \E\left[\nu_z(t)\mid a(t)\right].\label{eq:optimal}
\end{equation}

\subsection{Baseline Scheduler}\label{sec:baseline}
We start by defining our benchmark scheme, which aims to minimize the \gls{mse} between the true state $\mathbf{x}$ and the estimated state $\hat{\mathbf{x}}$. The query is then computed as in \eqref{eq:E_z} based on the \gls{mmse} state estimate.

The squared state estimation error can be expressed as
\begin{equation} \E\left[\nu_{\text{MSE}}(t)\right]=\E\left[(\mathbf{x}(t)-\hat{\mathbf{x}}(t))^T(\mathbf{x}(t)-\hat{\mathbf{x}}(t))\right].
\end{equation}
Because $(\mathbf{x}(t)-\hat{\mathbf{x}}(t))\sim\mathcal{N}(\mathbf{0},\bm{\psi}(t))$, the expression above is equivalent to the trace of the covariance matrix $\bm{\psi}$~\cite{mathai1992quadratic}:
\begin{equation}
  \E\left[\nu_{\text{MSE}}(t)\mid\bm{\psi}(t)\right]=\tr(\bm{\psi}(t)).
\end{equation}
We can then use~\eqref{eq:optimal} to compute the minimum \gls{mse} schedule.

\subsection{Sample Mean Scheduling}
We now consider the most basic statistic, the sample mean:
\begin{equation}
 z_{\text{avg}}(\mathbf{x}(t))=\frac{1}{N}\sum_{n=1}^N x_n(t).\label{eq:z_avg}
\end{equation}
The estimation error $\nu_{\text{avg}}(t)$ is equal to the square of the average difference between the true and the estimated entries of $\mathbf{x}$. Since the sum of all elements in $\mathbf{x}(t)-\hat{\mathbf{x}}(t)$ is a Gaussian random variable with zero mean and variance equal to the sum of all elements in $\bm{\psi}(t)$, we have:
\begin{equation}
 \E\left[\nu_{\text{avg}}(t)\mid \bm{\psi}(t)\right]=\frac{\sum_{i=1}^N\sum_{j=1}^N\psi^{(i,j)}(t)}{N^2},
\end{equation}
where $\psi^{(i,j)}(t)$ is entry $(i,j)$ of $\bm{\psi}(t)$. We can then use the result in~\eqref{eq:optimal} to derive the one-step optimal schedule.

\subsection{Sample Variance Scheduling}
Another important summary statistic is the sample variance, quantifying how much the state deviates from the mean:
\begin{equation}
z_{\text{var}}(\mathbf{x}(t))=\frac{1}{N-1}\sum_{n=1}^N\left(x_n(t)-\sum_{m=1}^N\frac{x_m(t)}{N}\right)^2.\label{eq:z_var}
\end{equation}
To derive the scheduling policy, it is convenient to express $z_{\text{var}}(\mathbf{x}(t))$ in quadratic form with matrix $\mathbf{M}=\mathbf{I}-1/N$:
\begin{equation}
z_{\text{var}}(\mathbf{x}(t))=\frac{(\mathbf{M}\mathbf{x}(t))^T\mathbf{M}\mathbf{x}(t)}{N-1}=\frac{\mathbf{x}(t)^T\mathbf{M}\mathbf{x}(t)}{N-1}.
\end{equation}
Taking into account the belief $p(\mathbf{x})\sim\mathcal{N}\left(\hat{\mathbf{x}}(t),\bm{\psi}(t)\right)$, the expected value and variance of the sample variance are known from the literature~\cite{mathai1992quadratic}:
\begin{align}
  \hat{z}_{\text{var}}(t)&= \frac{1}{N-1}\left(\tr\left(\mathbf{M}\bm{\psi}(t)\right) +  \hat{\mathbf{x}}(t)^T\mathbf{M}\hat{\mathbf{x}}(t) \right)\\
    \E\left[\nu_{\text{var}}(t)|\bm{\psi}(t)\right]&=\frac{2\tr\left(\mathbf{M}\bm{\psi}(t)^2\right)+ 4\hat{\mathbf{x}}(t)^T\mathbf{M}\bm{\psi}(t)\hat{\mathbf{x}}(t)}{(N-1)^2}.
\end{align}
As for the \gls{mse} and sample mean, we can now simply derive the scheduler by using this result in~\eqref{eq:optimal}.

\subsection{Statistic-aware Monte Carlo scheduling}\label{sec:montecarlo}
We can now consider a generic summary statistic $z$: in the general case, computing the expected \gls{mse} can be extremely complex, or even impossible in closed form. In order to still provide an approximate scheduler, we consider Monte Carlo sampling to estimate $\E\left[\nu_z(t)|a(t)\right]$. This method consists of drawing $M$ samples from the conditioned multivariate Gaussian distribution $p(\mathbf{x}|\hat{\mathbf{x}})$, with complexity $O(MN^2)$, and is guaranteed to converge to the correct estimate as $M\rightarrow\infty$ thanks to the law of large numbers~\cite{luengo2020survey}. This estimate can then be used in~\eqref{eq:optimal} to perform scheduling.

The operation of the scheduler is specified in Alg.~\ref{alg:montecarlo}: in order to estimate the expected \gls{mse} when selecting each sensor $n$, it samples from the posterior distribution of the query. First, the scheduler performs the prior update step from~\eqref{eq:blind} and~\eqref{eq:var_blind}, then it draws an outcome to simulate the transmission, with failure probability $\varepsilon_n$. If the simulated transmission was successful, an observation is randomly drawn from a Gaussian distribution with mean $\mathbf{1}_n\hat{\mathbf{x}}$ and variance $\mathbf{1}_n\mathbf{S}(t)\mathbf{1}_n^T$, and the posterior update is performed. We then have the parameters $\hat{\mathbf{x}}$ and $\bm{\psi}$ of the multivariate Gaussian belief distribution of the state, from which we can draw a sample $\mathbf{x}_m$ to compute $u_m=z(\mathbf{x}_m)$. The sample variance over vector $\mathbf{u}$ is then our estimate of $\E\left[\nu_z(t)|a(t)\right]$, and we can simply select the sensor that gives the minimum expected \gls{mse}.

\begin{algorithm}[t]
\scriptsize
\caption{Monte Carlo scheduling policy}\label{alg:montecarlo}
\begin{algorithmic}[1]
\Function {Schedule}{$\hat{\mathbf{x}}(t-1),\bm{\psi}(t-1),\mathbf{A},\mathbf{\Sigma}_{\mathbf{v}},\mathbf{\Sigma}_{\mathbf{w}},\bm{\varepsilon},z$}
\State $\nu\gets\mathbf{0}$
\For{$n\in\{1,\ldots,N\}$}
\State $\mathbf{u}\gets\mathbf{0}$
\For{$m\in\{1,\ldots,M\}$}
\State $\hat{\mathbf{x}}(t-1),\bm{\psi}(t-1),\mathbf{S}\gets$\Call{PriorUpdate}{$\hat{\mathbf{x}},\bm{\psi},\mathbf{A},\mathbf{\Sigma}_{\mathbf{v}}$}
\If{\Call{random}{$0,1$} $\geq\varepsilon_n$}\Comment{Update successful}
\State $y\gets$\Call{GaussianSample}{$\mathbf{1}_n\hat{\mathbf{x}},\mathbf{1}_n\mathbf{S}\mathbf{1}_n^T$}
\State $\hat{\mathbf{x}},\bm{\psi}\gets$\Call{PosteriorUpdate}{$\hat{\mathbf{x}},\bm{\psi},y,n,\mathbf{\Sigma}_{\mathbf{w}}$}
\EndIf
\State $\mathbf{x}_m\gets$\Call{GaussianSample}{$\hat{\mathbf{x}},\bm{\psi}$}
\State$u(n)\gets$\Call{z}{$\mathbf{x}_m$}\Comment{Compute query value}
\EndFor
\State$\nu(n)\gets$\Call{var}{$\mathbf{u}$} \Comment{Sample variance}
\EndFor
\State \Return $\argmin\nu$
\EndFunction
\end{algorithmic}
\end{algorithm}

\begin{figure*}
    \centering    
    \input{./fig/selection_bar_inv.tex}
    \input{./fig/selection_bar_inv2.tex}
    \caption{Sensor selection frequency for the considered policies.\vspace{-0.3cm}}
 \label{fig:selection}
\end{figure*}
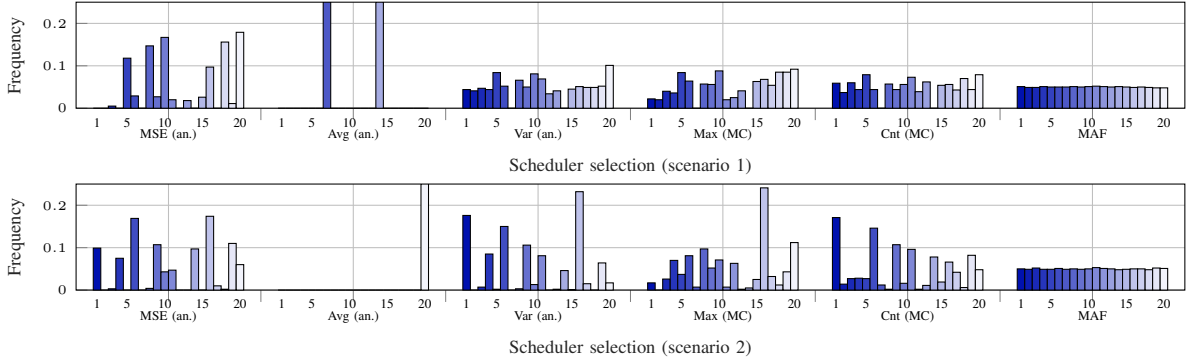

\section{Numerical Evaluation}\label{sec:results}

In the following, we show the effects of the sampling strategy on different statistics by simulation, using a Monte Carlo approach: we generate a synthetic process, then try to estimate it at the \gls{bs} using the different schedulers. The systems below represent two highly asymmetric examples, but the strategies we derived are optimal for all observable linear systems. The scenarios are constructed to be stable, i.e., the eigenvalues of the system matrices are all smaller than 1.

\subsection{Scenario and Settings}

We evolve the system for 100 episodes of 1000 samples each, and the Monte Carlo scheduler computes a total of $M=1000$ samples for each state. We consider two systems with $N=20$ sensors, in which the elements of the update matrix $\mathbf{A}$ are known. In the first scenario, the matrix $\mathbf{A}_1$ is given by:
\begin{equation}
  A_1^{(i,j)}=\begin{cases}
            \frac{3}{4}, &\text{if } i=j;\\
            -\frac{1}{8},&\text{if } i\neq j, \text{mod}(i-2j,7)=0,
          \end{cases}
\end{equation}
where $\text{mod}(m,n)$ is the integer modulo function, and the values are 0 everywhere else. On the other hand, in the second scenario, we have:
\begin{equation}
  A_2^{(i,j)}=\begin{cases}
            \frac{4}{5}, &\text{if } i=j;\\
            -\frac{1}{9},&\text{if } i\neq j, \text{mod}(\left\lceil i-2.3j\right\rceil,7)=0.
          \end{cases}
\end{equation}
The other parameters are the same in both scenarios. We also have $\mathbf{\Sigma_w}=\mathbf{I}$, while the process noise covariance is given by:
\begin{equation}
\Sigma_v^{(i,j)}=\begin{cases}
            \frac{11+\text{mod}(i,10)}{5}, &\text{if } i=j;\\
            1,&\text{if } i\neq j, \text{mod}(i-j,6)=0.
          \end{cases}
\end{equation}
Sensors with higher indices will have a slightly higher variance. The transmission error probabilities are $\varepsilon_n=0.02\left\lceil\frac{n-1}{10}\right\rceil$. The filter is initialized at step 0 with state $\hat{\mathbf{x}}(0)=\mathbf{x}(0)=0$, and $\bm{\psi}(0)=\mathbf{I}$.

In addition to the baseline scheduler, we also consider the well-known \gls{maf} scheduler as a benchmark. If we denote the age of the last received packet from sensor $n$ as $\Delta_n$, the scheduler always picks the sensor with the highest age:
\begin{equation}
  a_{\text{MAF}}^*(t)=\argmax_{n\in\{1,\ldots,N\}} \Delta_{n}(t).
\end{equation}

Finally, we consider four different summary statistics, as well as the state \gls{mse}: aside from the sample mean and variance, we consider the maximum and count statistics, denoted as $z_{\max}=\max_n x_n(t)$ and $z_{\text{cnt}}$, which is given by:
\begin{align}
  z_{\text{cnt}}(\mathbf{x})&=\sum_{n=1}^N \mathbbm{1}(x_n(t)-a)\mathbbm{1}(b-x_n(t)),
\end{align}
where $\mathbbm{1}(x)$ is the step function, equal to 1 if $x\geq0$ and 0 otherwise, and the count interval $[a,b]$. In other words, the count statistic is a simple count of the number of state components that are within $[a,b]$, which we set to $[-5,5]$.

\begin{figure*}
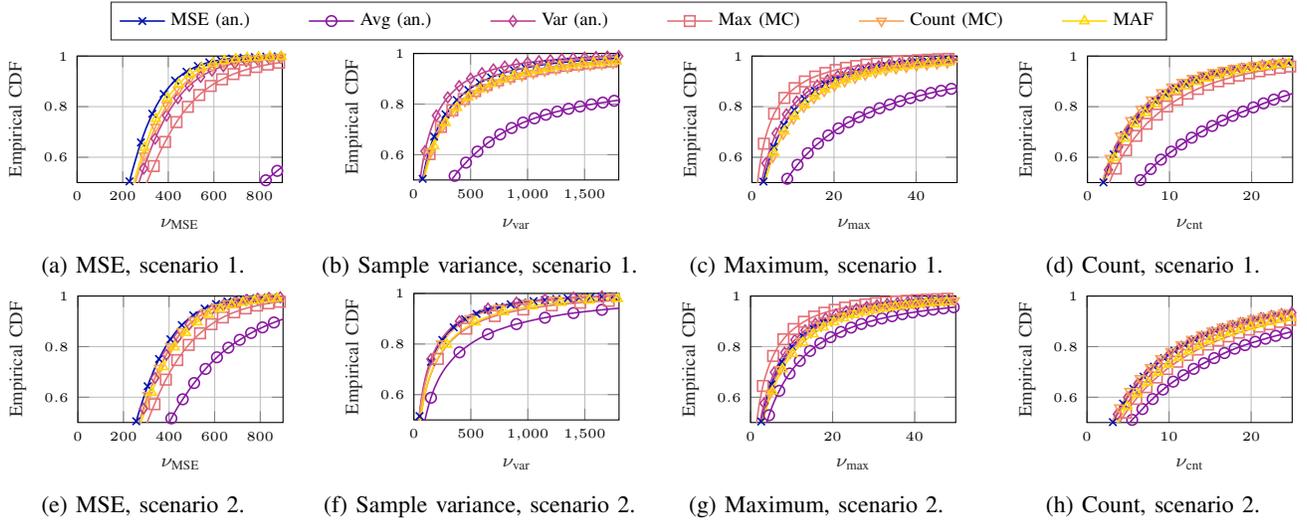

\centering
    \begin{subfigure}[b]{.9\linewidth}
	    \centering
        \input{./fig/legend_scheme.tex}
        \label{fig:legend}
    \end{subfigure}	\\
    \begin{subfigure}[b]{.24\linewidth}
	    \centering
        \input{./fig/mse_err.tex}
        \caption{MSE, scenario 1.}
        \label{fig:rmse}
    \end{subfigure}	
	\begin{subfigure}[b]{.24\linewidth}
	    \centering
        \input{./fig/var_err.tex}
        \caption{Sample variance, scenario 1.}
        \label{fig:var}
    \end{subfigure}		
    \begin{subfigure}[b]{.24\linewidth}
	    \centering
        \input{./fig/max_err.tex}
        \caption{Maximum, scenario 1.}
        \label{fig:max}
    \end{subfigure}		
    \begin{subfigure}[b]{.24\linewidth}
	    \centering
        \input{./fig/cnt_err.tex}
        \caption{Count, scenario 1.}
        \label{fig:cnt}
    \end{subfigure}	
    \centering
    \begin{subfigure}[b]{.24\linewidth}
	    \centering
        \input{./fig/mse_err2.tex}
        \caption{MSE, scenario 2.}
        \label{fig:mse2}
    \end{subfigure}	
	\begin{subfigure}[b]{.24\linewidth}
	    \centering
        \input{./fig/var_err2.tex}
        \caption{Sample variance, scenario 2.}
        \label{fig:var2}
    \end{subfigure}		
    \begin{subfigure}[b]{.24\linewidth}
	    \centering
        \input{./fig/max_err2.tex}
        \caption{Maximum, scenario 2.}
        \label{fig:max2}
    \end{subfigure}		
    \begin{subfigure}[b]{.24\linewidth}
	    \centering
        \input{./fig/cnt_err2.tex}
        \caption{Count, scenario 2.}
        \label{fig:cnt2}
    \end{subfigure}	
     \caption{Estimation error distributions with the different strategies.\vspace{-0.3cm}}
 \label{fig:err}
\end{figure*}

\subsection{Results}

We can first look at the choices \edit{of the schemes aimed at each target metric} in one of the episodes for each scenario, shown in Fig.~\ref{fig:selection}. As expected, the \gls{maf} scheduler selects sensors with a similar frequency in both scenarios: sensors with an index over 10 are selected slightly more frequently, as transmission errors occur more often, but the difference is small. On the other hand, the average scheduler only selects two sensors, 7 and 14, in the first scenario, and only the last, sensor 20, in the second: this holds throughout all episodes, independently from the state of the system. In the first scenario, alternating between these two sensors gives the best estimate of the overall average, as the state of each of these two sensors only depends on the other's. In the second scenario, no sensor is isolated, but sensor 20 is the one that affects the average the most. The average scheduler then gets the best estimate it can for the other values, concentrating on these sensors and actually getting a better average performance. Naturally, this results in a significantly worse performance when looking at any other summary statistic. We also remark that all other policies excluding \gls{maf} never choose sensors 7 and 14 in the first scenario: as errors compound for most of these summary statistics, it does not make sense to choose isolated sensors, as sensors that are more correlated to their neighbors have a better chance to reduce the overall error.
We can also note that, in the second scenario, the \gls{mse}, count, and sample variance schedulers often make similar choices, while they do not in the first scenario: this similarity is purely due to the specific features of the system, and cannot be relied upon for design.

The \glspl{cdf} of the quadratic estimation errors $\nu_z(t)$ obtained for the various summary statistics are shown in Fig.~\ref{fig:err}. As expected, optimizing the scheduling for a given summary statistic can reduce the error on it, for all the considered statistics. We did not plot the average statistic, as all policies had a similar performance, although, as expected, the average scheduler performed best for that statistic. However, $\pi^*_{\text{avg}}$ was almost always the worst policy when looking at other summary statistics, often by a wide margin, in both scenarios, for the reasons we explained above. The maximum scheduler $\pi^*_{\max}$ was also noticeably worse when looking at the state \gls{mse} or the interval count, as it tended to pick sensors with a high value, often accepting a larger error on other components of the state. As the maximum value was almost always far over the interval boundaries, the count statistic was also negatively affected. 
Finally, the similarity in the behavior of the count, sample variance, and the \gls{mse} scheduler in the second scenario has a similar performance, as Fig.~\ref{fig:mse2}-\subref{fig:cnt2} show.

In general, the average and count statistics tend to be relatively insensitive to the scheduling policy used, with most policies showing similar results: in both cases, estimation errors tend to compensate, and the error is relatively low. On the other hand, the gain from using the appropriate scheduling strategy is clearly noticeable when looking at the \gls{mse} and at the maximum statistic. In these cases, individual components of the state can have a disproportionate effects, and errors tend to compound rather than compensate each other. In general, while never being the optimum, the \gls{maf} scheduler is also never the worst, as it is a purely \gls{aoi}-oriented approach that does not consider the specific definition of \gls{voi}.

\section{Conclusion}\label{sec:concl}

In this letter, we have considered the optimization of a sensor polling strategy, using different statistics to define the \gls{voi}. The difference between the policies can be important, as errors tend to compensate each other in some cases and compound for other statistics, leading to different choices of sensors. Naturally, one-step optimization is a limited approach, and we plan to consider more complex schemes which can take long-term effects into account, as well as different statistics over the same process. \edit{Energy consumption is also another important metric, and we plan to compare \gls{voi}-based strategies to energy-efficient ones and try to find a balance between them.}

\bibliographystyle{IEEEtran}
\bibliography{bibliography}

\end{document}

%% file: fig/scenario.tex
\begin{tikzpicture}
  
\node (bs) at (0,-0.6) {\includegraphics[width=1.2cm]{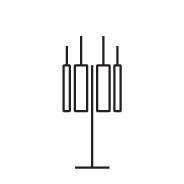}};
\node (srv1) at (0,-1.2) {\includegraphics[width=1.2cm]{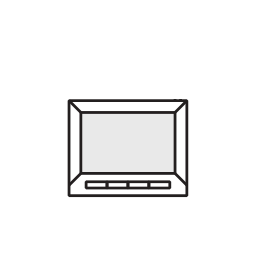}};
\node (srv2) at (3,-1.2) {\includegraphics[width=1.2cm]{fig/srv.png}};
\node (label1) at (0,-1.8) {\scriptsize Edge node estimator};
\node (label2) at (3,-1.8) {\scriptsize Remote server};
\node (label3) at (0,1.5) {\scriptsize Sensors};
\node (s1) at (-1,-0.5) {\includegraphics[width=0.8cm]{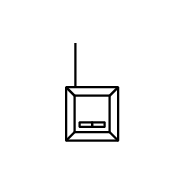}};
\node (s2) at (-1.5,0.5) {\includegraphics[width=0.8cm]{fig/sensor.png}};
\node (s3) at (0.4,1) {\includegraphics[width=0.8cm]{fig/sensor.png}};
\node (s4) at (1,0.6) {\includegraphics[width=0.8cm]{fig/sensor.png}};
\node (s5) at (1.2,-0.2) {\includegraphics[width=0.8cm]{fig/sensor.png}};
\node (s6) at (-0.5,0.9) {\includegraphics[width=0.8cm]{fig/sensor.png}};

\draw[->] ([xshift=0.3cm]srv2.west) to node[midway,above]{\scriptsize Queries} ([xshift=-0.3cm]srv1.east);

\end{tikzpicture}

%% file: fig/selection_bar_inv.tex
\pgfplotstableread{
1	0	0	0.00500000000000000	0	0.118000000000000	0.0290000000000000	0	0.147000000000000	0.0270000000000000	0.167000000000000	0.0200000000000000	0	0.0180000000000000	0	0.0260000000000000	0.0970000000000000	0	0.156000000000000	0.0110000000000000	0.179000000000000
2	0	0	0	0	0	0	0.25	0	0	0	0	0	0	0.25	0	0	0	0	0	0
3	0.0440000000000000	0.0410000000000000	0.0470000000000000	0.0440000000000000	0.0840000000000000	0.0520000000000000	0	0.0660000000000000	0.0500000000000000	0.0810000000000000	0.0690000000000000	0.0340000000000000	0.0410000000000000	0	0.0450000000000000	0.0510000000000000	0.0490000000000000	0.0490000000000000	0.0520000000000000	0.101000000000000
4	0.0220000000000000	0.0200000000000000	0.0400000000000000	0.0360000000000000	0.0840000000000000	0.0640000000000000	0	0.0570000000000000	0.0560000000000000	0.0880000000000000	0.0200000000000000	0.0250000000000000	0.0410000000000000	0	0.0630000000000000	0.0680000000000000	0.0540000000000000	0.0850000000000000	0.0850000000000000	0.0920000000000000
5	0.0590000000000000	0.0370000000000000	0.0600000000000000	0.0440000000000000	0.0790000000000000	0.0440000000000000	0	0.0570000000000000	0.0440000000000000	0.0560000000000000	0.0730000000000000	0.0390000000000000	0.0620000000000000	0	0.0540000000000000	0.0560000000000000	0.0430000000000000	0.0700000000000000	0.0440000000000000	0.0790000000000000
6	0.0510000000000000	0.0490000000000000	0.0490000000000000	0.0510000000000000	0.0500000000000000	0.0500000000000000	0.0500000000000000	0.0510000000000000	0.0500000000000000	0.0510000000000000	0.0520000000000000	0.0510000000000000	0.0500000000000000	0.0510000000000000	0.0500000000000000	0.0490000000000000	0.0500000000000000	0.0490000000000000	0.0480000000000000	0.0480000000000000
}\dataseta

\begin{tikzpicture}
\pgfplotsset{every tick label/.append style={font=\tiny}}
\tikzstyle{dotted}= [dash pattern=on \pgflinewidth off 0.5mm] 
\tikzstyle{dashed}= [dash pattern=on 7.5*0.8*0.8pt off 7.5*0.4*0.8pt]
\tikzstyle{dashdotted} = [dash pattern=on 7.5*0.8*0.6pt off 7.5*0.8*0.3pt on \the\pgflinewidth off 7.5*0.8*0.3pt]
\tikzstyle{dotted2} = [dash pattern=on 7.5*0.8*0.3pt off 7.5*0.8*0.2pt]

\begin{axis}[ybar=0pt,
        name=ax1,
        width=\fswidth,
        height=\fsheight,
        enlarge x limits=0.1, 
        ymin=0,
        tick pos=left,
        ymax=0.25,        
        xlabel style={font=\scriptsize\color{white!15!black}},
        ylabel near ticks,
        ylabel style={font=\scriptsize\color{white!15!black}},
        ylabel={Frequency},
        xlabel={Scheduler selection (scenario 1)},
        xtick=data,
        xticklabels={{MSE (an.)}, {Avg (an.)}, {Var (an.)}, {Max (MC)}, {Cnt (MC)}, {MAF}},
        ymajorgrids,
        xmajorgrids,
        minor x tick num=1,
        minor tick length=1ex,
        scatter/position=absolute,
        node near coords style={
            at={(axis cs:\pgfkeysvalueof{/data point/x},\pgfkeysvalueof{/pgfplots/ymin})},
            anchor=north
        },
        bar width=0.1cm,
        ]
\addplot[draw=black,fill=msecol,nodes near coords={\tiny 1}] table[x index=0,y index=1] \dataseta; 
\addplot[draw=black,fill=white!5!msecol] table[x index=0,y index=2] \dataseta; 
\addplot[draw=black,fill=white!10!msecol] table[x index=0,y index=3] \dataseta; 
\addplot[draw=black,fill=white!15!msecol] table[x index=0,y index=4] \dataseta; 
\addplot[draw=black,fill=white!20!msecol,nodes near coords={\tiny 5}] table[x index=0,y index=5] \dataseta; 
\addplot[draw=black,fill=white!25!msecol] table[x index=0,y index=6] \dataseta; 
\addplot[draw=black,fill=white!30!msecol] table[x index=0,y index=7] \dataseta; 
\addplot[draw=black,fill=white!35!msecol] table[x index=0,y index=8] \dataseta; 
\addplot[draw=black,fill=white!40!msecol] table[x index=0,y index=9] \dataseta; 
\addplot[draw=black,fill=white!45!msecol,nodes near coords={\tiny 10}] table[x index=0,y index=10] \dataseta; 
\addplot[draw=black,fill=white!50!msecol] table[x index=0,y index=11] \dataseta; 
\addplot[draw=black,fill=white!55!msecol] table[x index=0,y index=12] \dataseta; 
\addplot[draw=black,fill=white!60!msecol] table[x index=0,y index=13] \dataseta; 
\addplot[draw=black,fill=white!65!msecol] table[x index=0,y index=14] \dataseta; 
\addplot[draw=black,fill=white!70!msecol,nodes near coords={\tiny 15}] table[x index=0,y index=15] \dataseta; 
\addplot[draw=black,fill=white!75!msecol] table[x index=0,y index=16] \dataseta; 
\addplot[draw=black,fill=white!80!msecol] table[x index=0,y index=17] \dataseta; 
\addplot[draw=black,fill=white!85!msecol] table[x index=0,y index=18] \dataseta; 
\addplot[draw=black,fill=white!90!msecol] table[x index=0,y index=19] \dataseta; 
\addplot[draw=black,fill=white!95!msecol,nodes near coords={\tiny 20}] table[x index=0,y index=20] \dataseta; 

\end{axis}

\end{tikzpicture}

%% file: fig/selection_bar_inv2.tex
\pgfplotstableread{
1	0.0990000000000000	0	0.00300000000000000	0.0750000000000000	0	0.169000000000000	0	0.00400000000000000	0.107000000000000	0.0430000000000000	0.0470000000000000	0	0	0.0970000000000000	0	0.174000000000000	0.0100000000000000	0.00200000000000000	0.110000000000000	0.0600000000000000
2	0	0	0	0	0	0	0	0	0	0	0	0	0	0	0	0	0	0	0	0.25
3	0.176000000000000	0	0.00700000000000000	0.0850000000000000	0.00200000000000000	0.150000000000000	0	0.00300000000000000	0.106000000000000	0.0130000000000000	0.0810000000000000	0	0.00200000000000000	0.0460000000000000	0.00100000000000000	0.232000000000000	0.0150000000000000	0	0.0640000000000000	0.0170000000000000
4	0.0170000000000000	0	0.0260000000000000	0.0700000000000000	0.0370000000000000	0.0810000000000000	0.00700000000000000	0.0970000000000000	0.0520000000000000	0.0710000000000000	0.00700000000000000	0.0630000000000000	0.00200000000000000	0.00500000000000000	0.0250000000000000	0.241000000000000	0.0320000000000000	0.0120000000000000	0.0430000000000000	0.112000000000000
5	0.171000000000000	0.0140000000000000	0.0270000000000000	0.0280000000000000	0.0270000000000000	0.146000000000000	0.0120000000000000	0.00200000000000000	0.107000000000000	0.0160000000000000	0.0960000000000000	0.00200000000000000	0.0110000000000000	0.0780000000000000	0.0190000000000000	0.0660000000000000	0.0420000000000000	0.00600000000000000	0.0820000000000000	0.0480000000000000
6	0.0500000000000000	0.0490000000000000	0.0520000000000000	0.0490000000000000	0.0490000000000000	0.0510000000000000	0.0490000000000000	0.0500000000000000	0.0490000000000000	0.0500000000000000	0.0530000000000000	0.0510000000000000	0.0500000000000000	0.0480000000000000	0.0490000000000000	0.0500000000000000	0.0500000000000000	0.0480000000000000	0.0520000000000000	0.0510000000000000
}\dataseta

\begin{tikzpicture}
\pgfplotsset{every tick label/.append style={font=\tiny}}
\tikzstyle{dotted}= [dash pattern=on \pgflinewidth off 0.5mm] 
\tikzstyle{dashed}= [dash pattern=on 7.5*0.8*0.8pt off 7.5*0.4*0.8pt]
\tikzstyle{dashdotted} = [dash pattern=on 7.5*0.8*0.6pt off 7.5*0.8*0.3pt on \the\pgflinewidth off 7.5*0.8*0.3pt]
\tikzstyle{dotted2} = [dash pattern=on 7.5*0.8*0.3pt off 7.5*0.8*0.2pt]

\begin{axis}[ybar=0pt,
        name=ax1,
        width=\fswidth,
        height=\fsheight,
        enlarge x limits=0.1, 
        ymin=0,
        ymax=0.25,      
        tick pos=left,
        xlabel style={font=\scriptsize\color{white!15!black}},
        ylabel near ticks,
        ylabel style={font=\scriptsize\color{white!15!black}},
        ylabel={Frequency},
        xlabel={Scheduler selection (scenario 2)},
        xtick=data,
        xticklabels={{MSE (an.)}, {Avg (an.)}, {Var (an.)}, {Max (MC)}, {Cnt (MC)}, {MAF}},
        ymajorgrids,
        xmajorgrids,
        minor x tick num=1,
        minor tick length=1ex,
        scatter/position=absolute,
        node near coords style={
            at={(axis cs:\pgfkeysvalueof{/data point/x},\pgfkeysvalueof{/pgfplots/ymin})},
            anchor=north
        },
        bar width=0.1cm,
        ]
\addplot[draw=black,fill=msecol,nodes near coords={\tiny 1}] table[x index=0,y index=1] \dataseta; 
\addplot[draw=black,fill=white!5!msecol] table[x index=0,y index=2] \dataseta; 
\addplot[draw=black,fill=white!10!msecol] table[x index=0,y index=3] \dataseta; 
\addplot[draw=black,fill=white!15!msecol] table[x index=0,y index=4] \dataseta; 
\addplot[draw=black,fill=white!20!msecol,nodes near coords={\tiny 5}] table[x index=0,y index=5] \dataseta; 
\addplot[draw=black,fill=white!25!msecol] table[x index=0,y index=6] \dataseta; 
\addplot[draw=black,fill=white!30!msecol] table[x index=0,y index=7] \dataseta; 
\addplot[draw=black,fill=white!35!msecol] table[x index=0,y index=8] \dataseta; 
\addplot[draw=black,fill=white!40!msecol] table[x index=0,y index=9] \dataseta; 
\addplot[draw=black,fill=white!45!msecol,nodes near coords={\tiny 10}] table[x index=0,y index=10] \dataseta; 
\addplot[draw=black,fill=white!50!msecol] table[x index=0,y index=11] \dataseta; 
\addplot[draw=black,fill=white!55!msecol] table[x index=0,y index=12] \dataseta; 
\addplot[draw=black,fill=white!60!msecol] table[x index=0,y index=13] \dataseta; 
\addplot[draw=black,fill=white!65!msecol] table[x index=0,y index=14] \dataseta; 
\addplot[draw=black,fill=white!70!msecol,nodes near coords={\tiny 15}] table[x index=0,y index=15] \dataseta; 
\addplot[draw=black,fill=white!75!msecol] table[x index=0,y index=16] \dataseta; 
\addplot[draw=black,fill=white!80!msecol] table[x index=0,y index=17] \dataseta; 
\addplot[draw=black,fill=white!85!msecol] table[x index=0,y index=18] \dataseta; 
\addplot[draw=black,fill=white!90!msecol] table[x index=0,y index=19] \dataseta; 
\addplot[draw=black,fill=white!95!msecol,nodes near coords={\tiny 20}] table[x index=0,y index=20] \dataseta; 

\end{axis}

\end{tikzpicture}

%% file: fig/legend_scheme.tex
\begin{tikzpicture}

\begin{axis}[
    width=0,
    height=0,
    at={(0,0)},
    scale only axis,
    xmin=0,
    xmax=0,
    xtick={},
    ymin=0,
    ymax=0,
    ytick={},
    axis background/.style={fill=white},
    legend style={legend cell align=center, align=center, draw=white!15!black, font=\scriptsize, at={(0, 0)}, anchor=center, /tikz/every even column/.append style={column sep=2em}},
    legend columns=10,
]
\addplot [color=msecol, semithick,mark=x]
table {%
0 1
};
\addlegendentry{MSE (an.)}
\addplot [color=avgcol,semithick,mark=o]
table {%
0 1
};
\addlegendentry{Avg (an.)}
\addplot [color=varcol,semithick,mark=diamond]
table {%
0 1
};
\addlegendentry{Var (an.)}
\addplot [color=maxcol,semithick,mark=square]
table {%
0 1
};
\addlegendentry{Max (MC)}
\addplot [color=cntcol,semithick,mark=triangle,mark options={solid,rotate=180}]
table {%
0 1
};
\addlegendentry{Count (MC)}
\addplot [color=mafcol, semithick, mark=triangle]
table {%
0 1
};
\addlegendentry{MAF}

\end{axis}

\end{tikzpicture}

%% file: fig/var_err.tex
\begin{tikzpicture}
\pgfplotsset{every tick label/.append style={font=\tiny}}
\tikzstyle{dotted}= [dash pattern=on \pgflinewidth off 0.5mm] 
\tikzstyle{dashed}= [dash pattern=on 7.5*0.8*0.8pt off 7.5*0.4*0.8pt]
\tikzstyle{dashdotted} = [dash pattern=on 7.5*0.8*0.6pt off 7.5*0.8*0.3pt on \the\pgflinewidth off 7.5*0.8*0.3pt]
\tikzstyle{dotted2} = [dash pattern=on 7.5*0.8*0.3pt off 7.5*0.8*0.2pt]

\begin{axis}[%
width=\fwidth,
height=\fheight,
at={(0,0)},
xmin=0,
xmax=1800,
xlabel near ticks,
xlabel style={font=\scriptsize\color{white!15!black}},
xlabel={$\nu_{\text{var}}$},
ylabel near ticks,
ymin=0.5,
ymax=1,
ylabel style={font=\scriptsize\color{white!15!black}},
ylabel={Empirical CDF},axis background/.style={fill=white},
xmajorgrids,
ymajorgrids
]
\addplot [color=msecol, semithick,mark=x,mark repeat=10,mark phase=1]
  table[row sep=crcr]{%
0	0\\
10	0.19707\\
20	0.27606\\
30	0.33214\\
40	0.37765\\
50	0.41673\\
60	0.44924\\
70	0.47911\\
80	0.50458\\
90	0.52825\\
100	0.55048\\
110	0.57051\\
120	0.58837\\
130	0.6052\\
140	0.61991\\
150	0.63468\\
160	0.64813\\
170	0.66045\\
180	0.67176\\
190	0.68303\\
200	0.69357\\
210	0.70301\\
220	0.71278\\
230	0.72184\\
240	0.73052\\
250	0.7379\\
260	0.74559\\
270	0.7525\\
280	0.75962\\
290	0.76635\\
300	0.77271\\
310	0.77906\\
320	0.78513\\
330	0.79091\\
340	0.79651\\
350	0.80136\\
360	0.80678\\
370	0.81183\\
380	0.81636\\
390	0.82067\\
400	0.82497\\
410	0.82933\\
420	0.83345\\
430	0.83748\\
440	0.84142\\
450	0.8453\\
460	0.8487\\
470	0.85224\\
480	0.85568\\
490	0.85906\\
500	0.86247\\
510	0.86554\\
520	0.86831\\
530	0.87098\\
540	0.87383\\
550	0.87645\\
560	0.87906\\
570	0.88164\\
580	0.88424\\
590	0.88683\\
600	0.88918\\
610	0.89137\\
620	0.89368\\
630	0.89572\\
640	0.89767\\
650	0.89972\\
660	0.90182\\
670	0.90363\\
680	0.9052\\
690	0.90725\\
700	0.90907\\
710	0.91067\\
720	0.91236\\
730	0.91402\\
740	0.91556\\
750	0.91705\\
760	0.91854\\
770	0.91998\\
780	0.92155\\
790	0.92296\\
800	0.92411\\
810	0.92543\\
820	0.92666\\
830	0.928\\
840	0.92919\\
850	0.93035\\
860	0.93138\\
870	0.93257\\
880	0.93373\\
890	0.93484\\
900	0.9358\\
910	0.93671\\
920	0.9378\\
930	0.93879\\
940	0.93956\\
950	0.94046\\
960	0.94141\\
970	0.94221\\
980	0.9432\\
990	0.94405\\
1000	0.94488\\
1010	0.94572\\
1020	0.94654\\
1030	0.94748\\
1040	0.94826\\
1050	0.94916\\
1060	0.94999\\
1070	0.95066\\
1080	0.95151\\
1090	0.95229\\
1100	0.95308\\
1110	0.95374\\
1120	0.95443\\
1130	0.95502\\
1140	0.95575\\
1150	0.95643\\
1160	0.95685\\
1170	0.95739\\
1180	0.95799\\
1190	0.95852\\
1200	0.95912\\
1210	0.95967\\
1220	0.96032\\
1230	0.96083\\
1240	0.96129\\
1250	0.96173\\
1260	0.9622\\
1270	0.96264\\
1280	0.96312\\
1290	0.96359\\
1300	0.96412\\
1310	0.96462\\
1320	0.96508\\
1330	0.96559\\
1340	0.96607\\
1350	0.96653\\
1360	0.96699\\
1370	0.96742\\
1380	0.96779\\
1390	0.96825\\
1400	0.96857\\
1410	0.96901\\
1420	0.96955\\
1430	0.96998\\
1440	0.97035\\
1450	0.97075\\
1460	0.97102\\
1470	0.97135\\
1480	0.97173\\
1490	0.97217\\
1500	0.97248\\
1510	0.97277\\
1520	0.97315\\
1530	0.97349\\
1540	0.97379\\
1550	0.97411\\
1560	0.9745\\
1570	0.97484\\
1580	0.97509\\
1590	0.9754\\
1600	0.97571\\
1610	0.976\\
1620	0.97625\\
1630	0.97658\\
1640	0.97687\\
1650	0.97713\\
1660	0.97735\\
1670	0.97758\\
1680	0.97785\\
1690	0.97818\\
1700	0.97847\\
1710	0.9787\\
1720	0.97894\\
1730	0.97918\\
1740	0.97949\\
1750	0.9797\\
1760	0.97991\\
1770	0.98016\\
1780	0.98042\\
1790	0.98065\\
1800	0.98092\\
1810	0.98113\\
1820	0.98136\\
1830	0.98161\\
1840	0.98185\\
1850	0.98205\\
1860	0.98223\\
1870	0.98238\\
1880	0.98265\\
1890	0.98289\\
1900	0.9831\\
1910	0.98326\\
1920	0.98347\\
1930	0.98362\\
1940	0.98373\\
1950	0.9839\\
1960	0.98405\\
1970	0.98421\\
1980	0.9844\\
1990	0.98459\\
2000	0.98485\\
2010	0.98504\\
2020	0.98516\\
2030	0.98536\\
2040	0.98555\\
2050	0.98571\\
2060	0.98587\\
2070	0.98596\\
2080	0.98614\\
2090	0.98625\\
2100	0.98642\\
2110	0.98657\\
2120	0.98669\\
2130	0.98685\\
2140	0.98698\\
2150	0.98715\\
2160	0.98728\\
2170	0.9874\\
2180	0.9875\\
2190	0.98759\\
2200	0.98774\\
2210	0.98789\\
2220	0.98803\\
2230	0.98813\\
2240	0.98824\\
2250	0.98838\\
2260	0.98848\\
2270	0.98857\\
2280	0.98866\\
2290	0.98879\\
2300	0.98886\\
2310	0.98898\\
2320	0.98911\\
2330	0.98921\\
2340	0.98934\\
2350	0.98941\\
2360	0.98954\\
2370	0.98971\\
2380	0.98984\\
2390	0.98997\\
2400	0.99014\\
2410	0.99027\\
2420	0.99036\\
2430	0.99046\\
2440	0.99064\\
2450	0.99072\\
2460	0.99083\\
2470	0.99091\\
2480	0.99101\\
2490	0.9911\\
2500	0.99115\\
2510	0.9912\\
2520	0.99123\\
2530	0.99131\\
2540	0.99139\\
2550	0.99145\\
2560	0.99154\\
2570	0.99163\\
2580	0.99169\\
2590	0.9918\\
2600	0.99189\\
2610	0.99193\\
2620	0.992\\
2630	0.99207\\
2640	0.99215\\
2650	0.99227\\
2660	0.99235\\
2670	0.99241\\
2680	0.99246\\
2690	0.99251\\
2700	0.99258\\
2710	0.99264\\
2720	0.99271\\
2730	0.99278\\
2740	0.99286\\
2750	0.99291\\
2760	0.99296\\
2770	0.99304\\
2780	0.99307\\
2790	0.99313\\
2800	0.9932\\
2810	0.99322\\
2820	0.99328\\
2830	0.99332\\
2840	0.99337\\
2850	0.99342\\
2860	0.99349\\
2870	0.99353\\
2880	0.9936\\
2890	0.99363\\
2900	0.99369\\
2910	0.99374\\
2920	0.99379\\
2930	0.99383\\
2940	0.9939\\
2950	0.99392\\
2960	0.99398\\
2970	0.99402\\
2980	0.99406\\
2990	0.99413\\
3000	0.9942\\
3010	0.99426\\
3020	0.99429\\
3030	0.99439\\
3040	0.99445\\
3050	0.99451\\
3060	0.99453\\
3070	0.99458\\
3080	0.99459\\
3090	0.99463\\
3100	0.99465\\
3110	0.99466\\
3120	0.99473\\
3130	0.99477\\
3140	0.99484\\
3150	0.99489\\
3160	0.99496\\
3170	0.995\\
3180	0.99512\\
3190	0.99517\\
3200	0.99521\\
3210	0.99525\\
3220	0.99527\\
3230	0.99528\\
3240	0.99534\\
3250	0.99537\\
3260	0.99542\\
3270	0.99544\\
3280	0.99547\\
3290	0.99549\\
3300	0.99552\\
3310	0.99557\\
3320	0.99562\\
3330	0.99562\\
3340	0.99565\\
3350	0.9957\\
3360	0.99573\\
3370	0.99578\\
3380	0.99581\\
3390	0.99583\\
3400	0.99588\\
3410	0.99588\\
3420	0.9959\\
3430	0.99595\\
3440	0.99603\\
3450	0.99606\\
3460	0.9961\\
3470	0.99611\\
3480	0.99614\\
3490	0.9962\\
3500	0.99624\\
3510	0.99624\\
3520	0.99627\\
3530	0.99632\\
3540	0.99633\\
3550	0.99635\\
3560	0.99642\\
3570	0.99647\\
3580	0.99649\\
3590	0.99653\\
3600	0.99657\\
3610	0.99661\\
3620	0.99662\\
3630	0.99666\\
3640	0.99667\\
3650	0.99669\\
3660	0.99671\\
3670	0.99674\\
3680	0.99676\\
3690	0.99678\\
3700	0.9968\\
3710	0.99682\\
3720	0.99684\\
3730	0.99686\\
3740	0.9969\\
3750	0.99694\\
3760	0.99696\\
3770	0.99699\\
3780	0.99702\\
3790	0.99703\\
3800	0.99704\\
3810	0.99705\\
3820	0.99708\\
3830	0.9971\\
3840	0.99711\\
3850	0.99716\\
3860	0.99719\\
3870	0.99722\\
3880	0.99723\\
3890	0.99726\\
3900	0.99729\\
3910	0.9973\\
3920	0.99732\\
3930	0.99733\\
3940	0.99735\\
3950	0.99735\\
3960	0.99736\\
3970	0.99738\\
3980	0.99741\\
3990	0.99743\\
4000	0.99744\\
4010	0.99746\\
4020	0.99749\\
4030	0.99752\\
4040	0.99752\\
4050	0.99753\\
4060	0.99756\\
4070	0.99758\\
4080	0.99761\\
4090	0.99763\\
4100	0.99769\\
4110	0.99772\\
4120	0.99775\\
4130	0.99775\\
4140	0.9978\\
4150	0.99783\\
4160	0.99783\\
4170	0.99784\\
4180	0.99784\\
4190	0.99787\\
4200	0.99789\\
4210	0.9979\\
4220	0.99793\\
4230	0.99795\\
4240	0.99796\\
4250	0.99796\\
4260	0.99797\\
4270	0.99798\\
4280	0.998\\
4290	0.998\\
4300	0.99801\\
4310	0.99803\\
4320	0.99805\\
4330	0.99806\\
4340	0.99807\\
4350	0.99808\\
4360	0.99809\\
4370	0.99812\\
4380	0.99814\\
4390	0.99815\\
4400	0.99815\\
4410	0.99815\\
4420	0.99815\\
4430	0.99815\\
4440	0.99818\\
4450	0.99821\\
4460	0.99822\\
4470	0.99824\\
4480	0.99826\\
4490	0.99826\\
4500	0.99829\\
4510	0.99832\\
4520	0.99835\\
4530	0.99835\\
4540	0.99835\\
4550	0.99835\\
4560	0.99836\\
4570	0.99836\\
4580	0.99839\\
4590	0.99839\\
4600	0.99842\\
4610	0.99846\\
4620	0.99846\\
4630	0.99849\\
4640	0.9985\\
4650	0.9985\\
4660	0.99852\\
4670	0.99852\\
4680	0.99854\\
4690	0.99856\\
4700	0.99858\\
4710	0.99858\\
4720	0.99858\\
4730	0.99858\\
4740	0.99858\\
4750	0.99858\\
4760	0.99859\\
4770	0.9986\\
4780	0.99861\\
4790	0.99862\\
4800	0.99862\\
4810	0.99863\\
4820	0.99864\\
4830	0.99866\\
4840	0.99867\\
4850	0.99868\\
4860	0.9987\\
4870	0.9987\\
4880	0.99871\\
4890	0.99871\\
4900	0.99871\\
4910	0.99872\\
4920	0.99872\\
4930	0.99872\\
4940	0.99873\\
4950	0.99874\\
4960	0.99875\\
4970	0.99875\\
4980	0.99876\\
4990	0.99876\\
5000	0.99878\\
5010	0.99879\\
5020	0.9988\\
5030	0.9988\\
5040	0.9988\\
5050	0.99881\\
5060	0.99881\\
5070	0.99881\\
5080	0.99881\\
5090	0.99881\\
5100	0.99883\\
5110	0.99884\\
5120	0.99884\\
5130	0.99886\\
5140	0.99886\\
5150	0.99886\\
5160	0.99886\\
5170	0.99886\\
5180	0.99886\\
5190	0.99886\\
5200	1\\
};

\addplot [color=avgcol,semithick,mark=o,mark repeat=10,mark phase=3]
  table[row sep=crcr]{%
0	0\\
10	0.0944\\
20	0.13407\\
30	0.16347\\
40	0.1873\\
50	0.20869\\
60	0.2283\\
70	0.24517\\
80	0.26111\\
90	0.2763\\
100	0.29012\\
110	0.30432\\
120	0.31702\\
130	0.32906\\
140	0.34083\\
150	0.35163\\
160	0.36199\\
170	0.37241\\
180	0.38247\\
190	0.39187\\
200	0.40138\\
210	0.41036\\
220	0.41867\\
230	0.42677\\
240	0.43448\\
250	0.44227\\
260	0.44983\\
270	0.45742\\
280	0.46407\\
290	0.47097\\
300	0.47821\\
310	0.48494\\
320	0.49135\\
330	0.49764\\
340	0.5043\\
350	0.51048\\
360	0.51613\\
370	0.52261\\
380	0.52848\\
390	0.53394\\
400	0.53949\\
410	0.54411\\
420	0.54904\\
430	0.55409\\
440	0.55927\\
450	0.56401\\
460	0.56864\\
470	0.57409\\
480	0.5787\\
490	0.58334\\
500	0.58744\\
510	0.5917\\
520	0.59608\\
530	0.60006\\
540	0.60433\\
550	0.60829\\
560	0.61229\\
570	0.61634\\
580	0.62059\\
590	0.6243\\
600	0.62816\\
610	0.63184\\
620	0.63561\\
630	0.63882\\
640	0.64239\\
650	0.64596\\
660	0.64937\\
670	0.6528\\
680	0.65577\\
690	0.65929\\
700	0.66269\\
710	0.66527\\
720	0.66849\\
730	0.67138\\
740	0.67448\\
750	0.67748\\
760	0.68023\\
770	0.68319\\
780	0.68585\\
790	0.68853\\
800	0.69098\\
810	0.69329\\
820	0.69593\\
830	0.69814\\
840	0.70061\\
850	0.70317\\
860	0.70551\\
870	0.70773\\
880	0.7098\\
890	0.71192\\
900	0.71421\\
910	0.71626\\
920	0.71844\\
930	0.72045\\
940	0.72249\\
950	0.72473\\
960	0.72694\\
970	0.72874\\
980	0.7304\\
990	0.7319\\
1000	0.73355\\
1010	0.73533\\
1020	0.73706\\
1030	0.73859\\
1040	0.74034\\
1050	0.74201\\
1060	0.74369\\
1070	0.74544\\
1080	0.74698\\
1090	0.7483\\
1100	0.74972\\
1110	0.75121\\
1120	0.75261\\
1130	0.75405\\
1140	0.75528\\
1150	0.75641\\
1160	0.75788\\
1170	0.7592\\
1180	0.76064\\
1190	0.76205\\
1200	0.76314\\
1210	0.7645\\
1220	0.7657\\
1230	0.76691\\
1240	0.76811\\
1250	0.7693\\
1260	0.77048\\
1270	0.77143\\
1280	0.77242\\
1290	0.77376\\
1300	0.77502\\
1310	0.77614\\
1320	0.77716\\
1330	0.77826\\
1340	0.77943\\
1350	0.78042\\
1360	0.78136\\
1370	0.78238\\
1380	0.78331\\
1390	0.78415\\
1400	0.78496\\
1410	0.78587\\
1420	0.78682\\
1430	0.78762\\
1440	0.78862\\
1450	0.78961\\
1460	0.79044\\
1470	0.79126\\
1480	0.79198\\
1490	0.79308\\
1500	0.79384\\
1510	0.79475\\
1520	0.79553\\
1530	0.79638\\
1540	0.79725\\
1550	0.798\\
1560	0.79882\\
1570	0.79964\\
1580	0.80048\\
1590	0.80123\\
1600	0.80193\\
1610	0.80264\\
1620	0.80345\\
1630	0.8041\\
1640	0.80475\\
1650	0.80537\\
1660	0.80594\\
1670	0.80671\\
1680	0.80728\\
1690	0.80802\\
1700	0.8087\\
1710	0.80943\\
1720	0.8102\\
1730	0.81093\\
1740	0.81173\\
1750	0.81242\\
1760	0.81301\\
1770	0.81362\\
1780	0.81433\\
1790	0.815\\
1800	0.81576\\
1810	0.81636\\
1820	0.81708\\
1830	0.81778\\
1840	0.81847\\
1850	0.81919\\
1860	0.81975\\
1870	0.82027\\
1880	0.82095\\
1890	0.82151\\
1900	0.82201\\
1910	0.82262\\
1920	0.82311\\
1930	0.82369\\
1940	0.82417\\
1950	0.82493\\
1960	0.82562\\
1970	0.82635\\
1980	0.82695\\
1990	0.82749\\
2000	0.82812\\
2010	0.8287\\
2020	0.82934\\
2030	0.8299\\
2040	0.8306\\
2050	0.83108\\
2060	0.83167\\
2070	0.83229\\
2080	0.83291\\
2090	0.8335\\
2100	0.83423\\
2110	0.83486\\
2120	0.83548\\
2130	0.83601\\
2140	0.83658\\
2150	0.83716\\
2160	0.83766\\
2170	0.83824\\
2180	0.8388\\
2190	0.83948\\
2200	0.83991\\
2210	0.84048\\
2220	0.84105\\
2230	0.84157\\
2240	0.84208\\
2250	0.84262\\
2260	0.84317\\
2270	0.84371\\
2280	0.84423\\
2290	0.84463\\
2300	0.84505\\
2310	0.84559\\
2320	0.84616\\
2330	0.8466\\
2340	0.84711\\
2350	0.84767\\
2360	0.84813\\
2370	0.84864\\
2380	0.84911\\
2390	0.84962\\
2400	0.85009\\
2410	0.85066\\
2420	0.85112\\
2430	0.85162\\
2440	0.85206\\
2450	0.85259\\
2460	0.85315\\
2470	0.85361\\
2480	0.85403\\
2490	0.85438\\
2500	0.85484\\
2510	0.85529\\
2520	0.85574\\
2530	0.85618\\
2540	0.85659\\
2550	0.85708\\
2560	0.85781\\
2570	0.85832\\
2580	0.85869\\
2590	0.8591\\
2600	0.8595\\
2610	0.8599\\
2620	0.86025\\
2630	0.8607\\
2640	0.86113\\
2650	0.86153\\
2660	0.86201\\
2670	0.86248\\
2680	0.8629\\
2690	0.86331\\
2700	0.86372\\
2710	0.86421\\
2720	0.86462\\
2730	0.86514\\
2740	0.86559\\
2750	0.86593\\
2760	0.86644\\
2770	0.86678\\
2780	0.86726\\
2790	0.86765\\
2800	0.86802\\
2810	0.86839\\
2820	0.86884\\
2830	0.86931\\
2840	0.86962\\
2850	0.86991\\
2860	0.87024\\
2870	0.87058\\
2880	0.87093\\
2890	0.87135\\
2900	0.8717\\
2910	0.87202\\
2920	0.8724\\
2930	0.87284\\
2940	0.87316\\
2950	0.87352\\
2960	0.87391\\
2970	0.8743\\
2980	0.87472\\
2990	0.87507\\
3000	0.8755\\
3010	0.87593\\
3020	0.8763\\
3030	0.8767\\
3040	0.87706\\
3050	0.87744\\
3060	0.87788\\
3070	0.87817\\
3080	0.87853\\
3090	0.8789\\
3100	0.87921\\
3110	0.87955\\
3120	0.87986\\
3130	0.88021\\
3140	0.88058\\
3150	0.88099\\
3160	0.88125\\
3170	0.88161\\
3180	0.88186\\
3190	0.88224\\
3200	0.88251\\
3210	0.88291\\
3220	0.8833\\
3230	0.88369\\
3240	0.88405\\
3250	0.88443\\
3260	0.88482\\
3270	0.88515\\
3280	0.88551\\
3290	0.88581\\
3300	0.88612\\
3310	0.88652\\
3320	0.88679\\
3330	0.88713\\
3340	0.88742\\
3350	0.88769\\
3360	0.88815\\
3370	0.88854\\
3380	0.88894\\
3390	0.8892\\
3400	0.88945\\
3410	0.88977\\
3420	0.89003\\
3430	0.89039\\
3440	0.89068\\
3450	0.891\\
3460	0.89124\\
3470	0.8915\\
3480	0.89177\\
3490	0.89205\\
3500	0.89241\\
3510	0.89265\\
3520	0.89294\\
3530	0.89325\\
3540	0.89355\\
3550	0.89381\\
3560	0.89421\\
3570	0.89451\\
3580	0.89481\\
3590	0.89509\\
3600	0.89547\\
3610	0.89561\\
3620	0.89595\\
3630	0.89622\\
3640	0.8965\\
3650	0.89678\\
3660	0.89707\\
3670	0.89735\\
3680	0.89764\\
3690	0.89794\\
3700	0.89814\\
3710	0.89836\\
3720	0.89859\\
3730	0.89887\\
3740	0.89924\\
3750	0.89954\\
3760	0.89984\\
3770	0.90006\\
3780	0.90044\\
3790	0.90073\\
3800	0.90095\\
3810	0.90135\\
3820	0.90173\\
3830	0.90203\\
3840	0.90229\\
3850	0.90256\\
3860	0.90284\\
3870	0.90314\\
3880	0.90338\\
3890	0.90387\\
3900	0.90425\\
3910	0.90455\\
3920	0.90477\\
3930	0.90505\\
3940	0.90532\\
3950	0.90558\\
3960	0.90588\\
3970	0.90614\\
3980	0.90643\\
3990	0.90667\\
4000	0.90685\\
4010	0.90709\\
4020	0.90729\\
4030	0.90757\\
4040	0.9078\\
4050	0.90799\\
4060	0.90822\\
4070	0.90853\\
4080	0.90879\\
4090	0.90909\\
4100	0.90929\\
4110	0.90948\\
4120	0.90974\\
4130	0.91005\\
4140	0.91025\\
4150	0.91049\\
4160	0.91069\\
4170	0.91091\\
4180	0.9111\\
4190	0.91136\\
4200	0.91152\\
4210	0.91181\\
4220	0.91199\\
4230	0.9123\\
4240	0.91259\\
4250	0.91279\\
4260	0.91303\\
4270	0.91323\\
4280	0.91345\\
4290	0.91364\\
4300	0.91387\\
4310	0.91415\\
4320	0.91438\\
4330	0.91456\\
4340	0.9148\\
4350	0.91503\\
4360	0.91525\\
4370	0.91551\\
4380	0.91576\\
4390	0.91594\\
4400	0.91621\\
4410	0.91642\\
4420	0.91665\\
4430	0.91684\\
4440	0.91707\\
4450	0.91727\\
4460	0.91745\\
4470	0.9176\\
4480	0.91777\\
4490	0.91796\\
4500	0.91808\\
4510	0.91819\\
4520	0.91834\\
4530	0.91844\\
4540	0.91862\\
4550	0.91875\\
4560	0.91896\\
4570	0.91917\\
4580	0.91937\\
4590	0.91958\\
4600	0.91969\\
4610	0.91991\\
4620	0.92008\\
4630	0.92027\\
4640	0.92047\\
4650	0.92066\\
4660	0.9209\\
4670	0.92103\\
4680	0.92122\\
4690	0.9214\\
4700	0.92153\\
4710	0.92164\\
4720	0.92176\\
4730	0.9219\\
4740	0.92204\\
4750	0.92228\\
4760	0.92236\\
4770	0.92253\\
4780	0.92267\\
4790	0.92283\\
4800	0.92301\\
4810	0.92318\\
4820	0.92336\\
4830	0.9235\\
4840	0.92371\\
4850	0.92387\\
4860	0.92401\\
4870	0.92425\\
4880	0.92446\\
4890	0.9246\\
4900	0.92475\\
4910	0.92494\\
4920	0.92506\\
4930	0.92524\\
4940	0.92538\\
4950	0.92546\\
4960	0.92559\\
4970	0.92575\\
4980	0.92599\\
4990	0.92607\\
5000	0.92628\\
5010	0.92652\\
5020	0.92668\\
5030	0.92685\\
5040	0.92695\\
5050	0.92713\\
5060	0.92738\\
5070	0.92763\\
5080	0.92781\\
5090	0.92799\\
5100	0.92815\\
5110	0.92831\\
5120	0.92849\\
5130	0.92877\\
5140	0.92897\\
5150	0.92914\\
5160	0.9293\\
5170	0.92951\\
5180	0.92967\\
5190	0.92988\\
5200	1\\
};

\addplot [color=varcol,semithick,mark=diamond,mark repeat=10,mark phase=5]
  table[row sep=crcr]{%
0	0\\
10	0.22906\\
20	0.31871\\
30	0.38199\\
40	0.43272\\
50	0.47514\\
60	0.5102\\
70	0.54082\\
80	0.56811\\
90	0.59216\\
100	0.61396\\
110	0.63405\\
120	0.65185\\
130	0.66779\\
140	0.68366\\
150	0.69771\\
160	0.71112\\
170	0.72323\\
180	0.7339\\
190	0.74502\\
200	0.75475\\
210	0.76424\\
220	0.77296\\
230	0.78102\\
240	0.78868\\
250	0.79574\\
260	0.80246\\
270	0.80896\\
280	0.81527\\
290	0.82096\\
300	0.82649\\
310	0.83168\\
320	0.83696\\
330	0.842\\
340	0.84635\\
350	0.8508\\
360	0.85515\\
370	0.85942\\
380	0.86349\\
390	0.86762\\
400	0.87104\\
410	0.87469\\
420	0.87792\\
430	0.88113\\
440	0.88454\\
450	0.88752\\
460	0.89012\\
470	0.89293\\
480	0.89552\\
490	0.89829\\
500	0.9008\\
510	0.9033\\
520	0.90583\\
530	0.90802\\
540	0.91005\\
550	0.91204\\
560	0.91424\\
570	0.9164\\
580	0.9184\\
590	0.92023\\
600	0.92209\\
610	0.92388\\
620	0.9256\\
630	0.92721\\
640	0.92872\\
650	0.93013\\
660	0.93146\\
670	0.93279\\
680	0.93424\\
690	0.9356\\
700	0.93697\\
710	0.93846\\
720	0.93973\\
730	0.94078\\
740	0.94199\\
750	0.94315\\
760	0.94436\\
770	0.94537\\
780	0.94636\\
790	0.94752\\
800	0.94842\\
810	0.94945\\
820	0.95034\\
830	0.95125\\
840	0.95217\\
850	0.95306\\
860	0.95402\\
870	0.95482\\
880	0.95562\\
890	0.95639\\
900	0.95712\\
910	0.95808\\
920	0.95874\\
930	0.95929\\
940	0.96009\\
950	0.96079\\
960	0.96152\\
970	0.96226\\
980	0.96285\\
990	0.96345\\
1000	0.9641\\
1010	0.9647\\
1020	0.96521\\
1030	0.96573\\
1040	0.96644\\
1050	0.96706\\
1060	0.96751\\
1070	0.96807\\
1080	0.96851\\
1090	0.96904\\
1100	0.96962\\
1110	0.97015\\
1120	0.97058\\
1130	0.97097\\
1140	0.97151\\
1150	0.97198\\
1160	0.9724\\
1170	0.97288\\
1180	0.97324\\
1190	0.97363\\
1200	0.97407\\
1210	0.97456\\
1220	0.9749\\
1230	0.97522\\
1240	0.97564\\
1250	0.97606\\
1260	0.97642\\
1270	0.97684\\
1280	0.97716\\
1290	0.97742\\
1300	0.9778\\
1310	0.97808\\
1320	0.97843\\
1330	0.97873\\
1340	0.97898\\
1350	0.9793\\
1360	0.97965\\
1370	0.97995\\
1380	0.98029\\
1390	0.98053\\
1400	0.98077\\
1410	0.98111\\
1420	0.98141\\
1430	0.98174\\
1440	0.98201\\
1450	0.98239\\
1460	0.98272\\
1470	0.98297\\
1480	0.98332\\
1490	0.98353\\
1500	0.98378\\
1510	0.98397\\
1520	0.98417\\
1530	0.98444\\
1540	0.98467\\
1550	0.98484\\
1560	0.98496\\
1570	0.98524\\
1580	0.98541\\
1590	0.98571\\
1600	0.9859\\
1610	0.98607\\
1620	0.98623\\
1630	0.98641\\
1640	0.9866\\
1650	0.98681\\
1660	0.98701\\
1670	0.98716\\
1680	0.98735\\
1690	0.98756\\
1700	0.98774\\
1710	0.98786\\
1720	0.98805\\
1730	0.98819\\
1740	0.98832\\
1750	0.98857\\
1760	0.98873\\
1770	0.98889\\
1780	0.98902\\
1790	0.98912\\
1800	0.9893\\
1810	0.98951\\
1820	0.98969\\
1830	0.98986\\
1840	0.98993\\
1850	0.99008\\
1860	0.99018\\
1870	0.99032\\
1880	0.99044\\
1890	0.99054\\
1900	0.99065\\
1910	0.99074\\
1920	0.99085\\
1930	0.99091\\
1940	0.99104\\
1950	0.99118\\
1960	0.99129\\
1970	0.99138\\
1980	0.9915\\
1990	0.99163\\
2000	0.99171\\
2010	0.99184\\
2020	0.99191\\
2030	0.992\\
2040	0.99206\\
2050	0.99215\\
2060	0.99226\\
2070	0.99234\\
2080	0.99245\\
2090	0.99254\\
2100	0.99258\\
2110	0.99272\\
2120	0.99279\\
2130	0.99284\\
2140	0.9929\\
2150	0.99297\\
2160	0.99305\\
2170	0.99313\\
2180	0.9932\\
2190	0.99324\\
2200	0.99332\\
2210	0.99337\\
2220	0.99344\\
2230	0.99354\\
2240	0.99362\\
2250	0.99369\\
2260	0.99372\\
2270	0.99377\\
2280	0.99384\\
2290	0.99389\\
2300	0.99396\\
2310	0.99404\\
2320	0.99409\\
2330	0.99411\\
2340	0.99417\\
2350	0.99424\\
2360	0.9944\\
2370	0.99445\\
2380	0.99449\\
2390	0.99457\\
2400	0.99461\\
2410	0.99463\\
2420	0.99471\\
2430	0.99477\\
2440	0.99484\\
2450	0.99489\\
2460	0.99493\\
2470	0.99496\\
2480	0.99501\\
2490	0.99509\\
2500	0.99515\\
2510	0.99523\\
2520	0.99527\\
2530	0.99537\\
2540	0.99542\\
2550	0.99547\\
2560	0.99553\\
2570	0.99558\\
2580	0.99565\\
2590	0.99567\\
2600	0.99575\\
2610	0.99579\\
2620	0.99581\\
2630	0.99583\\
2640	0.99588\\
2650	0.99591\\
2660	0.99597\\
2670	0.99598\\
2680	0.99604\\
2690	0.99611\\
2700	0.99617\\
2710	0.99622\\
2720	0.99627\\
2730	0.99628\\
2740	0.99634\\
2750	0.99636\\
2760	0.99641\\
2770	0.99644\\
2780	0.99646\\
2790	0.99648\\
2800	0.99653\\
2810	0.99656\\
2820	0.99657\\
2830	0.99661\\
2840	0.99662\\
2850	0.99666\\
2860	0.99669\\
2870	0.9967\\
2880	0.99672\\
2890	0.99679\\
2900	0.99683\\
2910	0.99684\\
2920	0.99689\\
2930	0.99691\\
2940	0.99694\\
2950	0.99695\\
2960	0.99695\\
2970	0.99696\\
2980	0.99702\\
2990	0.99705\\
3000	0.99708\\
3010	0.99711\\
3020	0.99713\\
3030	0.99714\\
3040	0.99715\\
3050	0.99717\\
3060	0.99719\\
3070	0.99721\\
3080	0.99726\\
3090	0.99729\\
3100	0.9973\\
3110	0.9973\\
3120	0.99735\\
3130	0.99737\\
3140	0.99741\\
3150	0.99743\\
3160	0.99748\\
3170	0.99749\\
3180	0.99751\\
3190	0.99752\\
3200	0.99755\\
3210	0.99757\\
3220	0.9976\\
3230	0.99762\\
3240	0.99765\\
3250	0.99766\\
3260	0.99768\\
3270	0.99772\\
3280	0.99774\\
3290	0.99775\\
3300	0.99778\\
3310	0.9978\\
3320	0.99783\\
3330	0.99786\\
3340	0.99787\\
3350	0.99787\\
3360	0.99791\\
3370	0.99792\\
3380	0.99793\\
3390	0.99796\\
3400	0.99796\\
3410	0.99797\\
3420	0.99798\\
3430	0.99799\\
3440	0.99804\\
3450	0.99806\\
3460	0.99809\\
3470	0.99811\\
3480	0.99814\\
3490	0.99816\\
3500	0.99816\\
3510	0.99821\\
3520	0.99825\\
3530	0.99828\\
3540	0.9983\\
3550	0.99833\\
3560	0.99834\\
3570	0.99835\\
3580	0.99839\\
3590	0.99841\\
3600	0.99843\\
3610	0.99843\\
3620	0.99843\\
3630	0.99844\\
3640	0.99846\\
3650	0.99848\\
3660	0.99849\\
3670	0.9985\\
3680	0.99851\\
3690	0.99852\\
3700	0.99852\\
3710	0.99854\\
3720	0.99854\\
3730	0.99856\\
3740	0.99858\\
3750	0.99859\\
3760	0.9986\\
3770	0.99861\\
3780	0.99861\\
3790	0.99863\\
3800	0.99863\\
3810	0.99867\\
3820	0.99868\\
3830	0.9987\\
3840	0.99871\\
3850	0.99871\\
3860	0.99872\\
3870	0.99873\\
3880	0.99874\\
3890	0.99876\\
3900	0.99876\\
3910	0.99878\\
3920	0.9988\\
3930	0.9988\\
3940	0.99881\\
3950	0.99881\\
3960	0.99881\\
3970	0.99882\\
3980	0.99882\\
3990	0.99882\\
4000	0.99882\\
4010	0.99884\\
4020	0.99886\\
4030	0.99888\\
4040	0.99889\\
4050	0.99892\\
4060	0.99892\\
4070	0.99893\\
4080	0.99893\\
4090	0.99895\\
4100	0.99898\\
4110	0.99899\\
4120	0.99899\\
4130	0.99899\\
4140	0.999\\
4150	0.99902\\
4160	0.99903\\
4170	0.99904\\
4180	0.99904\\
4190	0.99904\\
4200	0.99906\\
4210	0.99906\\
4220	0.99907\\
4230	0.99909\\
4240	0.99909\\
4250	0.9991\\
4260	0.99912\\
4270	0.99913\\
4280	0.99913\\
4290	0.99913\\
4300	0.99914\\
4310	0.99915\\
4320	0.99916\\
4330	0.99916\\
4340	0.99916\\
4350	0.99917\\
4360	0.99918\\
4370	0.99918\\
4380	0.99918\\
4390	0.99919\\
4400	0.99919\\
4410	0.99919\\
4420	0.99919\\
4430	0.9992\\
4440	0.99923\\
4450	0.99923\\
4460	0.99925\\
4470	0.99925\\
4480	0.99926\\
4490	0.99927\\
4500	0.99928\\
4510	0.99928\\
4520	0.9993\\
4530	0.99931\\
4540	0.99931\\
4550	0.99931\\
4560	0.99932\\
4570	0.99932\\
4580	0.99933\\
4590	0.99934\\
4600	0.99934\\
4610	0.99936\\
4620	0.99937\\
4630	0.99937\\
4640	0.99938\\
4650	0.99939\\
4660	0.9994\\
4670	0.99941\\
4680	0.99941\\
4690	0.99941\\
4700	0.99942\\
4710	0.99943\\
4720	0.99943\\
4730	0.99944\\
4740	0.99944\\
4750	0.99944\\
4760	0.99945\\
4770	0.99945\\
4780	0.99946\\
4790	0.99946\\
4800	0.99946\\
4810	0.99949\\
4820	0.99949\\
4830	0.99949\\
4840	0.9995\\
4850	0.9995\\
4860	0.99951\\
4870	0.99951\\
4880	0.99951\\
4890	0.99951\\
4900	0.99951\\
4910	0.99951\\
4920	0.99952\\
4930	0.99952\\
4940	0.99952\\
4950	0.99955\\
4960	0.99955\\
4970	0.99955\\
4980	0.99956\\
4990	0.99957\\
5000	0.99957\\
5010	0.99957\\
5020	0.99957\\
5030	0.99957\\
5040	0.99957\\
5050	0.99957\\
5060	0.99957\\
5070	0.99958\\
5080	0.99958\\
5090	0.99959\\
5100	0.9996\\
5110	0.9996\\
5120	0.9996\\
5130	0.9996\\
5140	0.9996\\
5150	0.9996\\
5160	0.9996\\
5170	0.9996\\
5180	0.99961\\
5190	0.99962\\
5200	1\\
};

\addplot [color=maxcol,semithick,mark=square,mark options={solid},mark repeat=10,mark phase=6]
  table[row sep=crcr]{%
0	0\\
10	0.19508\\
20	0.27094\\
30	0.32625\\
40	0.36983\\
50	0.40637\\
60	0.43861\\
70	0.46694\\
80	0.49265\\
90	0.51546\\
100	0.53583\\
110	0.55481\\
120	0.57158\\
130	0.5874\\
140	0.60229\\
150	0.6161\\
160	0.62939\\
170	0.6415\\
180	0.65274\\
190	0.6636\\
200	0.67353\\
210	0.68294\\
220	0.69234\\
230	0.70154\\
240	0.70985\\
250	0.71782\\
260	0.72546\\
270	0.73302\\
280	0.7396\\
290	0.74629\\
300	0.75259\\
310	0.75858\\
320	0.7644\\
330	0.76982\\
340	0.77536\\
350	0.78057\\
360	0.78548\\
370	0.79023\\
380	0.79499\\
390	0.79924\\
400	0.80337\\
410	0.80767\\
420	0.8118\\
430	0.81617\\
440	0.81999\\
450	0.82408\\
460	0.82757\\
470	0.83118\\
480	0.83439\\
490	0.83734\\
500	0.84039\\
510	0.84333\\
520	0.84635\\
530	0.84906\\
540	0.85156\\
550	0.85432\\
560	0.8571\\
570	0.85972\\
580	0.86215\\
590	0.86429\\
600	0.86638\\
610	0.86881\\
620	0.87104\\
630	0.87326\\
640	0.87541\\
650	0.87734\\
660	0.87924\\
670	0.88117\\
680	0.88314\\
690	0.88507\\
700	0.88699\\
710	0.88866\\
720	0.8906\\
730	0.8924\\
740	0.89417\\
750	0.89575\\
760	0.89738\\
770	0.89875\\
780	0.90028\\
790	0.90174\\
800	0.90324\\
810	0.90446\\
820	0.90579\\
830	0.90715\\
840	0.90852\\
850	0.90976\\
860	0.91102\\
870	0.91231\\
880	0.91348\\
890	0.91476\\
900	0.91605\\
910	0.91717\\
920	0.91823\\
930	0.91922\\
940	0.92022\\
950	0.92142\\
960	0.92263\\
970	0.92381\\
980	0.92497\\
990	0.92594\\
1000	0.92683\\
1010	0.92757\\
1020	0.92853\\
1030	0.92942\\
1040	0.93036\\
1050	0.93112\\
1060	0.93203\\
1070	0.93292\\
1080	0.93384\\
1090	0.93476\\
1100	0.93554\\
1110	0.93623\\
1120	0.93704\\
1130	0.93791\\
1140	0.93859\\
1150	0.9392\\
1160	0.93983\\
1170	0.94048\\
1180	0.94114\\
1190	0.94178\\
1200	0.94242\\
1210	0.94305\\
1220	0.94368\\
1230	0.9443\\
1240	0.94493\\
1250	0.94577\\
1260	0.94633\\
1270	0.94685\\
1280	0.94759\\
1290	0.9482\\
1300	0.94883\\
1310	0.94949\\
1320	0.95008\\
1330	0.95054\\
1340	0.95113\\
1350	0.9517\\
1360	0.95218\\
1370	0.95265\\
1380	0.9532\\
1390	0.95374\\
1400	0.95421\\
1410	0.95471\\
1420	0.95528\\
1430	0.95574\\
1440	0.95612\\
1450	0.95651\\
1460	0.95704\\
1470	0.95747\\
1480	0.95792\\
1490	0.9583\\
1500	0.95864\\
1510	0.9591\\
1520	0.95951\\
1530	0.96\\
1540	0.96048\\
1550	0.96096\\
1560	0.96141\\
1570	0.96185\\
1580	0.9623\\
1590	0.96256\\
1600	0.96298\\
1610	0.96339\\
1620	0.96381\\
1630	0.96415\\
1640	0.96456\\
1650	0.96495\\
1660	0.96526\\
1670	0.9656\\
1680	0.96596\\
1690	0.96629\\
1700	0.96668\\
1710	0.96708\\
1720	0.96741\\
1730	0.96767\\
1740	0.96805\\
1750	0.9684\\
1760	0.96871\\
1770	0.96902\\
1780	0.96937\\
1790	0.96962\\
1800	0.96994\\
1810	0.97024\\
1820	0.97044\\
1830	0.97073\\
1840	0.97101\\
1850	0.97131\\
1860	0.97164\\
1870	0.97191\\
1880	0.97217\\
1890	0.97241\\
1900	0.97263\\
1910	0.97277\\
1920	0.97302\\
1930	0.97321\\
1940	0.97346\\
1950	0.97366\\
1960	0.97387\\
1970	0.9741\\
1980	0.9743\\
1990	0.97456\\
2000	0.97473\\
2010	0.97501\\
2020	0.9753\\
2030	0.97551\\
2040	0.9757\\
2050	0.97592\\
2060	0.97619\\
2070	0.97634\\
2080	0.97654\\
2090	0.97677\\
2100	0.97691\\
2110	0.97719\\
2120	0.97743\\
2130	0.97758\\
2140	0.97781\\
2150	0.97801\\
2160	0.97818\\
2170	0.97843\\
2180	0.97863\\
2190	0.9788\\
2200	0.97902\\
2210	0.97924\\
2220	0.97945\\
2230	0.97965\\
2240	0.97985\\
2250	0.98003\\
2260	0.98023\\
2270	0.98037\\
2280	0.9806\\
2290	0.98083\\
2300	0.98103\\
2310	0.98118\\
2320	0.98135\\
2330	0.98145\\
2340	0.9817\\
2350	0.98183\\
2360	0.98206\\
2370	0.9822\\
2380	0.98228\\
2390	0.98246\\
2400	0.98257\\
2410	0.98271\\
2420	0.98283\\
2430	0.98305\\
2440	0.98317\\
2450	0.98339\\
2460	0.98353\\
2470	0.98363\\
2480	0.9838\\
2490	0.98394\\
2500	0.98404\\
2510	0.98419\\
2520	0.98428\\
2530	0.98444\\
2540	0.98458\\
2550	0.98469\\
2560	0.98484\\
2570	0.98496\\
2580	0.98512\\
2590	0.98521\\
2600	0.98532\\
2610	0.98547\\
2620	0.98557\\
2630	0.98566\\
2640	0.98573\\
2650	0.98582\\
2660	0.98593\\
2670	0.98609\\
2680	0.98619\\
2690	0.98626\\
2700	0.98641\\
2710	0.98653\\
2720	0.98659\\
2730	0.9867\\
2740	0.98681\\
2750	0.98691\\
2760	0.98701\\
2770	0.98716\\
2780	0.9873\\
2790	0.98737\\
2800	0.98745\\
2810	0.98756\\
2820	0.98763\\
2830	0.98771\\
2840	0.98778\\
2850	0.98785\\
2860	0.98795\\
2870	0.98802\\
2880	0.9881\\
2890	0.98818\\
2900	0.9883\\
2910	0.9884\\
2920	0.98848\\
2930	0.98862\\
2940	0.9887\\
2950	0.98878\\
2960	0.98888\\
2970	0.98899\\
2980	0.98908\\
2990	0.9892\\
3000	0.98929\\
3010	0.98933\\
3020	0.98938\\
3030	0.98945\\
3040	0.98955\\
3050	0.98963\\
3060	0.9897\\
3070	0.98979\\
3080	0.98992\\
3090	0.99\\
3100	0.99008\\
3110	0.99015\\
3120	0.99019\\
3130	0.99027\\
3140	0.99031\\
3150	0.99039\\
3160	0.99044\\
3170	0.99051\\
3180	0.99059\\
3190	0.99064\\
3200	0.99073\\
3210	0.99075\\
3220	0.99086\\
3230	0.99098\\
3240	0.99105\\
3250	0.99113\\
3260	0.99119\\
3270	0.99125\\
3280	0.99129\\
3290	0.99134\\
3300	0.99138\\
3310	0.99142\\
3320	0.99148\\
3330	0.99156\\
3340	0.9916\\
3350	0.99166\\
3360	0.99174\\
3370	0.99181\\
3380	0.99192\\
3390	0.99199\\
3400	0.99205\\
3410	0.9921\\
3420	0.99217\\
3430	0.99223\\
3440	0.99228\\
3450	0.9923\\
3460	0.99236\\
3470	0.99242\\
3480	0.99243\\
3490	0.99251\\
3500	0.99262\\
3510	0.99268\\
3520	0.99271\\
3530	0.99278\\
3540	0.99283\\
3550	0.99292\\
3560	0.99297\\
3570	0.99304\\
3580	0.99308\\
3590	0.99312\\
3600	0.99317\\
3610	0.99326\\
3620	0.99333\\
3630	0.99335\\
3640	0.99341\\
3650	0.99349\\
3660	0.99354\\
3670	0.99358\\
3680	0.99362\\
3690	0.99366\\
3700	0.99372\\
3710	0.99379\\
3720	0.99381\\
3730	0.99386\\
3740	0.99391\\
3750	0.99394\\
3760	0.99398\\
3770	0.99402\\
3780	0.99405\\
3790	0.99414\\
3800	0.99419\\
3810	0.99425\\
3820	0.9943\\
3830	0.99433\\
3840	0.99442\\
3850	0.99444\\
3860	0.99445\\
3870	0.99451\\
3880	0.99453\\
3890	0.99457\\
3900	0.99461\\
3910	0.99462\\
3920	0.99465\\
3930	0.99469\\
3940	0.99474\\
3950	0.9948\\
3960	0.99482\\
3970	0.99484\\
3980	0.99487\\
3990	0.99489\\
4000	0.99491\\
4010	0.99494\\
4020	0.99498\\
4030	0.99503\\
4040	0.99504\\
4050	0.99505\\
4060	0.99507\\
4070	0.99512\\
4080	0.99513\\
4090	0.99516\\
4100	0.99523\\
4110	0.99525\\
4120	0.99528\\
4130	0.9953\\
4140	0.99533\\
4150	0.99536\\
4160	0.9954\\
4170	0.99542\\
4180	0.99544\\
4190	0.99545\\
4200	0.99549\\
4210	0.9955\\
4220	0.99552\\
4230	0.99555\\
4240	0.99563\\
4250	0.99568\\
4260	0.99568\\
4270	0.99572\\
4280	0.99576\\
4290	0.99578\\
4300	0.99578\\
4310	0.99579\\
4320	0.99584\\
4330	0.99588\\
4340	0.99592\\
4350	0.99594\\
4360	0.99594\\
4370	0.99597\\
4380	0.99599\\
4390	0.99603\\
4400	0.99607\\
4410	0.99608\\
4420	0.99609\\
4430	0.99613\\
4440	0.99615\\
4450	0.9962\\
4460	0.99624\\
4470	0.99625\\
4480	0.99627\\
4490	0.99634\\
4500	0.9964\\
4510	0.9964\\
4520	0.99647\\
4530	0.99648\\
4540	0.9965\\
4550	0.99652\\
4560	0.99654\\
4570	0.99657\\
4580	0.99657\\
4590	0.9966\\
4600	0.99661\\
4610	0.99662\\
4620	0.99662\\
4630	0.99666\\
4640	0.99666\\
4650	0.9967\\
4660	0.99673\\
4670	0.99674\\
4680	0.99677\\
4690	0.9968\\
4700	0.9968\\
4710	0.99682\\
4720	0.99685\\
4730	0.99688\\
4740	0.99689\\
4750	0.99689\\
4760	0.99691\\
4770	0.99692\\
4780	0.99696\\
4790	0.99697\\
4800	0.99697\\
4810	0.99697\\
4820	0.99698\\
4830	0.99701\\
4840	0.99703\\
4850	0.99705\\
4860	0.99705\\
4870	0.99706\\
4880	0.99706\\
4890	0.99706\\
4900	0.99708\\
4910	0.99708\\
4920	0.99709\\
4930	0.99713\\
4940	0.99716\\
4950	0.99718\\
4960	0.99719\\
4970	0.99722\\
4980	0.99724\\
4990	0.99724\\
5000	0.99726\\
5010	0.99728\\
5020	0.99729\\
5030	0.99731\\
5040	0.99731\\
5050	0.99735\\
5060	0.99735\\
5070	0.99737\\
5080	0.99739\\
5090	0.99739\\
5100	0.9974\\
5110	0.99743\\
5120	0.99747\\
5130	0.99747\\
5140	0.99747\\
5150	0.99747\\
5160	0.99747\\
5170	0.99747\\
5180	0.9975\\
5190	0.9975\\
5200	1\\
};

\addplot [color=cntcol,semithick,mark=triangle,mark options={solid,rotate=180},mark repeat=10,mark phase=7]
  table[row sep=crcr]{%
0	0\\
10	0.19546\\
20	0.27276\\
30	0.32727\\
40	0.37081\\
50	0.40716\\
60	0.43873\\
70	0.46516\\
80	0.48976\\
90	0.51217\\
100	0.5329\\
110	0.55095\\
120	0.56716\\
130	0.58266\\
140	0.59698\\
150	0.61084\\
160	0.62317\\
170	0.63471\\
180	0.646\\
190	0.65675\\
200	0.66661\\
210	0.67603\\
220	0.68475\\
230	0.69288\\
240	0.7007\\
250	0.70847\\
260	0.71595\\
270	0.7229\\
280	0.72952\\
290	0.73626\\
300	0.74215\\
310	0.74816\\
320	0.75394\\
330	0.75959\\
340	0.7648\\
350	0.76969\\
360	0.7747\\
370	0.77959\\
380	0.78411\\
390	0.7882\\
400	0.7929\\
410	0.79697\\
420	0.80097\\
430	0.80481\\
440	0.80863\\
450	0.81215\\
460	0.81572\\
470	0.81914\\
480	0.8225\\
490	0.8256\\
500	0.82882\\
510	0.83178\\
520	0.83459\\
530	0.83721\\
540	0.83997\\
550	0.84268\\
560	0.84528\\
570	0.84798\\
580	0.85043\\
590	0.85294\\
600	0.85521\\
610	0.85761\\
620	0.85971\\
630	0.86221\\
640	0.86437\\
650	0.86671\\
660	0.86876\\
670	0.87076\\
680	0.87252\\
690	0.87434\\
700	0.87633\\
710	0.8783\\
720	0.87995\\
730	0.88179\\
740	0.88351\\
750	0.8851\\
760	0.88664\\
770	0.88852\\
780	0.88988\\
790	0.89137\\
800	0.89308\\
810	0.89454\\
820	0.89588\\
830	0.89727\\
840	0.89861\\
850	0.89984\\
860	0.9013\\
870	0.9027\\
880	0.90402\\
890	0.90519\\
900	0.90627\\
910	0.90747\\
920	0.9087\\
930	0.90999\\
940	0.91108\\
950	0.9122\\
960	0.91335\\
970	0.91468\\
980	0.91591\\
990	0.9168\\
1000	0.9178\\
1010	0.91878\\
1020	0.91969\\
1030	0.92061\\
1040	0.9214\\
1050	0.92234\\
1060	0.92309\\
1070	0.92405\\
1080	0.92505\\
1090	0.92593\\
1100	0.92683\\
1110	0.92775\\
1120	0.92864\\
1130	0.92947\\
1140	0.93018\\
1150	0.93089\\
1160	0.93174\\
1170	0.93246\\
1180	0.93313\\
1190	0.93393\\
1200	0.93469\\
1210	0.93537\\
1220	0.93609\\
1230	0.93684\\
1240	0.93752\\
1250	0.9381\\
1260	0.93887\\
1270	0.93952\\
1280	0.94019\\
1290	0.94088\\
1300	0.94155\\
1310	0.94221\\
1320	0.94292\\
1330	0.94344\\
1340	0.9442\\
1350	0.94476\\
1360	0.94548\\
1370	0.94597\\
1380	0.94648\\
1390	0.94703\\
1400	0.9476\\
1410	0.94813\\
1420	0.94862\\
1430	0.94905\\
1440	0.94958\\
1450	0.95002\\
1460	0.95051\\
1470	0.95091\\
1480	0.95128\\
1490	0.95165\\
1500	0.95223\\
1510	0.95274\\
1520	0.95321\\
1530	0.95369\\
1540	0.95397\\
1550	0.95427\\
1560	0.95467\\
1570	0.95513\\
1580	0.95554\\
1590	0.95603\\
1600	0.95646\\
1610	0.95681\\
1620	0.95721\\
1630	0.9574\\
1640	0.95786\\
1650	0.95825\\
1660	0.95876\\
1670	0.95905\\
1680	0.95949\\
1690	0.95986\\
1700	0.96022\\
1710	0.96053\\
1720	0.96085\\
1730	0.96118\\
1740	0.9615\\
1750	0.96185\\
1760	0.9623\\
1770	0.96262\\
1780	0.96293\\
1790	0.96332\\
1800	0.96368\\
1810	0.96388\\
1820	0.96424\\
1830	0.96455\\
1840	0.96484\\
1850	0.96512\\
1860	0.9654\\
1870	0.96563\\
1880	0.96598\\
1890	0.9663\\
1900	0.96654\\
1910	0.96681\\
1920	0.96714\\
1930	0.96744\\
1940	0.96765\\
1950	0.96788\\
1960	0.96816\\
1970	0.96843\\
1980	0.96871\\
1990	0.96898\\
2000	0.96923\\
2010	0.96952\\
2020	0.96987\\
2030	0.97011\\
2040	0.97053\\
2050	0.97075\\
2060	0.97108\\
2070	0.97134\\
2080	0.9716\\
2090	0.97184\\
2100	0.97205\\
2110	0.97223\\
2120	0.97246\\
2130	0.97267\\
2140	0.9728\\
2150	0.97301\\
2160	0.97326\\
2170	0.97344\\
2180	0.97368\\
2190	0.97395\\
2200	0.97411\\
2210	0.97433\\
2220	0.97448\\
2230	0.97468\\
2240	0.97495\\
2250	0.97512\\
2260	0.97531\\
2270	0.97556\\
2280	0.9758\\
2290	0.97598\\
2300	0.97621\\
2310	0.97639\\
2320	0.9765\\
2330	0.97669\\
2340	0.97687\\
2350	0.97701\\
2360	0.97726\\
2370	0.97744\\
2380	0.97764\\
2390	0.97774\\
2400	0.97793\\
2410	0.97798\\
2420	0.97813\\
2430	0.97833\\
2440	0.97849\\
2450	0.97862\\
2460	0.97875\\
2470	0.97893\\
2480	0.97916\\
2490	0.97931\\
2500	0.97951\\
2510	0.97966\\
2520	0.97982\\
2530	0.97997\\
2540	0.98009\\
2550	0.98026\\
2560	0.98041\\
2570	0.98062\\
2580	0.9807\\
2590	0.98084\\
2600	0.98102\\
2610	0.98115\\
2620	0.9813\\
2630	0.98153\\
2640	0.98161\\
2650	0.98174\\
2660	0.98183\\
2670	0.98192\\
2680	0.98204\\
2690	0.98218\\
2700	0.98232\\
2710	0.98241\\
2720	0.98252\\
2730	0.98262\\
2740	0.98278\\
2750	0.98295\\
2760	0.98303\\
2770	0.98314\\
2780	0.98324\\
2790	0.98344\\
2800	0.98353\\
2810	0.98362\\
2820	0.98371\\
2830	0.98379\\
2840	0.98384\\
2850	0.98395\\
2860	0.98408\\
2870	0.98416\\
2880	0.98425\\
2890	0.98436\\
2900	0.98446\\
2910	0.98451\\
2920	0.98462\\
2930	0.98471\\
2940	0.98477\\
2950	0.98489\\
2960	0.98502\\
2970	0.98512\\
2980	0.98517\\
2990	0.98523\\
3000	0.98536\\
3010	0.98546\\
3020	0.98555\\
3030	0.98562\\
3040	0.98573\\
3050	0.98576\\
3060	0.98584\\
3070	0.9859\\
3080	0.98601\\
3090	0.98611\\
3100	0.98623\\
3110	0.98633\\
3120	0.9864\\
3130	0.98649\\
3140	0.98659\\
3150	0.98667\\
3160	0.98679\\
3170	0.98687\\
3180	0.98692\\
3190	0.98698\\
3200	0.98702\\
3210	0.98709\\
3220	0.9872\\
3230	0.98725\\
3240	0.98735\\
3250	0.98746\\
3260	0.98751\\
3270	0.98758\\
3280	0.98765\\
3290	0.9877\\
3300	0.98775\\
3310	0.98781\\
3320	0.9879\\
3330	0.98794\\
3340	0.988\\
3350	0.98806\\
3360	0.98814\\
3370	0.98821\\
3380	0.98829\\
3390	0.98839\\
3400	0.98847\\
3410	0.98855\\
3420	0.98858\\
3430	0.98865\\
3440	0.98873\\
3450	0.98882\\
3460	0.98885\\
3470	0.98892\\
3480	0.98898\\
3490	0.98902\\
3500	0.98913\\
3510	0.98918\\
3520	0.98928\\
3530	0.98935\\
3540	0.98948\\
3550	0.98953\\
3560	0.98958\\
3570	0.98962\\
3580	0.98969\\
3590	0.98978\\
3600	0.98983\\
3610	0.98985\\
3620	0.98993\\
3630	0.98998\\
3640	0.99002\\
3650	0.9901\\
3660	0.99019\\
3670	0.99031\\
3680	0.99032\\
3690	0.99036\\
3700	0.99039\\
3710	0.99045\\
3720	0.99055\\
3730	0.99063\\
3740	0.99067\\
3750	0.99074\\
3760	0.99079\\
3770	0.99082\\
3780	0.99088\\
3790	0.99091\\
3800	0.99097\\
3810	0.99101\\
3820	0.99109\\
3830	0.99112\\
3840	0.99116\\
3850	0.9912\\
3860	0.99122\\
3870	0.99129\\
3880	0.99135\\
3890	0.99138\\
3900	0.99141\\
3910	0.99146\\
3920	0.99148\\
3930	0.9915\\
3940	0.99156\\
3950	0.99164\\
3960	0.99169\\
3970	0.99171\\
3980	0.99176\\
3990	0.99178\\
4000	0.99183\\
4010	0.99187\\
4020	0.99193\\
4030	0.992\\
4040	0.99202\\
4050	0.99207\\
4060	0.99211\\
4070	0.99217\\
4080	0.9922\\
4090	0.99223\\
4100	0.99227\\
4110	0.99233\\
4120	0.99234\\
4130	0.99238\\
4140	0.99244\\
4150	0.99247\\
4160	0.99249\\
4170	0.99255\\
4180	0.99262\\
4190	0.99264\\
4200	0.99272\\
4210	0.99276\\
4220	0.99278\\
4230	0.99284\\
4240	0.9929\\
4250	0.99293\\
4260	0.99297\\
4270	0.99303\\
4280	0.99306\\
4290	0.9931\\
4300	0.99315\\
4310	0.99317\\
4320	0.99321\\
4330	0.99321\\
4340	0.99324\\
4350	0.99325\\
4360	0.99329\\
4370	0.99333\\
4380	0.99338\\
4390	0.9934\\
4400	0.99343\\
4410	0.99346\\
4420	0.9935\\
4430	0.99353\\
4440	0.99357\\
4450	0.9936\\
4460	0.99363\\
4470	0.99364\\
4480	0.99367\\
4490	0.99373\\
4500	0.99374\\
4510	0.99378\\
4520	0.99381\\
4530	0.99386\\
4540	0.9939\\
4550	0.99392\\
4560	0.99395\\
4570	0.99397\\
4580	0.99398\\
4590	0.994\\
4600	0.99405\\
4610	0.99411\\
4620	0.99413\\
4630	0.99415\\
4640	0.99417\\
4650	0.99419\\
4660	0.99419\\
4670	0.99419\\
4680	0.99421\\
4690	0.99424\\
4700	0.99428\\
4710	0.9943\\
4720	0.99431\\
4730	0.99436\\
4740	0.99438\\
4750	0.99442\\
4760	0.99443\\
4770	0.99443\\
4780	0.99447\\
4790	0.99452\\
4800	0.99455\\
4810	0.99458\\
4820	0.99462\\
4830	0.99466\\
4840	0.99466\\
4850	0.99468\\
4860	0.99469\\
4870	0.99471\\
4880	0.99473\\
4890	0.99474\\
4900	0.99474\\
4910	0.99478\\
4920	0.9948\\
4930	0.99482\\
4940	0.99485\\
4950	0.99486\\
4960	0.9949\\
4970	0.99492\\
4980	0.99495\\
4990	0.99498\\
5000	0.99498\\
5010	0.99501\\
5020	0.99502\\
5030	0.99504\\
5040	0.99507\\
5050	0.99511\\
5060	0.99515\\
5070	0.99515\\
5080	0.99516\\
5090	0.99517\\
5100	0.99517\\
5110	0.99522\\
5120	0.99522\\
5130	0.99523\\
5140	0.99523\\
5150	0.99524\\
5160	0.99527\\
5170	0.99528\\
5180	0.99531\\
5190	0.99532\\
5200	1\\
};

\addplot [color=mafcol, semithick, mark=triangle,mark repeat=10,mark phase=9]
  table[row sep=crcr]{%
0	0\\
10	0.18222\\
20	0.25509\\
30	0.30904\\
40	0.35248\\
50	0.38923\\
60	0.42097\\
70	0.44855\\
80	0.4731\\
90	0.49504\\
100	0.51559\\
110	0.53422\\
120	0.55164\\
130	0.56751\\
140	0.58251\\
150	0.5968\\
160	0.61053\\
170	0.62322\\
180	0.63456\\
190	0.64601\\
200	0.65717\\
210	0.66764\\
220	0.67752\\
230	0.68675\\
240	0.6956\\
250	0.70367\\
260	0.71178\\
270	0.71944\\
280	0.72646\\
290	0.73354\\
300	0.74033\\
310	0.74679\\
320	0.75295\\
330	0.75891\\
340	0.76537\\
350	0.77112\\
360	0.7763\\
370	0.78119\\
380	0.78611\\
390	0.79083\\
400	0.79534\\
410	0.79983\\
420	0.80392\\
430	0.8079\\
440	0.81206\\
450	0.81606\\
460	0.81962\\
470	0.82321\\
480	0.82692\\
490	0.83055\\
500	0.83393\\
510	0.83749\\
520	0.84073\\
530	0.84368\\
540	0.84685\\
550	0.84962\\
560	0.8526\\
570	0.85545\\
580	0.85796\\
590	0.8606\\
600	0.86338\\
610	0.86595\\
620	0.86799\\
630	0.87022\\
640	0.87258\\
650	0.87481\\
660	0.87681\\
670	0.87904\\
680	0.88089\\
690	0.88291\\
700	0.88471\\
710	0.88656\\
720	0.88818\\
730	0.89009\\
740	0.89169\\
750	0.89313\\
760	0.89487\\
770	0.89657\\
780	0.89826\\
790	0.89966\\
800	0.90122\\
810	0.90283\\
820	0.90421\\
830	0.90571\\
840	0.90701\\
850	0.90839\\
860	0.9096\\
870	0.9109\\
880	0.9122\\
890	0.91336\\
900	0.91467\\
910	0.91586\\
920	0.91692\\
930	0.91802\\
940	0.91917\\
950	0.92035\\
960	0.92133\\
970	0.92251\\
980	0.92354\\
990	0.92457\\
1000	0.92564\\
1010	0.92665\\
1020	0.92768\\
1030	0.92861\\
1040	0.92977\\
1050	0.93063\\
1060	0.93162\\
1070	0.93251\\
1080	0.93334\\
1090	0.93423\\
1100	0.93508\\
1110	0.93584\\
1120	0.93666\\
1130	0.93747\\
1140	0.93847\\
1150	0.93939\\
1160	0.94017\\
1170	0.94105\\
1180	0.94179\\
1190	0.94263\\
1200	0.94332\\
1210	0.94411\\
1220	0.94486\\
1230	0.9456\\
1240	0.94626\\
1250	0.94684\\
1260	0.94746\\
1270	0.94811\\
1280	0.94876\\
1290	0.94932\\
1300	0.94996\\
1310	0.95062\\
1320	0.9513\\
1330	0.9519\\
1340	0.95235\\
1350	0.95293\\
1360	0.95355\\
1370	0.95394\\
1380	0.95455\\
1390	0.95506\\
1400	0.95555\\
1410	0.95596\\
1420	0.95637\\
1430	0.9569\\
1440	0.95738\\
1450	0.95779\\
1460	0.95832\\
1470	0.95877\\
1480	0.95924\\
1490	0.95964\\
1500	0.96005\\
1510	0.96041\\
1520	0.96078\\
1530	0.96122\\
1540	0.96156\\
1550	0.96192\\
1560	0.96229\\
1570	0.96265\\
1580	0.9631\\
1590	0.9636\\
1600	0.96392\\
1610	0.96429\\
1620	0.96476\\
1630	0.9651\\
1640	0.9654\\
1650	0.96573\\
1660	0.9661\\
1670	0.96642\\
1680	0.96671\\
1690	0.96711\\
1700	0.96747\\
1710	0.96783\\
1720	0.96805\\
1730	0.96839\\
1740	0.96879\\
1750	0.96906\\
1760	0.96933\\
1770	0.96973\\
1780	0.97\\
1790	0.97024\\
1800	0.97047\\
1810	0.97064\\
1820	0.97092\\
1830	0.97116\\
1840	0.97137\\
1850	0.97164\\
1860	0.9719\\
1870	0.97213\\
1880	0.97247\\
1890	0.97277\\
1900	0.97296\\
1910	0.97315\\
1920	0.97334\\
1930	0.97359\\
1940	0.97387\\
1950	0.97416\\
1960	0.97452\\
1970	0.97472\\
1980	0.97496\\
1990	0.97518\\
2000	0.97538\\
2010	0.97557\\
2020	0.9758\\
2030	0.97604\\
2040	0.97624\\
2050	0.97651\\
2060	0.97673\\
2070	0.97694\\
2080	0.97718\\
2090	0.97737\\
2100	0.97758\\
2110	0.97774\\
2120	0.97792\\
2130	0.97809\\
2140	0.97824\\
2150	0.9784\\
2160	0.97852\\
2170	0.97871\\
2180	0.97887\\
2190	0.97912\\
2200	0.97925\\
2210	0.97942\\
2220	0.97957\\
2230	0.9798\\
2240	0.97999\\
2250	0.98022\\
2260	0.98041\\
2270	0.98059\\
2280	0.98075\\
2290	0.98089\\
2300	0.98104\\
2310	0.98117\\
2320	0.98135\\
2330	0.98148\\
2340	0.98168\\
2350	0.9819\\
2360	0.98198\\
2370	0.98216\\
2380	0.98224\\
2390	0.98235\\
2400	0.98244\\
2410	0.98254\\
2420	0.98268\\
2430	0.98279\\
2440	0.98286\\
2450	0.98291\\
2460	0.98305\\
2470	0.98324\\
2480	0.98334\\
2490	0.98347\\
2500	0.98365\\
2510	0.98376\\
2520	0.98389\\
2530	0.984\\
2540	0.98416\\
2550	0.9843\\
2560	0.98443\\
2570	0.98458\\
2580	0.98464\\
2590	0.98476\\
2600	0.98488\\
2610	0.98497\\
2620	0.98503\\
2630	0.98515\\
2640	0.9853\\
2650	0.98546\\
2660	0.98558\\
2670	0.98571\\
2680	0.98584\\
2690	0.98595\\
2700	0.98605\\
2710	0.98621\\
2720	0.98632\\
2730	0.98644\\
2740	0.98658\\
2750	0.98672\\
2760	0.98685\\
2770	0.98693\\
2780	0.98699\\
2790	0.98708\\
2800	0.98716\\
2810	0.98725\\
2820	0.98734\\
2830	0.98748\\
2840	0.98761\\
2850	0.98767\\
2860	0.98781\\
2870	0.98791\\
2880	0.98797\\
2890	0.98811\\
2900	0.98824\\
2910	0.98837\\
2920	0.98847\\
2930	0.98862\\
2940	0.98864\\
2950	0.98869\\
2960	0.98883\\
2970	0.98893\\
2980	0.98898\\
2990	0.98903\\
3000	0.98911\\
3010	0.98921\\
3020	0.98933\\
3030	0.98941\\
3040	0.98951\\
3050	0.9896\\
3060	0.98971\\
3070	0.98979\\
3080	0.9899\\
3090	0.99001\\
3100	0.99007\\
3110	0.99008\\
3120	0.99013\\
3130	0.99019\\
3140	0.99031\\
3150	0.9904\\
3160	0.99047\\
3170	0.99051\\
3180	0.99057\\
3190	0.99062\\
3200	0.99071\\
3210	0.99077\\
3220	0.99083\\
3230	0.99085\\
3240	0.99095\\
3250	0.99108\\
3260	0.99122\\
3270	0.99125\\
3280	0.99127\\
3290	0.99137\\
3300	0.99143\\
3310	0.99147\\
3320	0.99155\\
3330	0.99163\\
3340	0.99171\\
3350	0.99176\\
3360	0.99178\\
3370	0.99183\\
3380	0.99189\\
3390	0.99194\\
3400	0.99199\\
3410	0.99207\\
3420	0.99211\\
3430	0.99218\\
3440	0.99222\\
3450	0.99231\\
3460	0.99234\\
3470	0.99241\\
3480	0.99243\\
3490	0.99254\\
3500	0.9926\\
3510	0.99261\\
3520	0.99266\\
3530	0.99273\\
3540	0.99274\\
3550	0.99277\\
3560	0.9928\\
3570	0.99287\\
3580	0.99293\\
3590	0.99299\\
3600	0.99304\\
3610	0.99307\\
3620	0.99311\\
3630	0.99319\\
3640	0.99324\\
3650	0.99326\\
3660	0.99329\\
3670	0.99334\\
3680	0.9934\\
3690	0.99344\\
3700	0.99353\\
3710	0.9936\\
3720	0.99362\\
3730	0.99366\\
3740	0.99373\\
3750	0.99375\\
3760	0.9938\\
3770	0.9939\\
3780	0.99393\\
3790	0.99401\\
3800	0.99404\\
3810	0.99406\\
3820	0.99408\\
3830	0.99409\\
3840	0.99413\\
3850	0.99415\\
3860	0.99418\\
3870	0.99424\\
3880	0.99429\\
3890	0.99433\\
3900	0.99436\\
3910	0.99438\\
3920	0.99444\\
3930	0.99451\\
3940	0.99456\\
3950	0.99461\\
3960	0.99466\\
3970	0.99472\\
3980	0.99475\\
3990	0.99478\\
4000	0.9948\\
4010	0.99482\\
4020	0.99485\\
4030	0.99487\\
4040	0.99491\\
4050	0.99495\\
4060	0.99498\\
4070	0.99499\\
4080	0.995\\
4090	0.99501\\
4100	0.99508\\
4110	0.99516\\
4120	0.99518\\
4130	0.9952\\
4140	0.99524\\
4150	0.99527\\
4160	0.99527\\
4170	0.9953\\
4180	0.99534\\
4190	0.99537\\
4200	0.99539\\
4210	0.99542\\
4220	0.99547\\
4230	0.99552\\
4240	0.99556\\
4250	0.99557\\
4260	0.9956\\
4270	0.99564\\
4280	0.99565\\
4290	0.99568\\
4300	0.99571\\
4310	0.99572\\
4320	0.99576\\
4330	0.99578\\
4340	0.99583\\
4350	0.99588\\
4360	0.99592\\
4370	0.99594\\
4380	0.99596\\
4390	0.99596\\
4400	0.99604\\
4410	0.99609\\
4420	0.9961\\
4430	0.99612\\
4440	0.99614\\
4450	0.99616\\
4460	0.99617\\
4470	0.99619\\
4480	0.9962\\
4490	0.9962\\
4500	0.99624\\
4510	0.99635\\
4520	0.99639\\
4530	0.9964\\
4540	0.99641\\
4550	0.99642\\
4560	0.99644\\
4570	0.99645\\
4580	0.99645\\
4590	0.99647\\
4600	0.99649\\
4610	0.99651\\
4620	0.99655\\
4630	0.99656\\
4640	0.99656\\
4650	0.99658\\
4660	0.99661\\
4670	0.99663\\
4680	0.99666\\
4690	0.99668\\
4700	0.9967\\
4710	0.99672\\
4720	0.99674\\
4730	0.99676\\
4740	0.99677\\
4750	0.99677\\
4760	0.99677\\
4770	0.99679\\
4780	0.99679\\
4790	0.99681\\
4800	0.99681\\
4810	0.99685\\
4820	0.99686\\
4830	0.9969\\
4840	0.99692\\
4850	0.99695\\
4860	0.99699\\
4870	0.99701\\
4880	0.99704\\
4890	0.99706\\
4900	0.99708\\
4910	0.9971\\
4920	0.9971\\
4930	0.99714\\
4940	0.99716\\
4950	0.99716\\
4960	0.9972\\
4970	0.99721\\
4980	0.99725\\
4990	0.99725\\
5000	0.99727\\
5010	0.99731\\
5020	0.99732\\
5030	0.99734\\
5040	0.99735\\
5050	0.99736\\
5060	0.99738\\
5070	0.9974\\
5080	0.99742\\
5090	0.99745\\
5100	0.99747\\
5110	0.99749\\
5120	0.99752\\
5130	0.99754\\
5140	0.99755\\
5150	0.99756\\
5160	0.9976\\
5170	0.99762\\
5180	0.99762\\
5190	0.99764\\
5200	1\\
};

\end{axis}
\end{tikzpicture}%

%% file: fig/max_err.tex
\begin{tikzpicture}
\pgfplotsset{every tick label/.append style={font=\tiny}}
\tikzstyle{dotted}= [dash pattern=on \pgflinewidth off 0.5mm] 
\tikzstyle{dashed}= [dash pattern=on 7.5*0.8*0.8pt off 7.5*0.4*0.8pt]
\tikzstyle{dashdotted} = [dash pattern=on 7.5*0.8*0.6pt off 7.5*0.8*0.3pt on \the\pgflinewidth off 7.5*0.8*0.3pt]
\tikzstyle{dotted2} = [dash pattern=on 7.5*0.8*0.3pt off 7.5*0.8*0.2pt]

\begin{axis}[%
width=\fwidth,
height=\fheight,
at={(0,0)},
xmin=0,
xmax=50,
xlabel near ticks,
xlabel style={font=\scriptsize\color{white!15!black}},
xlabel={$\nu_{\text{max}}$},
ylabel near ticks,
ymin=0.5,
ymax=1,
ylabel style={font=\scriptsize\color{white!15!black}},
ylabel={Empirical CDF},axis background/.style={fill=white},
xmajorgrids,
ymajorgrids,
legend style={font=\tiny, at={(0.992,0.02)}, anchor=south east, legend cell align=left, align=left, fill opacity=0.8, draw opacity=1, text opacity=1, draw=white!80!black}
]
\addplot [color=msecol, semithick,mark=x,mark repeat=25,mark phase=1]
  table[row sep=crcr]{%
0	0\\
0.1	0.10531\\
0.2	0.14829\\
0.3	0.18095\\
0.4	0.20797\\
0.5	0.23181\\
0.6	0.25305\\
0.7	0.27191\\
0.8	0.28981\\
0.9	0.30607\\
1	0.32139\\
1.1	0.33598\\
1.2	0.34954\\
1.3	0.36238\\
1.4	0.37501\\
1.5	0.38698\\
1.6	0.39859\\
1.7	0.40896\\
1.8	0.41964\\
1.9	0.42949\\
2	0.43912\\
2.1	0.44746\\
2.2	0.45616\\
2.3	0.46512\\
2.4	0.47362\\
2.5	0.48111\\
2.6	0.48877\\
2.7	0.49666\\
2.8	0.50393\\
2.9	0.51092\\
3	0.51786\\
3.1	0.52487\\
3.2	0.53134\\
3.3	0.53803\\
3.4	0.54441\\
3.5	0.55047\\
3.6	0.55632\\
3.7	0.56208\\
3.8	0.56782\\
3.9	0.57359\\
4	0.57899\\
4.1	0.58455\\
4.2	0.58968\\
4.3	0.59463\\
4.4	0.59968\\
4.5	0.60447\\
4.6	0.6094\\
4.7	0.61401\\
4.8	0.61893\\
4.9	0.62377\\
5	0.62836\\
5.1	0.63242\\
5.2	0.63686\\
5.3	0.64128\\
5.4	0.64517\\
5.5	0.64903\\
5.6	0.65313\\
5.7	0.65672\\
5.8	0.66072\\
5.9	0.66459\\
6	0.66836\\
6.1	0.67172\\
6.2	0.67533\\
6.3	0.67872\\
6.4	0.68229\\
6.5	0.68586\\
6.6	0.68946\\
6.7	0.69292\\
6.8	0.69634\\
6.9	0.6998\\
7	0.70308\\
7.1	0.70625\\
7.2	0.70909\\
7.3	0.71215\\
7.4	0.71537\\
7.5	0.71848\\
7.6	0.72141\\
7.7	0.72424\\
7.8	0.72743\\
7.9	0.73051\\
8	0.73329\\
8.1	0.73607\\
8.2	0.73867\\
8.3	0.74123\\
8.4	0.74367\\
8.5	0.74646\\
8.6	0.74893\\
8.7	0.7512\\
8.8	0.7535\\
8.9	0.75576\\
9	0.75837\\
9.1	0.76079\\
9.2	0.76306\\
9.3	0.76537\\
9.4	0.76773\\
9.5	0.77014\\
9.6	0.77236\\
9.7	0.77471\\
9.8	0.77663\\
9.9	0.77872\\
10	0.78062\\
10.1	0.78297\\
10.2	0.78505\\
10.3	0.787\\
10.4	0.78901\\
10.5	0.79076\\
10.6	0.79272\\
10.7	0.79477\\
10.8	0.79675\\
10.9	0.79868\\
11	0.80071\\
11.1	0.80281\\
11.2	0.8046\\
11.3	0.80659\\
11.4	0.80838\\
11.5	0.81028\\
11.6	0.81204\\
11.7	0.81378\\
11.8	0.81537\\
11.9	0.81682\\
12	0.81827\\
12.1	0.81997\\
12.2	0.82175\\
12.3	0.82351\\
12.4	0.82512\\
12.5	0.82673\\
12.6	0.82825\\
12.7	0.82989\\
12.8	0.83124\\
12.9	0.83281\\
13	0.8344\\
13.1	0.83579\\
13.2	0.83712\\
13.3	0.83839\\
13.4	0.83988\\
13.5	0.84151\\
13.6	0.84286\\
13.7	0.84399\\
13.8	0.84528\\
13.9	0.84653\\
14	0.84785\\
14.1	0.84897\\
14.2	0.85023\\
14.3	0.8515\\
14.4	0.85266\\
14.5	0.85402\\
14.6	0.85534\\
14.7	0.85665\\
14.8	0.85776\\
14.9	0.85903\\
15	0.8601\\
15.1	0.86108\\
15.2	0.86226\\
15.3	0.86329\\
15.4	0.86444\\
15.5	0.86562\\
15.6	0.86668\\
15.7	0.86775\\
15.8	0.86885\\
15.9	0.86984\\
16	0.8709\\
16.1	0.87214\\
16.2	0.87309\\
16.3	0.87406\\
16.4	0.87516\\
16.5	0.87615\\
16.6	0.87713\\
16.7	0.87814\\
16.8	0.87903\\
16.9	0.87993\\
17	0.88086\\
17.1	0.88178\\
17.2	0.88274\\
17.3	0.88383\\
17.4	0.88474\\
17.5	0.88573\\
17.6	0.88666\\
17.7	0.88764\\
17.8	0.88858\\
17.9	0.88943\\
18	0.89027\\
18.1	0.89107\\
18.2	0.89203\\
18.3	0.89298\\
18.4	0.89377\\
18.5	0.8946\\
18.6	0.8956\\
18.7	0.89648\\
18.8	0.89718\\
18.9	0.89802\\
19	0.89891\\
19.1	0.89963\\
19.2	0.90045\\
19.3	0.90117\\
19.4	0.902\\
19.5	0.90285\\
19.6	0.90359\\
19.7	0.90431\\
19.8	0.90508\\
19.9	0.90585\\
20	0.9066\\
20.1	0.90727\\
20.2	0.90791\\
20.3	0.90845\\
20.4	0.90919\\
20.5	0.90979\\
20.6	0.91049\\
20.7	0.91121\\
20.8	0.91169\\
20.9	0.91243\\
21	0.91313\\
21.1	0.91372\\
21.2	0.9143\\
21.3	0.91494\\
21.4	0.91548\\
21.5	0.91618\\
21.6	0.91692\\
21.7	0.91756\\
21.8	0.91822\\
21.9	0.9188\\
22	0.9194\\
22.1	0.91995\\
22.2	0.92057\\
22.3	0.92118\\
22.4	0.92168\\
22.5	0.92232\\
22.6	0.92299\\
22.7	0.92366\\
22.8	0.92421\\
22.9	0.92477\\
23	0.92527\\
23.1	0.92578\\
23.2	0.92643\\
23.3	0.92697\\
23.4	0.92742\\
23.5	0.92796\\
23.6	0.9285\\
23.7	0.92908\\
23.8	0.92965\\
23.9	0.9301\\
24	0.9307\\
24.1	0.93112\\
24.2	0.93164\\
24.3	0.93208\\
24.4	0.93248\\
24.5	0.93288\\
24.6	0.93332\\
24.7	0.93387\\
24.8	0.93419\\
24.9	0.93468\\
25	0.93522\\
25.1	0.9358\\
25.2	0.93611\\
25.3	0.93657\\
25.4	0.93707\\
25.5	0.93737\\
25.6	0.93775\\
25.7	0.93826\\
25.8	0.93858\\
25.9	0.93898\\
26	0.93935\\
26.1	0.93971\\
26.2	0.94008\\
26.3	0.94038\\
26.4	0.94075\\
26.5	0.94112\\
26.6	0.94159\\
26.7	0.94205\\
26.8	0.94239\\
26.9	0.94282\\
27	0.94314\\
27.1	0.94353\\
27.2	0.94395\\
27.3	0.94436\\
27.4	0.94488\\
27.5	0.94522\\
27.6	0.94555\\
27.7	0.94591\\
27.8	0.94627\\
27.9	0.94667\\
28	0.94698\\
28.1	0.94727\\
28.2	0.94757\\
28.3	0.94785\\
28.4	0.94823\\
28.5	0.94853\\
28.6	0.94888\\
28.7	0.94932\\
28.8	0.94965\\
28.9	0.94998\\
29	0.95015\\
29.1	0.95048\\
29.2	0.95084\\
29.3	0.95108\\
29.4	0.95136\\
29.5	0.95175\\
29.6	0.952\\
29.7	0.95241\\
29.8	0.95274\\
29.9	0.95293\\
30	0.95322\\
30.1	0.95352\\
30.2	0.95385\\
30.3	0.95421\\
30.4	0.95445\\
30.5	0.95476\\
30.6	0.95504\\
30.7	0.9553\\
30.8	0.95569\\
30.9	0.95597\\
31	0.95627\\
31.1	0.95648\\
31.2	0.95681\\
31.3	0.95707\\
31.4	0.9573\\
31.5	0.95751\\
31.6	0.95775\\
31.7	0.95794\\
31.8	0.95827\\
31.9	0.95843\\
32	0.95869\\
32.1	0.95895\\
32.2	0.95914\\
32.3	0.95945\\
32.4	0.95965\\
32.5	0.95981\\
32.6	0.96016\\
32.7	0.96047\\
32.8	0.96068\\
32.9	0.96088\\
33	0.96123\\
33.1	0.96145\\
33.2	0.96168\\
33.3	0.96187\\
33.4	0.96201\\
33.5	0.96224\\
33.6	0.96251\\
33.7	0.96278\\
33.8	0.96303\\
33.9	0.96326\\
34	0.96345\\
34.1	0.96369\\
34.2	0.96391\\
34.3	0.96416\\
34.4	0.96433\\
34.5	0.96456\\
34.6	0.96481\\
34.7	0.96501\\
34.8	0.96521\\
34.9	0.96547\\
35	0.96557\\
35.1	0.96576\\
35.2	0.96614\\
35.3	0.96629\\
35.4	0.96644\\
35.5	0.96667\\
35.6	0.96689\\
35.7	0.96713\\
35.8	0.96735\\
35.9	0.96752\\
36	0.96773\\
36.1	0.96791\\
36.2	0.96808\\
36.3	0.96826\\
36.4	0.96841\\
36.5	0.96864\\
36.6	0.96884\\
36.7	0.96903\\
36.8	0.96923\\
36.9	0.96946\\
37	0.96963\\
37.1	0.96985\\
37.2	0.97005\\
37.3	0.97027\\
37.4	0.97047\\
37.5	0.97067\\
37.6	0.97083\\
37.7	0.97102\\
37.8	0.97119\\
37.9	0.97137\\
38	0.97155\\
38.1	0.97164\\
38.2	0.97184\\
38.3	0.97202\\
38.4	0.9722\\
38.5	0.9724\\
38.6	0.97254\\
38.7	0.97267\\
38.8	0.97288\\
38.9	0.97309\\
39	0.97326\\
39.1	0.97338\\
39.2	0.9735\\
39.3	0.97362\\
39.4	0.9737\\
39.5	0.97389\\
39.6	0.97401\\
39.7	0.97429\\
39.8	0.97444\\
39.9	0.97459\\
40	0.97482\\
40.1	0.97493\\
40.2	0.97503\\
40.3	0.97521\\
40.4	0.97533\\
40.5	0.97548\\
40.6	0.9756\\
40.7	0.97572\\
40.8	0.97584\\
40.9	0.97596\\
41	0.97605\\
41.1	0.97621\\
41.2	0.97636\\
41.3	0.97652\\
41.4	0.97658\\
41.5	0.97675\\
41.6	0.97688\\
41.7	0.97704\\
41.8	0.9772\\
41.9	0.97733\\
42	0.97753\\
42.1	0.97765\\
42.2	0.97778\\
42.3	0.97791\\
42.4	0.978\\
42.5	0.97816\\
42.6	0.97829\\
42.7	0.97842\\
42.8	0.97857\\
42.9	0.97876\\
43	0.97891\\
43.1	0.97903\\
43.2	0.97917\\
43.3	0.97933\\
43.4	0.97947\\
43.5	0.97956\\
43.6	0.97965\\
43.7	0.97973\\
43.8	0.97989\\
43.9	0.97994\\
44	0.98007\\
44.1	0.98022\\
44.2	0.98039\\
44.3	0.98049\\
44.4	0.98063\\
44.5	0.98073\\
44.6	0.98081\\
44.7	0.98094\\
44.8	0.98103\\
44.9	0.98117\\
45	0.98129\\
45.1	0.98135\\
45.2	0.98141\\
45.3	0.9815\\
45.4	0.98157\\
45.5	0.98162\\
45.6	0.9817\\
45.7	0.98173\\
45.8	0.98183\\
45.9	0.98189\\
46	0.98198\\
46.1	0.98213\\
46.2	0.98222\\
46.3	0.98235\\
46.4	0.9824\\
46.5	0.98254\\
46.6	0.98264\\
46.7	0.98273\\
46.8	0.98284\\
46.9	0.98294\\
47	0.98306\\
47.1	0.98316\\
47.2	0.98329\\
47.3	0.98337\\
47.4	0.98346\\
47.5	0.98355\\
47.6	0.98369\\
47.7	0.98375\\
47.8	0.9838\\
47.9	0.98388\\
48	0.98397\\
48.1	0.98403\\
48.2	0.98414\\
48.3	0.9842\\
48.4	0.98424\\
48.5	0.98436\\
48.6	0.98447\\
48.7	0.98458\\
48.8	0.98465\\
48.9	0.98473\\
49	0.98484\\
49.1	0.9849\\
49.2	0.98501\\
49.3	0.98512\\
49.4	0.98525\\
49.5	0.98534\\
49.6	0.98541\\
49.7	0.98548\\
49.8	0.98555\\
49.9	0.98569\\
50	0.98577\\
50.1	0.98587\\
50.2	0.98596\\
50.3	0.98599\\
50.4	0.98606\\
50.5	0.98609\\
50.6	0.98615\\
50.7	0.9862\\
50.8	0.98629\\
50.9	0.98638\\
51	1\\
};

\addplot [color=avgcol,semithick,mark=o,mark repeat=25,mark phase=6]
  table[row sep=crcr]{%
0	0\\
0.1	0.06138\\
0.2	0.08671\\
0.3	0.1061\\
0.4	0.12228\\
0.5	0.1364\\
0.6	0.1499\\
0.7	0.16152\\
0.8	0.17294\\
0.9	0.18291\\
1	0.1928\\
1.1	0.20116\\
1.2	0.21007\\
1.3	0.21765\\
1.4	0.22506\\
1.5	0.23297\\
1.6	0.24098\\
1.7	0.24773\\
1.8	0.25466\\
1.9	0.26145\\
2	0.26783\\
2.1	0.27416\\
2.2	0.28003\\
2.3	0.28581\\
2.4	0.29163\\
2.5	0.29737\\
2.6	0.30319\\
2.7	0.309\\
2.8	0.31442\\
2.9	0.31969\\
3	0.32473\\
3.1	0.32944\\
3.2	0.33461\\
3.3	0.33925\\
3.4	0.34388\\
3.5	0.34835\\
3.6	0.35312\\
3.7	0.35784\\
3.8	0.36241\\
3.9	0.3668\\
4	0.3712\\
4.1	0.37486\\
4.2	0.37901\\
4.3	0.38302\\
4.4	0.38708\\
4.5	0.39084\\
4.6	0.39479\\
4.7	0.39872\\
4.8	0.40229\\
4.9	0.40601\\
5	0.40939\\
5.1	0.41292\\
5.2	0.41628\\
5.3	0.41967\\
5.4	0.4233\\
5.5	0.42667\\
5.6	0.43038\\
5.7	0.43382\\
5.8	0.43699\\
5.9	0.4406\\
6	0.44406\\
6.1	0.44712\\
6.2	0.45021\\
6.3	0.45296\\
6.4	0.45589\\
6.5	0.45892\\
6.6	0.46181\\
6.7	0.46459\\
6.8	0.46756\\
6.9	0.47037\\
7	0.47303\\
7.1	0.47586\\
7.2	0.47836\\
7.3	0.48108\\
7.4	0.48384\\
7.5	0.48651\\
7.6	0.48908\\
7.7	0.49212\\
7.8	0.49459\\
7.9	0.49712\\
8	0.49967\\
8.1	0.5025\\
8.2	0.50513\\
8.3	0.50753\\
8.4	0.50981\\
8.5	0.51217\\
8.6	0.51496\\
8.7	0.51744\\
8.8	0.51965\\
8.9	0.5224\\
9	0.52472\\
9.1	0.52696\\
9.2	0.52945\\
9.3	0.53192\\
9.4	0.53407\\
9.5	0.53651\\
9.6	0.53894\\
9.7	0.54133\\
9.8	0.54355\\
9.9	0.54564\\
10	0.54756\\
10.1	0.54947\\
10.2	0.55157\\
10.3	0.55365\\
10.4	0.55585\\
10.5	0.55792\\
10.6	0.55993\\
10.7	0.56192\\
10.8	0.56413\\
10.9	0.56604\\
11	0.56807\\
11.1	0.57018\\
11.2	0.57224\\
11.3	0.57418\\
11.4	0.57615\\
11.5	0.57811\\
11.6	0.58008\\
11.7	0.58173\\
11.8	0.58375\\
11.9	0.58556\\
12	0.58742\\
12.1	0.58945\\
12.2	0.59137\\
12.3	0.59319\\
12.4	0.59516\\
12.5	0.59721\\
12.6	0.599\\
12.7	0.60082\\
12.8	0.60266\\
12.9	0.60442\\
13	0.60625\\
13.1	0.60808\\
13.2	0.60958\\
13.3	0.61105\\
13.4	0.61263\\
13.5	0.61436\\
13.6	0.61621\\
13.7	0.61777\\
13.8	0.61917\\
13.9	0.62097\\
14	0.6228\\
14.1	0.62459\\
14.2	0.62593\\
14.3	0.6277\\
14.4	0.62922\\
14.5	0.63083\\
14.6	0.63225\\
14.7	0.63402\\
14.8	0.63572\\
14.9	0.63707\\
15	0.63845\\
15.1	0.63989\\
15.2	0.64148\\
15.3	0.64297\\
15.4	0.64447\\
15.5	0.64581\\
15.6	0.64712\\
15.7	0.6486\\
15.8	0.65009\\
15.9	0.65142\\
16	0.65288\\
16.1	0.65418\\
16.2	0.65562\\
16.3	0.6569\\
16.4	0.65824\\
16.5	0.65971\\
16.6	0.66089\\
16.7	0.66235\\
16.8	0.66361\\
16.9	0.66505\\
17	0.66633\\
17.1	0.6677\\
17.2	0.66903\\
17.3	0.67035\\
17.4	0.67161\\
17.5	0.67282\\
17.6	0.6742\\
17.7	0.67536\\
17.8	0.67648\\
17.9	0.6776\\
18	0.67895\\
18.1	0.68018\\
18.2	0.6814\\
18.3	0.68281\\
18.4	0.68421\\
18.5	0.6853\\
18.6	0.6863\\
18.7	0.68759\\
18.8	0.68897\\
18.9	0.69003\\
19	0.69119\\
19.1	0.69239\\
19.2	0.69355\\
19.3	0.69458\\
19.4	0.69576\\
19.5	0.69702\\
19.6	0.69813\\
19.7	0.69935\\
19.8	0.70038\\
19.9	0.70153\\
20	0.70258\\
20.1	0.70355\\
20.2	0.70463\\
20.3	0.70566\\
20.4	0.7067\\
20.5	0.70785\\
20.6	0.70901\\
20.7	0.71002\\
20.8	0.71104\\
20.9	0.71211\\
21	0.71319\\
21.1	0.71435\\
21.2	0.71532\\
21.3	0.71639\\
21.4	0.71742\\
21.5	0.71853\\
21.6	0.71956\\
21.7	0.72035\\
21.8	0.7215\\
21.9	0.72243\\
22	0.72342\\
22.1	0.72434\\
22.2	0.7254\\
22.3	0.72643\\
22.4	0.72756\\
22.5	0.72854\\
22.6	0.72949\\
22.7	0.73037\\
22.8	0.7313\\
22.9	0.73243\\
23	0.73364\\
23.1	0.73459\\
23.2	0.73555\\
23.3	0.73629\\
23.4	0.73713\\
23.5	0.73791\\
23.6	0.73881\\
23.7	0.73972\\
23.8	0.74055\\
23.9	0.74137\\
24	0.74214\\
24.1	0.74291\\
24.2	0.74379\\
24.3	0.74471\\
24.4	0.74543\\
24.5	0.74621\\
24.6	0.74709\\
24.7	0.74778\\
24.8	0.74857\\
24.9	0.74937\\
25	0.75033\\
25.1	0.75111\\
25.2	0.75195\\
25.3	0.75272\\
25.4	0.75346\\
25.5	0.75432\\
25.6	0.75503\\
25.7	0.75582\\
25.8	0.75652\\
25.9	0.75725\\
26	0.75788\\
26.1	0.75864\\
26.2	0.75944\\
26.3	0.76038\\
26.4	0.76116\\
26.5	0.7619\\
26.6	0.76264\\
26.7	0.76335\\
26.8	0.76398\\
26.9	0.76485\\
27	0.76549\\
27.1	0.76625\\
27.2	0.76704\\
27.3	0.7679\\
27.4	0.76859\\
27.5	0.76933\\
27.6	0.77006\\
27.7	0.77077\\
27.8	0.77155\\
27.9	0.77225\\
28	0.77298\\
28.1	0.77363\\
28.2	0.77432\\
28.3	0.77504\\
28.4	0.77579\\
28.5	0.77645\\
28.6	0.77711\\
28.7	0.77774\\
28.8	0.77852\\
28.9	0.77916\\
29	0.77987\\
29.1	0.78065\\
29.2	0.78138\\
29.3	0.78213\\
29.4	0.7829\\
29.5	0.78367\\
29.6	0.78438\\
29.7	0.78508\\
29.8	0.78581\\
29.9	0.78629\\
30	0.78689\\
30.1	0.78768\\
30.2	0.78833\\
30.3	0.78906\\
30.4	0.78991\\
30.5	0.79049\\
30.6	0.79103\\
30.7	0.79163\\
30.8	0.79223\\
30.9	0.79289\\
31	0.79344\\
31.1	0.79394\\
31.2	0.79452\\
31.3	0.7952\\
31.4	0.7959\\
31.5	0.79651\\
31.6	0.79715\\
31.7	0.79768\\
31.8	0.79826\\
31.9	0.79886\\
32	0.79934\\
32.1	0.79993\\
32.2	0.80062\\
32.3	0.80131\\
32.4	0.80186\\
32.5	0.80239\\
32.6	0.80301\\
32.7	0.80358\\
32.8	0.80406\\
32.9	0.80461\\
33	0.80532\\
33.1	0.80585\\
33.2	0.80642\\
33.3	0.80701\\
33.4	0.80753\\
33.5	0.80803\\
33.6	0.80854\\
33.7	0.80907\\
33.8	0.80955\\
33.9	0.80998\\
34	0.81069\\
34.1	0.8112\\
34.2	0.81172\\
34.3	0.81236\\
34.4	0.8129\\
34.5	0.81339\\
34.6	0.81397\\
34.7	0.81448\\
34.8	0.81503\\
34.9	0.81554\\
35	0.81613\\
35.1	0.81682\\
35.2	0.81737\\
35.3	0.81788\\
35.4	0.81853\\
35.5	0.81896\\
35.6	0.81943\\
35.7	0.81991\\
35.8	0.82051\\
35.9	0.82101\\
36	0.8216\\
36.1	0.82205\\
36.2	0.82254\\
36.3	0.82298\\
36.4	0.82341\\
36.5	0.82389\\
36.6	0.82438\\
36.7	0.82484\\
36.8	0.82538\\
36.9	0.82584\\
37	0.82632\\
37.1	0.82671\\
37.2	0.82721\\
37.3	0.82758\\
37.4	0.82813\\
37.5	0.8286\\
37.6	0.82907\\
37.7	0.8295\\
37.8	0.82994\\
37.9	0.83049\\
38	0.83096\\
38.1	0.83146\\
38.2	0.83196\\
38.3	0.83237\\
38.4	0.8329\\
38.5	0.83334\\
38.6	0.83381\\
38.7	0.83419\\
38.8	0.83457\\
38.9	0.83496\\
39	0.83541\\
39.1	0.83582\\
39.2	0.8363\\
39.3	0.83668\\
39.4	0.83718\\
39.5	0.8376\\
39.6	0.83799\\
39.7	0.83843\\
39.8	0.83885\\
39.9	0.83926\\
40	0.83974\\
40.1	0.84019\\
40.2	0.84077\\
40.3	0.84117\\
40.4	0.84152\\
40.5	0.84195\\
40.6	0.84231\\
40.7	0.84279\\
40.8	0.84324\\
40.9	0.84365\\
41	0.84391\\
41.1	0.84434\\
41.2	0.84482\\
41.3	0.84518\\
41.4	0.84564\\
41.5	0.8459\\
41.6	0.84629\\
41.7	0.84674\\
41.8	0.84719\\
41.9	0.84754\\
42	0.84797\\
42.1	0.84838\\
42.2	0.84878\\
42.3	0.84923\\
42.4	0.84969\\
42.5	0.85002\\
42.6	0.85049\\
42.7	0.85088\\
42.8	0.85123\\
42.9	0.85154\\
43	0.85192\\
43.1	0.85238\\
43.2	0.85274\\
43.3	0.85309\\
43.4	0.85358\\
43.5	0.85396\\
43.6	0.85429\\
43.7	0.85471\\
43.8	0.85502\\
43.9	0.85538\\
44	0.85577\\
44.1	0.85614\\
44.2	0.85647\\
44.3	0.85692\\
44.4	0.85725\\
44.5	0.85758\\
44.6	0.85791\\
44.7	0.85822\\
44.8	0.85862\\
44.9	0.85894\\
45	0.8593\\
45.1	0.85965\\
45.2	0.85994\\
45.3	0.86026\\
45.4	0.86061\\
45.5	0.86094\\
45.6	0.86137\\
45.7	0.86163\\
45.8	0.8619\\
45.9	0.86228\\
46	0.86256\\
46.1	0.86298\\
46.2	0.86337\\
46.3	0.86376\\
46.4	0.86415\\
46.5	0.86438\\
46.6	0.86468\\
46.7	0.86502\\
46.8	0.86529\\
46.9	0.86555\\
47	0.86585\\
47.1	0.86618\\
47.2	0.86655\\
47.3	0.86689\\
47.4	0.86719\\
47.5	0.86752\\
47.6	0.86783\\
47.7	0.86813\\
47.8	0.86848\\
47.9	0.86872\\
48	0.86899\\
48.1	0.86932\\
48.2	0.86956\\
48.3	0.86986\\
48.4	0.87013\\
48.5	0.87044\\
48.6	0.87068\\
48.7	0.87115\\
48.8	0.87138\\
48.9	0.87172\\
49	0.87209\\
49.1	0.87242\\
49.2	0.87277\\
49.3	0.87308\\
49.4	0.87336\\
49.5	0.87363\\
49.6	0.8739\\
49.7	0.87416\\
49.8	0.87449\\
49.9	0.87478\\
50	0.8752\\
50.1	0.87554\\
50.2	0.87582\\
50.3	0.87598\\
50.4	0.87629\\
50.5	0.87658\\
50.6	0.87685\\
50.7	0.87721\\
50.8	0.87752\\
50.9	0.8778\\
51	1\\
};

\addplot [color=varcol,semithick,mark=diamond,mark repeat=25,mark phase=11]
  table[row sep=crcr]{%
0	0\\
0.1	0.11563\\
0.2	0.16234\\
0.3	0.19753\\
0.4	0.22684\\
0.5	0.25259\\
0.6	0.27426\\
0.7	0.29491\\
0.8	0.31405\\
0.9	0.33083\\
1	0.34757\\
1.1	0.36294\\
1.2	0.37692\\
1.3	0.39012\\
1.4	0.40284\\
1.5	0.41514\\
1.6	0.42603\\
1.7	0.43713\\
1.8	0.44807\\
1.9	0.45825\\
2	0.46832\\
2.1	0.47727\\
2.2	0.48647\\
2.3	0.49528\\
2.4	0.50376\\
2.5	0.51224\\
2.6	0.52059\\
2.7	0.52862\\
2.8	0.53629\\
2.9	0.54344\\
3	0.55068\\
3.1	0.55743\\
3.2	0.5641\\
3.3	0.57011\\
3.4	0.57658\\
3.5	0.58267\\
3.6	0.58877\\
3.7	0.59422\\
3.8	0.60019\\
3.9	0.60596\\
4	0.6109\\
4.1	0.61616\\
4.2	0.62123\\
4.3	0.62637\\
4.4	0.63141\\
4.5	0.63645\\
4.6	0.64119\\
4.7	0.6457\\
4.8	0.65042\\
4.9	0.65498\\
5	0.65931\\
5.1	0.66361\\
5.2	0.66782\\
5.3	0.67178\\
5.4	0.67595\\
5.5	0.67992\\
5.6	0.68349\\
5.7	0.6873\\
5.8	0.69107\\
5.9	0.69485\\
6	0.69852\\
6.1	0.70187\\
6.2	0.7057\\
6.3	0.70937\\
6.4	0.71297\\
6.5	0.71607\\
6.6	0.71958\\
6.7	0.7229\\
6.8	0.726\\
6.9	0.7287\\
7	0.73183\\
7.1	0.73465\\
7.2	0.73748\\
7.3	0.74042\\
7.4	0.74315\\
7.5	0.74583\\
7.6	0.74912\\
7.7	0.75207\\
7.8	0.75479\\
7.9	0.75727\\
8	0.75994\\
8.1	0.76238\\
8.2	0.76496\\
8.3	0.76737\\
8.4	0.76976\\
8.5	0.77203\\
8.6	0.77438\\
8.7	0.77672\\
8.8	0.77936\\
8.9	0.78174\\
9	0.7838\\
9.1	0.78596\\
9.2	0.78818\\
9.3	0.79029\\
9.4	0.7924\\
9.5	0.7942\\
9.6	0.7964\\
9.7	0.79836\\
9.8	0.80039\\
9.9	0.80239\\
10	0.80443\\
10.1	0.80636\\
10.2	0.80825\\
10.3	0.81022\\
10.4	0.812\\
10.5	0.81375\\
10.6	0.81542\\
10.7	0.81701\\
10.8	0.81888\\
10.9	0.82046\\
11	0.82215\\
11.1	0.82372\\
11.2	0.82537\\
11.3	0.82686\\
11.4	0.82847\\
11.5	0.82993\\
11.6	0.83163\\
11.7	0.83326\\
11.8	0.83475\\
11.9	0.83606\\
12	0.83755\\
12.1	0.83887\\
12.2	0.84032\\
12.3	0.84167\\
12.4	0.84288\\
12.5	0.84442\\
12.6	0.84578\\
12.7	0.84724\\
12.8	0.8486\\
12.9	0.84989\\
13	0.85134\\
13.1	0.85251\\
13.2	0.8539\\
13.3	0.85518\\
13.4	0.85637\\
13.5	0.85771\\
13.6	0.85895\\
13.7	0.86026\\
13.8	0.86143\\
13.9	0.86251\\
14	0.86391\\
14.1	0.86502\\
14.2	0.86639\\
14.3	0.86758\\
14.4	0.86866\\
14.5	0.86985\\
14.6	0.87093\\
14.7	0.87211\\
14.8	0.87302\\
14.9	0.87409\\
15	0.87502\\
15.1	0.87595\\
15.2	0.87701\\
15.3	0.87803\\
15.4	0.87899\\
15.5	0.87999\\
15.6	0.88102\\
15.7	0.882\\
15.8	0.88302\\
15.9	0.88401\\
16	0.88501\\
16.1	0.88586\\
16.2	0.88686\\
16.3	0.88772\\
16.4	0.88875\\
16.5	0.88972\\
16.6	0.89056\\
16.7	0.89132\\
16.8	0.8922\\
16.9	0.8931\\
17	0.89382\\
17.1	0.89478\\
17.2	0.89557\\
17.3	0.89639\\
17.4	0.89706\\
17.5	0.89782\\
17.6	0.89877\\
17.7	0.89953\\
17.8	0.90036\\
17.9	0.90118\\
18	0.90188\\
18.1	0.90255\\
18.2	0.9034\\
18.3	0.90422\\
18.4	0.90492\\
18.5	0.90562\\
18.6	0.90636\\
18.7	0.9071\\
18.8	0.90777\\
18.9	0.90841\\
19	0.90913\\
19.1	0.90982\\
19.2	0.91051\\
19.3	0.9112\\
19.4	0.91191\\
19.5	0.9126\\
19.6	0.9132\\
19.7	0.91389\\
19.8	0.9145\\
19.9	0.91517\\
20	0.91579\\
20.1	0.91642\\
20.2	0.91708\\
20.3	0.91774\\
20.4	0.91823\\
20.5	0.91879\\
20.6	0.91944\\
20.7	0.91991\\
20.8	0.92053\\
20.9	0.92108\\
21	0.92155\\
21.1	0.92213\\
21.2	0.92271\\
21.3	0.9234\\
21.4	0.92413\\
21.5	0.92467\\
21.6	0.92518\\
21.7	0.92556\\
21.8	0.92618\\
21.9	0.92664\\
22	0.9272\\
22.1	0.92767\\
22.2	0.92833\\
22.3	0.9288\\
22.4	0.9293\\
22.5	0.92972\\
22.6	0.93026\\
22.7	0.93075\\
22.8	0.93133\\
22.9	0.93185\\
23	0.93224\\
23.1	0.93269\\
23.2	0.93324\\
23.3	0.93374\\
23.4	0.93421\\
23.5	0.93473\\
23.6	0.93508\\
23.7	0.93548\\
23.8	0.93595\\
23.9	0.93646\\
24	0.93683\\
24.1	0.93731\\
24.2	0.93768\\
24.3	0.93808\\
24.4	0.93853\\
24.5	0.93899\\
24.6	0.93933\\
24.7	0.93973\\
24.8	0.94012\\
24.9	0.94049\\
25	0.94089\\
25.1	0.94134\\
25.2	0.9418\\
25.3	0.94215\\
25.4	0.94249\\
25.5	0.94274\\
25.6	0.94334\\
25.7	0.94367\\
25.8	0.94408\\
25.9	0.94451\\
26	0.94481\\
26.1	0.94517\\
26.2	0.94555\\
26.3	0.94602\\
26.4	0.94643\\
26.5	0.94673\\
26.6	0.9471\\
26.7	0.94729\\
26.8	0.94757\\
26.9	0.94787\\
27	0.94825\\
27.1	0.94862\\
27.2	0.94894\\
27.3	0.94927\\
27.4	0.94956\\
27.5	0.94993\\
27.6	0.95022\\
27.7	0.95056\\
27.8	0.95087\\
27.9	0.95127\\
28	0.95159\\
28.1	0.95194\\
28.2	0.9523\\
28.3	0.95255\\
28.4	0.95278\\
28.5	0.953\\
28.6	0.95334\\
28.7	0.9537\\
28.8	0.95396\\
28.9	0.95429\\
29	0.95451\\
29.1	0.95476\\
29.2	0.95506\\
29.3	0.95535\\
29.4	0.9557\\
29.5	0.95593\\
29.6	0.95629\\
29.7	0.9566\\
29.8	0.95686\\
29.9	0.95715\\
30	0.95743\\
30.1	0.95772\\
30.2	0.95805\\
30.3	0.95819\\
30.4	0.95847\\
30.5	0.95873\\
30.6	0.95894\\
30.7	0.95925\\
30.8	0.95963\\
30.9	0.95998\\
31	0.96022\\
31.1	0.96054\\
31.2	0.96068\\
31.3	0.96092\\
31.4	0.96115\\
31.5	0.96131\\
31.6	0.96155\\
31.7	0.96172\\
31.8	0.96197\\
31.9	0.96224\\
32	0.96248\\
32.1	0.96264\\
32.2	0.96281\\
32.3	0.9631\\
32.4	0.96334\\
32.5	0.96354\\
32.6	0.96381\\
32.7	0.964\\
32.8	0.96423\\
32.9	0.96446\\
33	0.96463\\
33.1	0.96493\\
33.2	0.96504\\
33.3	0.96521\\
33.4	0.96546\\
33.5	0.96569\\
33.6	0.96585\\
33.7	0.96606\\
33.8	0.96627\\
33.9	0.96644\\
34	0.96665\\
34.1	0.96684\\
34.2	0.96708\\
34.3	0.9673\\
34.4	0.96756\\
34.5	0.96771\\
34.6	0.96792\\
34.7	0.96809\\
34.8	0.96827\\
34.9	0.96845\\
35	0.96861\\
35.1	0.96885\\
35.2	0.96902\\
35.3	0.96924\\
35.4	0.96941\\
35.5	0.96958\\
35.6	0.96975\\
35.7	0.9699\\
35.8	0.97012\\
35.9	0.97034\\
36	0.97055\\
36.1	0.97075\\
36.2	0.97092\\
36.3	0.97107\\
36.4	0.97122\\
36.5	0.9715\\
36.6	0.97164\\
36.7	0.97178\\
36.8	0.97204\\
36.9	0.97223\\
37	0.97237\\
37.1	0.97254\\
37.2	0.97266\\
37.3	0.97278\\
37.4	0.97292\\
37.5	0.97311\\
37.6	0.97318\\
37.7	0.97334\\
37.8	0.97347\\
37.9	0.97363\\
38	0.97382\\
38.1	0.97398\\
38.2	0.9742\\
38.3	0.97434\\
38.4	0.97447\\
38.5	0.97467\\
38.6	0.97488\\
38.7	0.97511\\
38.8	0.97527\\
38.9	0.9754\\
39	0.9755\\
39.1	0.9757\\
39.2	0.9758\\
39.3	0.97595\\
39.4	0.97603\\
39.5	0.97618\\
39.6	0.97634\\
39.7	0.97651\\
39.8	0.9767\\
39.9	0.97688\\
40	0.97704\\
40.1	0.97715\\
40.2	0.97733\\
40.3	0.97741\\
40.4	0.97759\\
40.5	0.9777\\
40.6	0.97782\\
40.7	0.97794\\
40.8	0.97812\\
40.9	0.97826\\
41	0.97836\\
41.1	0.97857\\
41.2	0.97869\\
41.3	0.97884\\
41.4	0.97895\\
41.5	0.97908\\
41.6	0.97914\\
41.7	0.97924\\
41.8	0.97939\\
41.9	0.97948\\
42	0.97962\\
42.1	0.97974\\
42.2	0.97989\\
42.3	0.97995\\
42.4	0.98003\\
42.5	0.98017\\
42.6	0.98031\\
42.7	0.98044\\
42.8	0.98051\\
42.9	0.98058\\
43	0.9807\\
43.1	0.98078\\
43.2	0.98089\\
43.3	0.98102\\
43.4	0.98114\\
43.5	0.9813\\
43.6	0.98143\\
43.7	0.98151\\
43.8	0.98161\\
43.9	0.98173\\
44	0.9818\\
44.1	0.98189\\
44.2	0.98199\\
44.3	0.98207\\
44.4	0.98216\\
44.5	0.98228\\
44.6	0.98241\\
44.7	0.98254\\
44.8	0.98263\\
44.9	0.98271\\
45	0.98282\\
45.1	0.98291\\
45.2	0.98302\\
45.3	0.98311\\
45.4	0.98318\\
45.5	0.98329\\
45.6	0.98341\\
45.7	0.98351\\
45.8	0.98361\\
45.9	0.98373\\
46	0.9838\\
46.1	0.9839\\
46.2	0.984\\
46.3	0.98408\\
46.4	0.98416\\
46.5	0.98425\\
46.6	0.98439\\
46.7	0.9845\\
46.8	0.98458\\
46.9	0.98468\\
47	0.98476\\
47.1	0.98482\\
47.2	0.98494\\
47.3	0.98504\\
47.4	0.98511\\
47.5	0.98523\\
47.6	0.98533\\
47.7	0.98536\\
47.8	0.98546\\
47.9	0.98553\\
48	0.98564\\
48.1	0.9857\\
48.2	0.9858\\
48.3	0.98586\\
48.4	0.9859\\
48.5	0.98596\\
48.6	0.98604\\
48.7	0.98614\\
48.8	0.98624\\
48.9	0.98627\\
49	0.98635\\
49.1	0.98647\\
49.2	0.98656\\
49.3	0.98662\\
49.4	0.98672\\
49.5	0.98679\\
49.6	0.9869\\
49.7	0.98693\\
49.8	0.98699\\
49.9	0.98707\\
50	0.98714\\
50.1	0.98723\\
50.2	0.9873\\
50.3	0.98738\\
50.4	0.98744\\
50.5	0.98751\\
50.6	0.98758\\
50.7	0.98767\\
50.8	0.98771\\
50.9	0.98777\\
51	1\\
};

\addplot [color=maxcol,semithick,mark=square,mark options={solid},mark repeat=25,mark phase=16]
  table[row sep=crcr]{%
0	0\\
0.1	0.15217\\
0.2	0.21193\\
0.3	0.25688\\
0.4	0.29423\\
0.5	0.32627\\
0.6	0.35408\\
0.7	0.379\\
0.8	0.40191\\
0.9	0.42267\\
1	0.44166\\
1.1	0.45879\\
1.2	0.47556\\
1.3	0.49092\\
1.4	0.50631\\
1.5	0.51952\\
1.6	0.53256\\
1.7	0.54466\\
1.8	0.55638\\
1.9	0.56776\\
2	0.57804\\
2.1	0.58834\\
2.2	0.59756\\
2.3	0.607\\
2.4	0.61543\\
2.5	0.62342\\
2.6	0.63162\\
2.7	0.63907\\
2.8	0.64672\\
2.9	0.65374\\
3	0.66051\\
3.1	0.66678\\
3.2	0.67294\\
3.3	0.67907\\
3.4	0.68499\\
3.5	0.69047\\
3.6	0.696\\
3.7	0.70134\\
3.8	0.7065\\
3.9	0.71173\\
4	0.71643\\
4.1	0.7212\\
4.2	0.72555\\
4.3	0.73\\
4.4	0.73434\\
4.5	0.73799\\
4.6	0.74252\\
4.7	0.74678\\
4.8	0.75075\\
4.9	0.75473\\
5	0.75849\\
5.1	0.76155\\
5.2	0.76543\\
5.3	0.76856\\
5.4	0.77217\\
5.5	0.77545\\
5.6	0.77839\\
5.7	0.78133\\
5.8	0.78403\\
5.9	0.78684\\
6	0.78989\\
6.1	0.79274\\
6.2	0.7956\\
6.3	0.79817\\
6.4	0.80076\\
6.5	0.80333\\
6.6	0.80604\\
6.7	0.80859\\
6.8	0.81091\\
6.9	0.81355\\
7	0.81587\\
7.1	0.81796\\
7.2	0.82024\\
7.3	0.82244\\
7.4	0.82469\\
7.5	0.82664\\
7.6	0.82877\\
7.7	0.8309\\
7.8	0.83276\\
7.9	0.83457\\
8	0.83642\\
8.1	0.83823\\
8.2	0.84002\\
8.3	0.84196\\
8.4	0.84375\\
8.5	0.84567\\
8.6	0.84732\\
8.7	0.84906\\
8.8	0.85061\\
8.9	0.85232\\
9	0.85387\\
9.1	0.85542\\
9.2	0.85699\\
9.3	0.85839\\
9.4	0.85977\\
9.5	0.86107\\
9.6	0.86247\\
9.7	0.8641\\
9.8	0.86539\\
9.9	0.8669\\
10	0.86817\\
10.1	0.86954\\
10.2	0.87078\\
10.3	0.87211\\
10.4	0.87326\\
10.5	0.87455\\
10.6	0.8759\\
10.7	0.87705\\
10.8	0.87819\\
10.9	0.87926\\
11	0.88026\\
11.1	0.88151\\
11.2	0.88268\\
11.3	0.88398\\
11.4	0.88516\\
11.5	0.88633\\
11.6	0.88734\\
11.7	0.88831\\
11.8	0.88935\\
11.9	0.89032\\
12	0.89121\\
12.1	0.89217\\
12.2	0.89302\\
12.3	0.89391\\
12.4	0.89481\\
12.5	0.89573\\
12.6	0.89671\\
12.7	0.89771\\
12.8	0.89864\\
12.9	0.89972\\
13	0.90065\\
13.1	0.90164\\
13.2	0.90256\\
13.3	0.90352\\
13.4	0.90434\\
13.5	0.90521\\
13.6	0.90607\\
13.7	0.90687\\
13.8	0.90773\\
13.9	0.9085\\
14	0.90917\\
14.1	0.91001\\
14.2	0.91074\\
14.3	0.91153\\
14.4	0.91219\\
14.5	0.91287\\
14.6	0.91354\\
14.7	0.91438\\
14.8	0.91513\\
14.9	0.91582\\
15	0.91638\\
15.1	0.91713\\
15.2	0.91788\\
15.3	0.91847\\
15.4	0.91908\\
15.5	0.91964\\
15.6	0.92043\\
15.7	0.921\\
15.8	0.92154\\
15.9	0.92221\\
16	0.9229\\
16.1	0.92348\\
16.2	0.92402\\
16.3	0.92462\\
16.4	0.9253\\
16.5	0.92589\\
16.6	0.92633\\
16.7	0.92688\\
16.8	0.92743\\
16.9	0.928\\
17	0.92851\\
17.1	0.92909\\
17.2	0.92967\\
17.3	0.9302\\
17.4	0.9306\\
17.5	0.93106\\
17.6	0.93156\\
17.7	0.9321\\
17.8	0.93271\\
17.9	0.93322\\
18	0.93368\\
18.1	0.93422\\
18.2	0.9348\\
18.3	0.93525\\
18.4	0.93573\\
18.5	0.93627\\
18.6	0.93689\\
18.7	0.93737\\
18.8	0.93777\\
18.9	0.93821\\
19	0.93871\\
19.1	0.93917\\
19.2	0.93955\\
19.3	0.94008\\
19.4	0.94053\\
19.5	0.94097\\
19.6	0.94135\\
19.7	0.94176\\
19.8	0.94225\\
19.9	0.94281\\
20	0.94327\\
20.1	0.94375\\
20.2	0.94412\\
20.3	0.9445\\
20.4	0.94498\\
20.5	0.94539\\
20.6	0.9457\\
20.7	0.94605\\
20.8	0.94657\\
20.9	0.94699\\
21	0.94731\\
21.1	0.9476\\
21.2	0.94803\\
21.3	0.94838\\
21.4	0.94871\\
21.5	0.94899\\
21.6	0.94928\\
21.7	0.94965\\
21.8	0.94996\\
21.9	0.95022\\
22	0.95047\\
22.1	0.95084\\
22.2	0.95123\\
22.3	0.95155\\
22.4	0.95186\\
22.5	0.95229\\
22.6	0.95255\\
22.7	0.95293\\
22.8	0.95327\\
22.9	0.9536\\
23	0.95394\\
23.1	0.95432\\
23.2	0.95469\\
23.3	0.95497\\
23.4	0.95529\\
23.5	0.95559\\
23.6	0.9558\\
23.7	0.95602\\
23.8	0.9563\\
23.9	0.95667\\
24	0.95703\\
24.1	0.95721\\
24.2	0.95754\\
24.3	0.95792\\
24.4	0.95819\\
24.5	0.95839\\
24.6	0.95871\\
24.7	0.95899\\
24.8	0.95925\\
24.9	0.95958\\
25	0.95987\\
25.1	0.96013\\
25.2	0.96035\\
25.3	0.96057\\
25.4	0.96081\\
25.5	0.96104\\
25.6	0.96126\\
25.7	0.96158\\
25.8	0.96174\\
25.9	0.96201\\
26	0.9623\\
26.1	0.9625\\
26.2	0.96268\\
26.3	0.96292\\
26.4	0.96301\\
26.5	0.96322\\
26.6	0.9634\\
26.7	0.96365\\
26.8	0.96382\\
26.9	0.96401\\
27	0.96428\\
27.1	0.96448\\
27.2	0.96476\\
27.3	0.96495\\
27.4	0.96517\\
27.5	0.96533\\
27.6	0.96543\\
27.7	0.96566\\
27.8	0.96587\\
27.9	0.96608\\
28	0.96635\\
28.1	0.96654\\
28.2	0.96674\\
28.3	0.96697\\
28.4	0.96716\\
28.5	0.96736\\
28.6	0.96758\\
28.7	0.96782\\
28.8	0.96805\\
28.9	0.9683\\
29	0.96853\\
29.1	0.96874\\
29.2	0.96888\\
29.3	0.96897\\
29.4	0.96921\\
29.5	0.96933\\
29.6	0.96945\\
29.7	0.96966\\
29.8	0.96984\\
29.9	0.96999\\
30	0.97015\\
30.1	0.97029\\
30.2	0.97047\\
30.3	0.97061\\
30.4	0.97081\\
30.5	0.97099\\
30.6	0.97108\\
30.7	0.97118\\
30.8	0.97133\\
30.9	0.97147\\
31	0.97172\\
31.1	0.9718\\
31.2	0.97195\\
31.3	0.97213\\
31.4	0.97229\\
31.5	0.97241\\
31.6	0.97258\\
31.7	0.97278\\
31.8	0.97289\\
31.9	0.9731\\
32	0.97324\\
32.1	0.97344\\
32.2	0.97361\\
32.3	0.97379\\
32.4	0.97385\\
32.5	0.97398\\
32.6	0.9741\\
32.7	0.97425\\
32.8	0.9744\\
32.9	0.97454\\
33	0.9747\\
33.1	0.97478\\
33.2	0.975\\
33.3	0.97515\\
33.4	0.9753\\
33.5	0.97547\\
33.6	0.9756\\
33.7	0.97573\\
33.8	0.9759\\
33.9	0.97602\\
34	0.97615\\
34.1	0.97627\\
34.2	0.97642\\
34.3	0.97658\\
34.4	0.97672\\
34.5	0.97693\\
34.6	0.97711\\
34.7	0.97731\\
34.8	0.97745\\
34.9	0.97754\\
35	0.9777\\
35.1	0.97785\\
35.2	0.97799\\
35.3	0.9781\\
35.4	0.97822\\
35.5	0.97836\\
35.6	0.97844\\
35.7	0.9785\\
35.8	0.97857\\
35.9	0.97871\\
36	0.97884\\
36.1	0.97899\\
36.2	0.97912\\
36.3	0.9792\\
36.4	0.97933\\
36.5	0.97948\\
36.6	0.97957\\
36.7	0.97971\\
36.8	0.97985\\
36.9	0.97992\\
37	0.98005\\
37.1	0.98023\\
37.2	0.9803\\
37.3	0.9804\\
37.4	0.98049\\
37.5	0.98064\\
37.6	0.98073\\
37.7	0.98082\\
37.8	0.98089\\
37.9	0.98095\\
38	0.98103\\
38.1	0.98111\\
38.2	0.98127\\
38.3	0.98134\\
38.4	0.98146\\
38.5	0.98159\\
38.6	0.98168\\
38.7	0.98178\\
38.8	0.98189\\
38.9	0.982\\
39	0.98209\\
39.1	0.98216\\
39.2	0.98226\\
39.3	0.9824\\
39.4	0.9825\\
39.5	0.98258\\
39.6	0.98272\\
39.7	0.98286\\
39.8	0.98292\\
39.9	0.98304\\
40	0.98315\\
40.1	0.98328\\
40.2	0.98342\\
40.3	0.98347\\
40.4	0.98351\\
40.5	0.98362\\
40.6	0.98371\\
40.7	0.98376\\
40.8	0.9839\\
40.9	0.98398\\
41	0.98408\\
41.1	0.98422\\
41.2	0.98433\\
41.3	0.98438\\
41.4	0.98445\\
41.5	0.98452\\
41.6	0.98461\\
41.7	0.98468\\
41.8	0.98473\\
41.9	0.9848\\
42	0.98484\\
42.1	0.98492\\
42.2	0.98498\\
42.3	0.98502\\
42.4	0.98506\\
42.5	0.9852\\
42.6	0.98528\\
42.7	0.98536\\
42.8	0.98544\\
42.9	0.98557\\
43	0.98566\\
43.1	0.98572\\
43.2	0.98579\\
43.3	0.98589\\
43.4	0.98597\\
43.5	0.98609\\
43.6	0.98617\\
43.7	0.98622\\
43.8	0.98626\\
43.9	0.98634\\
44	0.98641\\
44.1	0.98649\\
44.2	0.98658\\
44.3	0.98668\\
44.4	0.98677\\
44.5	0.98684\\
44.6	0.9869\\
44.7	0.98698\\
44.8	0.98707\\
44.9	0.98715\\
45	0.98722\\
45.1	0.98728\\
45.2	0.98734\\
45.3	0.98739\\
45.4	0.98745\\
45.5	0.9875\\
45.6	0.98752\\
45.7	0.98761\\
45.8	0.98769\\
45.9	0.98778\\
46	0.98787\\
46.1	0.98792\\
46.2	0.98795\\
46.3	0.988\\
46.4	0.98803\\
46.5	0.98812\\
46.6	0.98816\\
46.7	0.98826\\
46.8	0.98828\\
46.9	0.98835\\
47	0.98845\\
47.1	0.98848\\
47.2	0.98849\\
47.3	0.9886\\
47.4	0.98867\\
47.5	0.98874\\
47.6	0.98878\\
47.7	0.9888\\
47.8	0.98883\\
47.9	0.98886\\
48	0.98891\\
48.1	0.98897\\
48.2	0.98904\\
48.3	0.98909\\
48.4	0.98917\\
48.5	0.98922\\
48.6	0.98934\\
48.7	0.98937\\
48.8	0.98944\\
48.9	0.98955\\
49	0.98961\\
49.1	0.98968\\
49.2	0.98975\\
49.3	0.98978\\
49.4	0.98984\\
49.5	0.98989\\
49.6	0.98992\\
49.7	0.98996\\
49.8	0.99004\\
49.9	0.99012\\
50	0.99017\\
50.1	0.99021\\
50.2	0.99028\\
50.3	0.99033\\
50.4	0.99035\\
50.5	0.99041\\
50.6	0.99049\\
50.7	0.99058\\
50.8	0.99068\\
50.9	0.99071\\
51	1\\
};

\addplot [color=cntcol,semithick,mark=triangle,mark options={solid,rotate=180},mark repeat=25,mark phase=20]
  table[row sep=crcr]{%
0	0\\
0.1	0.09895\\
0.2	0.13976\\
0.3	0.16956\\
0.4	0.19503\\
0.5	0.21698\\
0.6	0.23689\\
0.7	0.25515\\
0.8	0.27189\\
0.9	0.288\\
1	0.30282\\
1.1	0.31587\\
1.2	0.32816\\
1.3	0.34078\\
1.4	0.35215\\
1.5	0.36339\\
1.6	0.37393\\
1.7	0.38418\\
1.8	0.39437\\
1.9	0.40348\\
2	0.41322\\
2.1	0.4221\\
2.2	0.43063\\
2.3	0.4384\\
2.4	0.44611\\
2.5	0.45407\\
2.6	0.46148\\
2.7	0.46846\\
2.8	0.47586\\
2.9	0.48327\\
3	0.49043\\
3.1	0.49711\\
3.2	0.50382\\
3.3	0.50988\\
3.4	0.51593\\
3.5	0.52162\\
3.6	0.52723\\
3.7	0.53279\\
3.8	0.53819\\
3.9	0.54405\\
4	0.54896\\
4.1	0.55446\\
4.2	0.55918\\
4.3	0.56449\\
4.4	0.56976\\
4.5	0.57454\\
4.6	0.57916\\
4.7	0.58352\\
4.8	0.58802\\
4.9	0.5929\\
5	0.59737\\
5.1	0.60152\\
5.2	0.60579\\
5.3	0.6101\\
5.4	0.61424\\
5.5	0.61823\\
5.6	0.62221\\
5.7	0.6262\\
5.8	0.63016\\
5.9	0.63436\\
6	0.63802\\
6.1	0.64191\\
6.2	0.64535\\
6.3	0.64893\\
6.4	0.65269\\
6.5	0.65619\\
6.6	0.65961\\
6.7	0.66275\\
6.8	0.66602\\
6.9	0.66945\\
7	0.67282\\
7.1	0.6762\\
7.2	0.67947\\
7.3	0.68273\\
7.4	0.68611\\
7.5	0.68908\\
7.6	0.69195\\
7.7	0.69469\\
7.8	0.69749\\
7.9	0.70049\\
8	0.70305\\
8.1	0.70578\\
8.2	0.70848\\
8.3	0.71135\\
8.4	0.71398\\
8.5	0.71645\\
8.6	0.71884\\
8.7	0.72119\\
8.8	0.72362\\
8.9	0.7258\\
9	0.728\\
9.1	0.73025\\
9.2	0.73239\\
9.3	0.73486\\
9.4	0.73694\\
9.5	0.73932\\
9.6	0.74151\\
9.7	0.74392\\
9.8	0.74625\\
9.9	0.74813\\
10	0.75039\\
10.1	0.75255\\
10.2	0.75471\\
10.3	0.75688\\
10.4	0.75909\\
10.5	0.76107\\
10.6	0.76326\\
10.7	0.76521\\
10.8	0.76721\\
10.9	0.76911\\
11	0.77112\\
11.1	0.77289\\
11.2	0.77465\\
11.3	0.7765\\
11.4	0.77856\\
11.5	0.78057\\
11.6	0.78246\\
11.7	0.78444\\
11.8	0.7861\\
11.9	0.78761\\
12	0.78936\\
12.1	0.791\\
12.2	0.79273\\
12.3	0.79439\\
12.4	0.79596\\
12.5	0.79768\\
12.6	0.79956\\
12.7	0.80111\\
12.8	0.80287\\
12.9	0.80445\\
13	0.80602\\
13.1	0.80769\\
13.2	0.80928\\
13.3	0.8106\\
13.4	0.81217\\
13.5	0.81351\\
13.6	0.81482\\
13.7	0.81593\\
13.8	0.81741\\
13.9	0.81863\\
14	0.82009\\
14.1	0.8215\\
14.2	0.82288\\
14.3	0.8243\\
14.4	0.82573\\
14.5	0.82693\\
14.6	0.82818\\
14.7	0.82943\\
14.8	0.83063\\
14.9	0.83194\\
15	0.833\\
15.1	0.83421\\
15.2	0.83548\\
15.3	0.83675\\
15.4	0.83789\\
15.5	0.83906\\
15.6	0.84032\\
15.7	0.84155\\
15.8	0.84244\\
15.9	0.84358\\
16	0.84483\\
16.1	0.84592\\
16.2	0.84693\\
16.3	0.84795\\
16.4	0.84888\\
16.5	0.84993\\
16.6	0.8509\\
16.7	0.85187\\
16.8	0.85284\\
16.9	0.85386\\
17	0.85487\\
17.1	0.85606\\
17.2	0.85726\\
17.3	0.85825\\
17.4	0.8592\\
17.5	0.86015\\
17.6	0.86128\\
17.7	0.86223\\
17.8	0.86306\\
17.9	0.8639\\
18	0.86472\\
18.1	0.86584\\
18.2	0.86691\\
18.3	0.86777\\
18.4	0.86877\\
18.5	0.86964\\
18.6	0.87038\\
18.7	0.87125\\
18.8	0.87207\\
18.9	0.87299\\
19	0.87383\\
19.1	0.87458\\
19.2	0.87538\\
19.3	0.8763\\
19.4	0.87714\\
19.5	0.87812\\
19.6	0.87898\\
19.7	0.87981\\
19.8	0.88074\\
19.9	0.88152\\
20	0.88222\\
20.1	0.88318\\
20.2	0.88399\\
20.3	0.8848\\
20.4	0.88557\\
20.5	0.88616\\
20.6	0.88691\\
20.7	0.88764\\
20.8	0.8883\\
20.9	0.88906\\
21	0.88983\\
21.1	0.89046\\
21.2	0.8912\\
21.3	0.89186\\
21.4	0.89263\\
21.5	0.89339\\
21.6	0.89409\\
21.7	0.89481\\
21.8	0.89543\\
21.9	0.89595\\
22	0.89655\\
22.1	0.89706\\
22.2	0.8977\\
22.3	0.89833\\
22.4	0.89893\\
22.5	0.89957\\
22.6	0.90012\\
22.7	0.9008\\
22.8	0.90145\\
22.9	0.90186\\
23	0.90245\\
23.1	0.90317\\
23.2	0.90374\\
23.3	0.90435\\
23.4	0.90483\\
23.5	0.90545\\
23.6	0.90603\\
23.7	0.90661\\
23.8	0.90716\\
23.9	0.90774\\
24	0.90822\\
24.1	0.90871\\
24.2	0.90929\\
24.3	0.90984\\
24.4	0.9104\\
24.5	0.9109\\
24.6	0.91143\\
24.7	0.91203\\
24.8	0.91249\\
24.9	0.91302\\
25	0.91357\\
25.1	0.91412\\
25.2	0.91461\\
25.3	0.91517\\
25.4	0.91578\\
25.5	0.91632\\
25.6	0.91669\\
25.7	0.91713\\
25.8	0.91758\\
25.9	0.9181\\
26	0.91864\\
26.1	0.919\\
26.2	0.91955\\
26.3	0.92003\\
26.4	0.92049\\
26.5	0.92106\\
26.6	0.92152\\
26.7	0.92214\\
26.8	0.92263\\
26.9	0.92322\\
27	0.92356\\
27.1	0.92407\\
27.2	0.9245\\
27.3	0.92495\\
27.4	0.92546\\
27.5	0.92591\\
27.6	0.92627\\
27.7	0.92673\\
27.8	0.92722\\
27.9	0.92764\\
28	0.92796\\
28.1	0.92843\\
28.2	0.92884\\
28.3	0.9293\\
28.4	0.92962\\
28.5	0.93008\\
28.6	0.93059\\
28.7	0.93106\\
28.8	0.93153\\
28.9	0.93198\\
29	0.9324\\
29.1	0.93277\\
29.2	0.93309\\
29.3	0.93351\\
29.4	0.93398\\
29.5	0.93437\\
29.6	0.93477\\
29.7	0.93505\\
29.8	0.93553\\
29.9	0.93591\\
30	0.93612\\
30.1	0.93652\\
30.2	0.93696\\
30.3	0.93724\\
30.4	0.93757\\
30.5	0.93798\\
30.6	0.93828\\
30.7	0.93873\\
30.8	0.93904\\
30.9	0.93937\\
31	0.9397\\
31.1	0.94013\\
31.2	0.94045\\
31.3	0.94083\\
31.4	0.94107\\
31.5	0.94141\\
31.6	0.94183\\
31.7	0.94207\\
31.8	0.94238\\
31.9	0.94266\\
32	0.94292\\
32.1	0.94321\\
32.2	0.94349\\
32.3	0.94384\\
32.4	0.94405\\
32.5	0.94442\\
32.6	0.94478\\
32.7	0.9451\\
32.8	0.94541\\
32.9	0.94562\\
33	0.94591\\
33.1	0.94621\\
33.2	0.9465\\
33.3	0.94669\\
33.4	0.94703\\
33.5	0.94731\\
33.6	0.9476\\
33.7	0.94782\\
33.8	0.94806\\
33.9	0.94835\\
34	0.94864\\
34.1	0.94893\\
34.2	0.94918\\
34.3	0.94953\\
34.4	0.94973\\
34.5	0.95005\\
34.6	0.95027\\
34.7	0.95056\\
34.8	0.95082\\
34.9	0.95112\\
35	0.95138\\
35.1	0.95158\\
35.2	0.9518\\
35.3	0.95212\\
35.4	0.95242\\
35.5	0.95266\\
35.6	0.95291\\
35.7	0.95317\\
35.8	0.95336\\
35.9	0.95359\\
36	0.9538\\
36.1	0.95401\\
36.2	0.95425\\
36.3	0.95446\\
36.4	0.95467\\
36.5	0.95488\\
36.6	0.95501\\
36.7	0.95519\\
36.8	0.95545\\
36.9	0.95571\\
37	0.95599\\
37.1	0.95623\\
37.2	0.95648\\
37.3	0.95673\\
37.4	0.95693\\
37.5	0.9571\\
37.6	0.95732\\
37.7	0.95746\\
37.8	0.95767\\
37.9	0.95787\\
38	0.95808\\
38.1	0.95832\\
38.2	0.95852\\
38.3	0.95881\\
38.4	0.95894\\
38.5	0.95915\\
38.6	0.95926\\
38.7	0.9594\\
38.8	0.95962\\
38.9	0.95981\\
39	0.96004\\
39.1	0.96025\\
39.2	0.96045\\
39.3	0.96062\\
39.4	0.96086\\
39.5	0.96109\\
39.6	0.96123\\
39.7	0.96147\\
39.8	0.96164\\
39.9	0.96184\\
40	0.96204\\
40.1	0.96217\\
40.2	0.96233\\
40.3	0.9625\\
40.4	0.96267\\
40.5	0.96281\\
40.6	0.96304\\
40.7	0.96328\\
40.8	0.96352\\
40.9	0.96371\\
41	0.96388\\
41.1	0.96412\\
41.2	0.96425\\
41.3	0.96448\\
41.4	0.96465\\
41.5	0.9648\\
41.6	0.965\\
41.7	0.96514\\
41.8	0.96536\\
41.9	0.96551\\
42	0.96561\\
42.1	0.96578\\
42.2	0.96593\\
42.3	0.96605\\
42.4	0.96626\\
42.5	0.96645\\
42.6	0.96659\\
42.7	0.96672\\
42.8	0.96692\\
42.9	0.96709\\
43	0.96722\\
43.1	0.96742\\
43.2	0.96753\\
43.3	0.96764\\
43.4	0.96784\\
43.5	0.96792\\
43.6	0.96812\\
43.7	0.96829\\
43.8	0.96837\\
43.9	0.96849\\
44	0.96865\\
44.1	0.96884\\
44.2	0.96901\\
44.3	0.96908\\
44.4	0.96923\\
44.5	0.96941\\
44.6	0.96954\\
44.7	0.96968\\
44.8	0.96981\\
44.9	0.96998\\
45	0.9701\\
45.1	0.97024\\
45.2	0.97038\\
45.3	0.97055\\
45.4	0.97079\\
45.5	0.97091\\
45.6	0.97102\\
45.7	0.9712\\
45.8	0.97142\\
45.9	0.97156\\
46	0.97168\\
46.1	0.97181\\
46.2	0.97189\\
46.3	0.97211\\
46.4	0.97223\\
46.5	0.97233\\
46.6	0.97243\\
46.7	0.97257\\
46.8	0.97273\\
46.9	0.97281\\
47	0.97297\\
47.1	0.97315\\
47.2	0.97329\\
47.3	0.9734\\
47.4	0.97352\\
47.5	0.97361\\
47.6	0.97373\\
47.7	0.97384\\
47.8	0.974\\
47.9	0.97421\\
48	0.97434\\
48.1	0.97444\\
48.2	0.97449\\
48.3	0.97465\\
48.4	0.97473\\
48.5	0.97483\\
48.6	0.97501\\
48.7	0.97508\\
48.8	0.97516\\
48.9	0.97525\\
49	0.97539\\
49.1	0.97545\\
49.2	0.97554\\
49.3	0.97565\\
49.4	0.97576\\
49.5	0.97581\\
49.6	0.97591\\
49.7	0.97598\\
49.8	0.97608\\
49.9	0.97614\\
50	0.97631\\
50.1	0.97643\\
50.2	0.97655\\
50.3	0.97667\\
50.4	0.97676\\
50.5	0.97686\\
50.6	0.97694\\
50.7	0.97706\\
50.8	0.97718\\
50.9	0.97733\\
51	1\\
};

\addplot [color=mafcol, semithick, mark=triangle,mark repeat=25,mark phase=24]
  table[row sep=crcr]{%
0	0\\
0.1	0.09765\\
0.2	0.1387\\
0.3	0.16909\\
0.4	0.1953\\
0.5	0.21817\\
0.6	0.23785\\
0.7	0.25561\\
0.8	0.27269\\
0.9	0.28828\\
1	0.30334\\
1.1	0.31777\\
1.2	0.33105\\
1.3	0.34324\\
1.4	0.35495\\
1.5	0.36574\\
1.6	0.37613\\
1.7	0.38672\\
1.8	0.3971\\
1.9	0.40656\\
2	0.41579\\
2.1	0.42495\\
2.2	0.43341\\
2.3	0.44153\\
2.4	0.44947\\
2.5	0.45745\\
2.6	0.46473\\
2.7	0.47228\\
2.8	0.47963\\
2.9	0.48674\\
3	0.49375\\
3.1	0.50034\\
3.2	0.50664\\
3.3	0.51225\\
3.4	0.5184\\
3.5	0.52446\\
3.6	0.5299\\
3.7	0.53544\\
3.8	0.54124\\
3.9	0.54658\\
4	0.55204\\
4.1	0.55719\\
4.2	0.56279\\
4.3	0.56794\\
4.4	0.57288\\
4.5	0.57778\\
4.6	0.58283\\
4.7	0.58747\\
4.8	0.59226\\
4.9	0.59661\\
5	0.60107\\
5.1	0.60518\\
5.2	0.60939\\
5.3	0.61346\\
5.4	0.61765\\
5.5	0.62179\\
5.6	0.62571\\
5.7	0.62951\\
5.8	0.6334\\
5.9	0.63713\\
6	0.64096\\
6.1	0.64433\\
6.2	0.64796\\
6.3	0.652\\
6.4	0.65552\\
6.5	0.65908\\
6.6	0.66269\\
6.7	0.66615\\
6.8	0.6697\\
6.9	0.67299\\
7	0.6763\\
7.1	0.67945\\
7.2	0.68278\\
7.3	0.68599\\
7.4	0.68897\\
7.5	0.69178\\
7.6	0.69471\\
7.7	0.69782\\
7.8	0.70066\\
7.9	0.70324\\
8	0.70557\\
8.1	0.70828\\
8.2	0.71088\\
8.3	0.71346\\
8.4	0.71623\\
8.5	0.71898\\
8.6	0.72156\\
8.7	0.72422\\
8.8	0.72681\\
8.9	0.72941\\
9	0.73186\\
9.1	0.73417\\
9.2	0.73643\\
9.3	0.73866\\
9.4	0.74091\\
9.5	0.74298\\
9.6	0.74528\\
9.7	0.7476\\
9.8	0.74976\\
9.9	0.75211\\
10	0.75429\\
10.1	0.75637\\
10.2	0.75846\\
10.3	0.7604\\
10.4	0.76246\\
10.5	0.76448\\
10.6	0.76669\\
10.7	0.76877\\
10.8	0.7707\\
10.9	0.77252\\
11	0.77431\\
11.1	0.77598\\
11.2	0.77757\\
11.3	0.7793\\
11.4	0.7812\\
11.5	0.7829\\
11.6	0.78472\\
11.7	0.78655\\
11.8	0.78817\\
11.9	0.7899\\
12	0.79166\\
12.1	0.79335\\
12.2	0.79506\\
12.3	0.79671\\
12.4	0.79832\\
12.5	0.79986\\
12.6	0.80164\\
12.7	0.80332\\
12.8	0.80494\\
12.9	0.80658\\
13	0.80822\\
13.1	0.8101\\
13.2	0.81145\\
13.3	0.8129\\
13.4	0.81437\\
13.5	0.81568\\
13.6	0.81688\\
13.7	0.8181\\
13.8	0.81943\\
13.9	0.82111\\
14	0.82257\\
14.1	0.82396\\
14.2	0.82509\\
14.3	0.82643\\
14.4	0.82776\\
14.5	0.82922\\
14.6	0.83065\\
14.7	0.83198\\
14.8	0.8334\\
14.9	0.83493\\
15	0.83622\\
15.1	0.83731\\
15.2	0.83869\\
15.3	0.83996\\
15.4	0.84123\\
15.5	0.84237\\
15.6	0.84348\\
15.7	0.84448\\
15.8	0.84559\\
15.9	0.84693\\
16	0.84794\\
16.1	0.84921\\
16.2	0.85017\\
16.3	0.85121\\
16.4	0.85239\\
16.5	0.85342\\
16.6	0.85458\\
16.7	0.85571\\
16.8	0.85682\\
16.9	0.85801\\
17	0.85901\\
17.1	0.85999\\
17.2	0.86086\\
17.3	0.86185\\
17.4	0.86288\\
17.5	0.86395\\
17.6	0.86475\\
17.7	0.86588\\
17.8	0.86682\\
17.9	0.8679\\
18	0.86893\\
18.1	0.86967\\
18.2	0.87062\\
18.3	0.87161\\
18.4	0.87235\\
18.5	0.87331\\
18.6	0.87432\\
18.7	0.87524\\
18.8	0.87615\\
18.9	0.87708\\
19	0.87807\\
19.1	0.87909\\
19.2	0.87991\\
19.3	0.88088\\
19.4	0.88186\\
19.5	0.88259\\
19.6	0.88345\\
19.7	0.88436\\
19.8	0.88511\\
19.9	0.88584\\
20	0.8867\\
20.1	0.88755\\
20.2	0.88833\\
20.3	0.88904\\
20.4	0.8897\\
20.5	0.89034\\
20.6	0.89102\\
20.7	0.89177\\
20.8	0.89242\\
20.9	0.89313\\
21	0.89378\\
21.1	0.89443\\
21.2	0.89519\\
21.3	0.89585\\
21.4	0.89648\\
21.5	0.89718\\
21.6	0.89776\\
21.7	0.89841\\
21.8	0.8991\\
21.9	0.89986\\
22	0.90047\\
22.1	0.90113\\
22.2	0.90179\\
22.3	0.90241\\
22.4	0.90302\\
22.5	0.90364\\
22.6	0.90424\\
22.7	0.90493\\
22.8	0.90562\\
22.9	0.90637\\
23	0.90693\\
23.1	0.9077\\
23.2	0.90831\\
23.3	0.90886\\
23.4	0.90946\\
23.5	0.91005\\
23.6	0.91058\\
23.7	0.91104\\
23.8	0.91157\\
23.9	0.91212\\
24	0.91258\\
24.1	0.91322\\
24.2	0.91379\\
24.3	0.91428\\
24.4	0.91487\\
24.5	0.9155\\
24.6	0.91606\\
24.7	0.91646\\
24.8	0.91699\\
24.9	0.91749\\
25	0.91803\\
25.1	0.91845\\
25.2	0.919\\
25.3	0.91947\\
25.4	0.91985\\
25.5	0.92031\\
25.6	0.92079\\
25.7	0.92122\\
25.8	0.92175\\
25.9	0.9223\\
26	0.92289\\
26.1	0.92331\\
26.2	0.9238\\
26.3	0.92423\\
26.4	0.92468\\
26.5	0.92515\\
26.6	0.92566\\
26.7	0.92619\\
26.8	0.92664\\
26.9	0.9271\\
27	0.92758\\
27.1	0.92799\\
27.2	0.92838\\
27.3	0.9287\\
27.4	0.92917\\
27.5	0.92961\\
27.6	0.93009\\
27.7	0.93051\\
27.8	0.931\\
27.9	0.93135\\
28	0.93175\\
28.1	0.93213\\
28.2	0.93255\\
28.3	0.93297\\
28.4	0.93337\\
28.5	0.93382\\
28.6	0.93415\\
28.7	0.93465\\
28.8	0.93503\\
28.9	0.9354\\
29	0.93573\\
29.1	0.93606\\
29.2	0.93645\\
29.3	0.93682\\
29.4	0.93724\\
29.5	0.9377\\
29.6	0.93809\\
29.7	0.93848\\
29.8	0.93874\\
29.9	0.93899\\
30	0.93943\\
30.1	0.93979\\
30.2	0.94003\\
30.3	0.94036\\
30.4	0.94065\\
30.5	0.94108\\
30.6	0.94145\\
30.7	0.94174\\
30.8	0.94208\\
30.9	0.94237\\
31	0.94268\\
31.1	0.94298\\
31.2	0.94337\\
31.3	0.94372\\
31.4	0.9441\\
31.5	0.94444\\
31.6	0.94488\\
31.7	0.94519\\
31.8	0.9456\\
31.9	0.94587\\
32	0.94618\\
32.1	0.9465\\
32.2	0.94682\\
32.3	0.94713\\
32.4	0.94743\\
32.5	0.9478\\
32.6	0.94807\\
32.7	0.94839\\
32.8	0.94874\\
32.9	0.949\\
33	0.94932\\
33.1	0.94962\\
33.2	0.94983\\
33.3	0.95004\\
33.4	0.95026\\
33.5	0.95048\\
33.6	0.95079\\
33.7	0.95109\\
33.8	0.9513\\
33.9	0.95153\\
34	0.95183\\
34.1	0.95211\\
34.2	0.95237\\
34.3	0.95263\\
34.4	0.95294\\
34.5	0.95323\\
34.6	0.95351\\
34.7	0.95377\\
34.8	0.95405\\
34.9	0.95434\\
35	0.95452\\
35.1	0.95473\\
35.2	0.95501\\
35.3	0.95527\\
35.4	0.95545\\
35.5	0.95576\\
35.6	0.95592\\
35.7	0.9562\\
35.8	0.95632\\
35.9	0.95665\\
36	0.95679\\
36.1	0.95703\\
36.2	0.95726\\
36.3	0.95748\\
36.4	0.95768\\
36.5	0.95795\\
36.6	0.95818\\
36.7	0.95842\\
36.8	0.95871\\
36.9	0.95893\\
37	0.95916\\
37.1	0.95937\\
37.2	0.95951\\
37.3	0.95974\\
37.4	0.95996\\
37.5	0.96015\\
37.6	0.96035\\
37.7	0.9606\\
37.8	0.96086\\
37.9	0.96112\\
38	0.96131\\
38.1	0.96155\\
38.2	0.96183\\
38.3	0.96199\\
38.4	0.9621\\
38.5	0.96238\\
38.6	0.96266\\
38.7	0.9628\\
38.8	0.96297\\
38.9	0.96322\\
39	0.96348\\
39.1	0.96367\\
39.2	0.96396\\
39.3	0.96416\\
39.4	0.96427\\
39.5	0.96444\\
39.6	0.96458\\
39.7	0.96472\\
39.8	0.96488\\
39.9	0.96501\\
40	0.96526\\
40.1	0.96541\\
40.2	0.96559\\
40.3	0.96581\\
40.4	0.96592\\
40.5	0.96615\\
40.6	0.96629\\
40.7	0.96646\\
40.8	0.96656\\
40.9	0.9667\\
41	0.96689\\
41.1	0.96714\\
41.2	0.96725\\
41.3	0.96745\\
41.4	0.9676\\
41.5	0.96776\\
41.6	0.96792\\
41.7	0.96813\\
41.8	0.96833\\
41.9	0.96847\\
42	0.96858\\
42.1	0.96876\\
42.2	0.96892\\
42.3	0.96904\\
42.4	0.96921\\
42.5	0.96938\\
42.6	0.96951\\
42.7	0.96967\\
42.8	0.96976\\
42.9	0.96985\\
43	0.97001\\
43.1	0.97019\\
43.2	0.97036\\
43.3	0.97047\\
43.4	0.97062\\
43.5	0.97076\\
43.6	0.97087\\
43.7	0.97106\\
43.8	0.97117\\
43.9	0.97135\\
44	0.97147\\
44.1	0.9716\\
44.2	0.97179\\
44.3	0.97193\\
44.4	0.97209\\
44.5	0.9722\\
44.6	0.97238\\
44.7	0.97256\\
44.8	0.9727\\
44.9	0.97286\\
45	0.97298\\
45.1	0.97313\\
45.2	0.97323\\
45.3	0.97336\\
45.4	0.97345\\
45.5	0.97362\\
45.6	0.97376\\
45.7	0.9739\\
45.8	0.97399\\
45.9	0.97421\\
46	0.9744\\
46.1	0.97452\\
46.2	0.97463\\
46.3	0.97478\\
46.4	0.9749\\
46.5	0.97504\\
46.6	0.97521\\
46.7	0.97528\\
46.8	0.97539\\
46.9	0.97554\\
47	0.97566\\
47.1	0.97575\\
47.2	0.97587\\
47.3	0.97602\\
47.4	0.9761\\
47.5	0.97624\\
47.6	0.9764\\
47.7	0.97652\\
47.8	0.97661\\
47.9	0.97674\\
48	0.97685\\
48.1	0.97698\\
48.2	0.9771\\
48.3	0.97719\\
48.4	0.9773\\
48.5	0.97745\\
48.6	0.97756\\
48.7	0.9777\\
48.8	0.97782\\
48.9	0.97794\\
49	0.978\\
49.1	0.97809\\
49.2	0.97823\\
49.3	0.97833\\
49.4	0.9784\\
49.5	0.97848\\
49.6	0.97861\\
49.7	0.97866\\
49.8	0.97878\\
49.9	0.9789\\
50	0.97895\\
50.1	0.97906\\
50.2	0.97915\\
50.3	0.97929\\
50.4	0.97938\\
50.5	0.97948\\
50.6	0.97958\\
50.7	0.97971\\
50.8	0.97977\\
50.9	0.97998\\
51	1\\
};

\end{axis}
\end{tikzpicture}%

%% file: fig/var_err2.tex
\begin{tikzpicture}
\pgfplotsset{every tick label/.append style={font=\tiny}}
\tikzstyle{dotted}= [dash pattern=on \pgflinewidth off 0.5mm] 
\tikzstyle{dashed}= [dash pattern=on 7.5*0.8*0.8pt off 7.5*0.4*0.8pt]
\tikzstyle{dashdotted} = [dash pattern=on 7.5*0.8*0.6pt off 7.5*0.8*0.3pt on \the\pgflinewidth off 7.5*0.8*0.3pt]
\tikzstyle{dotted2} = [dash pattern=on 7.5*0.8*0.3pt off 7.5*0.8*0.2pt]

\begin{axis}[%
width=\fwidth,
height=\fheight,
at={(0,0)},
xmin=0,
xmax=1800,
xlabel near ticks,
xlabel style={font=\scriptsize\color{white!15!black}},
xlabel={$\nu_{\text{var}}$},
ylabel near ticks,
ymin=0.5,
ymax=1,
ylabel style={font=\scriptsize\color{white!15!black}},
ylabel={Empirical CDF},axis background/.style={fill=white},
xmajorgrids,
ymajorgrids
]
\addplot [color=msecol, semithick,mark=x,mark repeat=10,mark phase=1]
  table[row sep=crcr]{%
0	0\\
10	0.25123\\
20	0.34948\\
30	0.41782\\
40	0.47111\\
50	0.5151\\
60	0.55252\\
70	0.58301\\
80	0.60937\\
90	0.63271\\
100	0.65359\\
110	0.67165\\
120	0.68783\\
130	0.70304\\
140	0.7166\\
150	0.72976\\
160	0.74112\\
170	0.75201\\
180	0.76211\\
190	0.77136\\
200	0.77976\\
210	0.78816\\
220	0.79569\\
230	0.80303\\
240	0.80965\\
250	0.81628\\
260	0.82246\\
270	0.82778\\
280	0.83324\\
290	0.8385\\
300	0.84357\\
310	0.84857\\
320	0.85297\\
330	0.85775\\
340	0.86167\\
350	0.86593\\
360	0.86978\\
370	0.87356\\
380	0.87746\\
390	0.88076\\
400	0.88379\\
410	0.88698\\
420	0.88975\\
430	0.89259\\
440	0.89531\\
450	0.89802\\
460	0.9003\\
470	0.90265\\
480	0.90493\\
490	0.90701\\
500	0.90903\\
510	0.91128\\
520	0.91329\\
530	0.91526\\
540	0.91726\\
550	0.91905\\
560	0.92083\\
570	0.92273\\
580	0.92457\\
590	0.9263\\
600	0.92799\\
610	0.92965\\
620	0.93127\\
630	0.93274\\
640	0.93402\\
650	0.93543\\
660	0.93678\\
670	0.93791\\
680	0.93907\\
690	0.94016\\
700	0.94123\\
710	0.94241\\
720	0.94359\\
730	0.94468\\
740	0.9459\\
750	0.94715\\
760	0.94826\\
770	0.94922\\
780	0.95023\\
790	0.95103\\
800	0.95215\\
810	0.95336\\
820	0.95421\\
830	0.95507\\
840	0.95605\\
850	0.95681\\
860	0.95762\\
870	0.95838\\
880	0.95917\\
890	0.9599\\
900	0.96042\\
910	0.96121\\
920	0.96189\\
930	0.96245\\
940	0.963\\
950	0.96357\\
960	0.96426\\
970	0.96498\\
980	0.96568\\
990	0.96622\\
1000	0.96669\\
1010	0.96725\\
1020	0.96786\\
1030	0.9685\\
1040	0.96901\\
1050	0.96952\\
1060	0.9701\\
1070	0.97052\\
1080	0.97098\\
1090	0.97148\\
1100	0.97189\\
1110	0.97234\\
1120	0.97282\\
1130	0.97325\\
1140	0.97368\\
1150	0.97413\\
1160	0.97464\\
1170	0.97521\\
1180	0.97556\\
1190	0.97599\\
1200	0.97636\\
1210	0.97669\\
1220	0.97711\\
1230	0.97739\\
1240	0.97771\\
1250	0.97811\\
1260	0.97832\\
1270	0.97866\\
1280	0.97896\\
1290	0.97917\\
1300	0.97948\\
1310	0.97987\\
1320	0.98019\\
1330	0.98043\\
1340	0.98076\\
1350	0.981\\
1360	0.98119\\
1370	0.98142\\
1380	0.98168\\
1390	0.982\\
1400	0.98225\\
1410	0.98251\\
1420	0.98273\\
1430	0.98299\\
1440	0.98321\\
1450	0.9834\\
1460	0.98363\\
1470	0.98386\\
1480	0.98412\\
1490	0.9843\\
1500	0.98455\\
1510	0.98473\\
1520	0.98496\\
1530	0.9852\\
1540	0.98543\\
1550	0.98567\\
1560	0.98589\\
1570	0.9861\\
1580	0.98627\\
1590	0.98646\\
1600	0.9866\\
1610	0.98681\\
1620	0.987\\
1630	0.98716\\
1640	0.9874\\
1650	0.98756\\
1660	0.98775\\
1670	0.98792\\
1680	0.98806\\
1690	0.98821\\
1700	0.98835\\
1710	0.98851\\
1720	0.98871\\
1730	0.98894\\
1740	0.98907\\
1750	0.98914\\
1760	0.98936\\
1770	0.98946\\
1780	0.98963\\
1790	0.98973\\
1800	0.98987\\
1810	0.99\\
1820	0.99008\\
1830	0.99017\\
1840	0.99025\\
1850	0.9904\\
1860	0.99051\\
1870	0.99062\\
1880	0.99077\\
1890	0.99088\\
1900	0.99099\\
1910	0.99109\\
1920	0.99121\\
1930	0.99128\\
1940	0.99137\\
1950	0.99146\\
1960	0.99157\\
1970	0.99169\\
1980	0.9918\\
1990	0.99189\\
2000	0.992\\
2010	0.99215\\
2020	0.99225\\
2030	0.99238\\
2040	0.99251\\
2050	0.9926\\
2060	0.99273\\
2070	0.99281\\
2080	0.99291\\
2090	0.993\\
2100	0.99305\\
2110	0.99311\\
2120	0.99318\\
2130	0.99323\\
2140	0.99331\\
2150	0.99341\\
2160	0.99347\\
2170	0.99349\\
2180	0.99356\\
2190	0.99365\\
2200	0.9937\\
2210	0.99382\\
2220	0.9939\\
2230	0.99393\\
2240	0.99401\\
2250	0.99409\\
2260	0.99413\\
2270	0.99419\\
2280	0.99427\\
2290	0.99432\\
2300	0.9944\\
2310	0.99443\\
2320	0.99447\\
2330	0.99456\\
2340	0.99462\\
2350	0.99467\\
2360	0.99472\\
2370	0.99477\\
2380	0.9948\\
2390	0.99484\\
2400	0.99488\\
2410	0.99489\\
2420	0.995\\
2430	0.99506\\
2440	0.9951\\
2450	0.9951\\
2460	0.99514\\
2470	0.99519\\
2480	0.99522\\
2490	0.99528\\
2500	0.99529\\
2510	0.99534\\
2520	0.99535\\
2530	0.99542\\
2540	0.99548\\
2550	0.9955\\
2560	0.99554\\
2570	0.99556\\
2580	0.99562\\
2590	0.99565\\
2600	0.99568\\
2610	0.99573\\
2620	0.99576\\
2630	0.99578\\
2640	0.99582\\
2650	0.99587\\
2660	0.99594\\
2670	0.99598\\
2680	0.996\\
2690	0.99602\\
2700	0.99607\\
2710	0.99613\\
2720	0.9962\\
2730	0.99626\\
2740	0.99628\\
2750	0.99632\\
2760	0.99635\\
2770	0.99638\\
2780	0.99645\\
2790	0.9965\\
2800	0.99655\\
2810	0.99657\\
2820	0.9966\\
2830	0.99665\\
2840	0.99668\\
2850	0.99672\\
2860	0.99674\\
2870	0.99677\\
2880	0.99679\\
2890	0.99684\\
2900	0.99688\\
2910	0.99691\\
2920	0.99693\\
2930	0.99695\\
2940	0.99697\\
2950	0.997\\
2960	0.99701\\
2970	0.99702\\
2980	0.99706\\
2990	0.99709\\
3000	0.99712\\
3010	0.99718\\
3020	0.99721\\
3030	0.99723\\
3040	0.99725\\
3050	0.99729\\
3060	0.99731\\
3070	0.99732\\
3080	0.99739\\
3090	0.99741\\
3100	0.99742\\
3110	0.99745\\
3120	0.99747\\
3130	0.9975\\
3140	0.99754\\
3150	0.99755\\
3160	0.99756\\
3170	0.99763\\
3180	0.99766\\
3190	0.99768\\
3200	0.99768\\
3210	0.9977\\
3220	0.99773\\
3230	0.99779\\
3240	0.99781\\
3250	0.99783\\
3260	0.99784\\
3270	0.99786\\
3280	0.99787\\
3290	0.99789\\
3300	0.99796\\
3310	0.99799\\
3320	0.998\\
3330	0.99802\\
3340	0.99804\\
3350	0.99804\\
3360	0.99805\\
3370	0.9981\\
3380	0.9981\\
3390	0.99813\\
3400	0.99816\\
3410	0.99817\\
3420	0.99817\\
3430	0.99819\\
3440	0.99823\\
3450	0.99826\\
3460	0.99826\\
3470	0.99827\\
3480	0.99827\\
3490	0.99827\\
3500	0.99829\\
3510	0.99829\\
3520	0.99831\\
3530	0.99834\\
3540	0.99835\\
3550	0.99839\\
3560	0.99841\\
3570	0.99842\\
3580	0.99845\\
3590	0.9985\\
3600	0.99851\\
3610	0.99852\\
3620	0.99853\\
3630	0.99855\\
3640	0.99858\\
3650	0.99858\\
3660	0.99858\\
3670	0.99858\\
3680	0.9986\\
3690	0.99862\\
3700	0.99863\\
3710	0.99865\\
3720	0.99865\\
3730	0.99866\\
3740	0.99867\\
3750	0.99867\\
3760	0.99867\\
3770	0.99867\\
3780	0.99868\\
3790	0.99871\\
3800	0.99873\\
3810	0.99873\\
3820	0.99874\\
3830	0.99874\\
3840	0.99876\\
3850	0.99877\\
3860	0.99878\\
3870	0.99879\\
3880	0.99879\\
3890	0.9988\\
3900	0.99883\\
3910	0.99884\\
3920	0.99887\\
3930	0.99887\\
3940	0.99887\\
3950	0.99887\\
3960	0.99888\\
3970	0.99893\\
3980	0.99893\\
3990	0.99894\\
4000	0.99895\\
4010	0.99896\\
4020	0.99896\\
4030	0.99897\\
4040	0.99897\\
4050	0.99898\\
4060	0.99899\\
4070	0.99899\\
4080	0.99902\\
4090	0.99902\\
4100	0.99904\\
4110	0.99905\\
4120	0.99906\\
4130	0.99906\\
4140	0.99906\\
4150	0.99906\\
4160	0.99907\\
4170	0.99907\\
4180	0.9991\\
4190	0.99911\\
4200	0.99912\\
4210	0.99913\\
4220	0.99914\\
4230	0.99914\\
4240	0.99915\\
4250	0.99916\\
4260	0.99917\\
4270	0.99917\\
4280	0.99917\\
4290	0.99918\\
4300	0.99919\\
4310	0.99919\\
4320	0.99919\\
4330	0.99919\\
4340	0.99919\\
4350	0.99919\\
4360	0.99921\\
4370	0.99921\\
4380	0.99922\\
4390	0.99924\\
4400	0.99925\\
4410	0.99925\\
4420	0.99925\\
4430	0.99925\\
4440	0.99925\\
4450	0.99925\\
4460	0.99927\\
4470	0.99927\\
4480	0.99927\\
4490	0.99927\\
4500	0.99927\\
4510	0.99928\\
4520	0.99929\\
4530	0.99929\\
4540	0.9993\\
4550	0.99931\\
4560	0.99932\\
4570	0.99932\\
4580	0.99932\\
4590	0.99932\\
4600	0.99932\\
4610	0.99932\\
4620	0.99934\\
4630	0.99934\\
4640	0.99934\\
4650	0.99934\\
4660	0.99937\\
4670	0.99937\\
4680	0.99937\\
4690	0.99937\\
4700	0.99937\\
4710	0.99937\\
4720	0.99937\\
4730	0.99938\\
4740	0.99938\\
4750	0.99938\\
4760	0.99938\\
4770	0.99939\\
4780	0.99939\\
4790	0.9994\\
4800	0.99941\\
4810	0.99941\\
4820	0.99942\\
4830	0.99943\\
4840	0.99943\\
4850	0.99943\\
4860	0.99943\\
4870	0.99943\\
4880	0.99944\\
4890	0.99944\\
4900	0.99944\\
4910	0.99945\\
4920	0.99945\\
4930	0.99945\\
4940	0.99945\\
4950	0.99945\\
4960	0.99946\\
4970	0.99946\\
4980	0.99946\\
4990	0.99946\\
5000	0.99946\\
5010	0.99947\\
5020	0.99947\\
5030	0.99948\\
5040	0.99949\\
5050	0.9995\\
5060	0.9995\\
5070	0.9995\\
5080	0.99951\\
5090	0.99951\\
5100	0.99951\\
5110	0.99951\\
5120	0.99952\\
5130	0.99954\\
5140	0.99955\\
5150	0.99955\\
5160	0.99955\\
5170	0.99956\\
5180	0.99956\\
5190	0.99956\\
5200	1\\
};

\addplot [color=avgcol,semithick,mark=o,mark repeat=25,mark phase=6]
  table[row sep=crcr]{%
0	0\\
10	0.17106\\
20	0.24212\\
30	0.29445\\
40	0.33681\\
50	0.37361\\
60	0.40606\\
70	0.43427\\
80	0.46004\\
90	0.4834\\
100	0.50464\\
110	0.52409\\
120	0.5414\\
130	0.55838\\
140	0.5733\\
150	0.58765\\
160	0.60054\\
170	0.61346\\
180	0.62499\\
190	0.63547\\
200	0.64538\\
210	0.65487\\
220	0.66395\\
230	0.67247\\
240	0.67995\\
250	0.68763\\
260	0.6947\\
270	0.70147\\
280	0.70755\\
290	0.7137\\
300	0.71926\\
310	0.72482\\
320	0.72999\\
330	0.73536\\
340	0.74043\\
350	0.74568\\
360	0.75039\\
370	0.75495\\
380	0.75937\\
390	0.7632\\
400	0.76722\\
410	0.77136\\
420	0.7755\\
430	0.77896\\
440	0.7824\\
450	0.78533\\
460	0.78875\\
470	0.79197\\
480	0.79527\\
490	0.79837\\
500	0.80172\\
510	0.80483\\
520	0.80752\\
530	0.81019\\
540	0.81313\\
550	0.8156\\
560	0.81829\\
570	0.82086\\
580	0.82349\\
590	0.82586\\
600	0.82812\\
610	0.83037\\
620	0.83261\\
630	0.83459\\
640	0.83668\\
650	0.8388\\
660	0.84081\\
670	0.8428\\
680	0.84492\\
690	0.84694\\
700	0.84865\\
710	0.85059\\
720	0.85242\\
730	0.85434\\
740	0.85601\\
750	0.85761\\
760	0.85914\\
770	0.86045\\
780	0.86191\\
790	0.86356\\
800	0.86502\\
810	0.86664\\
820	0.86804\\
830	0.86951\\
840	0.8708\\
850	0.87225\\
860	0.87369\\
870	0.87493\\
880	0.87613\\
890	0.87747\\
900	0.87872\\
910	0.87995\\
920	0.88114\\
930	0.88218\\
940	0.88312\\
950	0.88412\\
960	0.88542\\
970	0.88654\\
980	0.88769\\
990	0.88878\\
1000	0.88974\\
1010	0.89089\\
1020	0.89193\\
1030	0.89297\\
1040	0.89398\\
1050	0.89479\\
1060	0.89581\\
1070	0.89681\\
1080	0.89768\\
1090	0.89874\\
1100	0.8996\\
1110	0.90056\\
1120	0.90152\\
1130	0.90238\\
1140	0.90318\\
1150	0.90405\\
1160	0.90486\\
1170	0.90561\\
1180	0.90651\\
1190	0.90751\\
1200	0.90821\\
1210	0.90899\\
1220	0.90977\\
1230	0.91051\\
1240	0.91143\\
1250	0.91196\\
1260	0.91284\\
1270	0.91359\\
1280	0.91439\\
1290	0.91532\\
1300	0.91599\\
1310	0.91665\\
1320	0.91745\\
1330	0.91821\\
1340	0.91887\\
1350	0.91942\\
1360	0.92008\\
1370	0.92084\\
1380	0.92131\\
1390	0.92185\\
1400	0.92239\\
1410	0.92309\\
1420	0.92367\\
1430	0.92419\\
1440	0.92479\\
1450	0.92538\\
1460	0.92584\\
1470	0.92642\\
1480	0.92695\\
1490	0.9275\\
1500	0.92802\\
1510	0.92853\\
1520	0.92898\\
1530	0.92943\\
1540	0.92989\\
1550	0.93042\\
1560	0.93087\\
1570	0.93136\\
1580	0.93187\\
1590	0.9323\\
1600	0.93277\\
1610	0.93331\\
1620	0.93373\\
1630	0.93419\\
1640	0.93469\\
1650	0.93524\\
1660	0.93563\\
1670	0.93617\\
1680	0.93661\\
1690	0.937\\
1700	0.93743\\
1710	0.93789\\
1720	0.93826\\
1730	0.93866\\
1740	0.93908\\
1750	0.93947\\
1760	0.9399\\
1770	0.94024\\
1780	0.94059\\
1790	0.94095\\
1800	0.94129\\
1810	0.94168\\
1820	0.94208\\
1830	0.94252\\
1840	0.9427\\
1850	0.94305\\
1860	0.94361\\
1870	0.94403\\
1880	0.94445\\
1890	0.94479\\
1900	0.9452\\
1910	0.94557\\
1920	0.94594\\
1930	0.94627\\
1940	0.94662\\
1950	0.94698\\
1960	0.94738\\
1970	0.94762\\
1980	0.94799\\
1990	0.94825\\
2000	0.94854\\
2010	0.9489\\
2020	0.94927\\
2030	0.9495\\
2040	0.94977\\
2050	0.95002\\
2060	0.95033\\
2070	0.95061\\
2080	0.95096\\
2090	0.95131\\
2100	0.95156\\
2110	0.95176\\
2120	0.95216\\
2130	0.95241\\
2140	0.95266\\
2150	0.95296\\
2160	0.9532\\
2170	0.95348\\
2180	0.95378\\
2190	0.95402\\
2200	0.95428\\
2210	0.95458\\
2220	0.95482\\
2230	0.95506\\
2240	0.95532\\
2250	0.95553\\
2260	0.95579\\
2270	0.95604\\
2280	0.95628\\
2290	0.95645\\
2300	0.95671\\
2310	0.95694\\
2320	0.95719\\
2330	0.95736\\
2340	0.95768\\
2350	0.95791\\
2360	0.95817\\
2370	0.95835\\
2380	0.95852\\
2390	0.95869\\
2400	0.95888\\
2410	0.95913\\
2420	0.95938\\
2430	0.95967\\
2440	0.95985\\
2450	0.96012\\
2460	0.96025\\
2470	0.96046\\
2480	0.96064\\
2490	0.96087\\
2500	0.96097\\
2510	0.96122\\
2520	0.96141\\
2530	0.96166\\
2540	0.96179\\
2550	0.96204\\
2560	0.96219\\
2570	0.96235\\
2580	0.96252\\
2590	0.96276\\
2600	0.96297\\
2610	0.9632\\
2620	0.96345\\
2630	0.96363\\
2640	0.96379\\
2650	0.96395\\
2660	0.96418\\
2670	0.96438\\
2680	0.96457\\
2690	0.96471\\
2700	0.96489\\
2710	0.96505\\
2720	0.96518\\
2730	0.96534\\
2740	0.96551\\
2750	0.96562\\
2760	0.96577\\
2770	0.96602\\
2780	0.96626\\
2790	0.96639\\
2800	0.9666\\
2810	0.96683\\
2820	0.96699\\
2830	0.96715\\
2840	0.9673\\
2850	0.96747\\
2860	0.96761\\
2870	0.96775\\
2880	0.96795\\
2890	0.96801\\
2900	0.9682\\
2910	0.96839\\
2920	0.96853\\
2930	0.96865\\
2940	0.96877\\
2950	0.96895\\
2960	0.96913\\
2970	0.96929\\
2980	0.96946\\
2990	0.96965\\
3000	0.96979\\
3010	0.96992\\
3020	0.97002\\
3030	0.97021\\
3040	0.97025\\
3050	0.97038\\
3060	0.9705\\
3070	0.97067\\
3080	0.97083\\
3090	0.97091\\
3100	0.97105\\
3110	0.97116\\
3120	0.97129\\
3130	0.97139\\
3140	0.97152\\
3150	0.97162\\
3160	0.97174\\
3170	0.9718\\
3180	0.97195\\
3190	0.97215\\
3200	0.97228\\
3210	0.97242\\
3220	0.97256\\
3230	0.97266\\
3240	0.97285\\
3250	0.97297\\
3260	0.97312\\
3270	0.97326\\
3280	0.97341\\
3290	0.97352\\
3300	0.97361\\
3310	0.9737\\
3320	0.9738\\
3330	0.97392\\
3340	0.97403\\
3350	0.97417\\
3360	0.97432\\
3370	0.97443\\
3380	0.97459\\
3390	0.97475\\
3400	0.97485\\
3410	0.97494\\
3420	0.97503\\
3430	0.97519\\
3440	0.97532\\
3450	0.9754\\
3460	0.97545\\
3470	0.97556\\
3480	0.97565\\
3490	0.97571\\
3500	0.97583\\
3510	0.97596\\
3520	0.97601\\
3530	0.97605\\
3540	0.97613\\
3550	0.97621\\
3560	0.97627\\
3570	0.97634\\
3580	0.97644\\
3590	0.97651\\
3600	0.97661\\
3610	0.97676\\
3620	0.97688\\
3630	0.97698\\
3640	0.97708\\
3650	0.97719\\
3660	0.97724\\
3670	0.97737\\
3680	0.97742\\
3690	0.97747\\
3700	0.9776\\
3710	0.97767\\
3720	0.97775\\
3730	0.97782\\
3740	0.97789\\
3750	0.97798\\
3760	0.97806\\
3770	0.97813\\
3780	0.97824\\
3790	0.9783\\
3800	0.97839\\
3810	0.97845\\
3820	0.97854\\
3830	0.97863\\
3840	0.97871\\
3850	0.97883\\
3860	0.97894\\
3870	0.97906\\
3880	0.97913\\
3890	0.97917\\
3900	0.97927\\
3910	0.97936\\
3920	0.97948\\
3930	0.97954\\
3940	0.97972\\
3950	0.9798\\
3960	0.97988\\
3970	0.98003\\
3980	0.98007\\
3990	0.98019\\
4000	0.98027\\
4010	0.98033\\
4020	0.98036\\
4030	0.98043\\
4040	0.98053\\
4050	0.9806\\
4060	0.98064\\
4070	0.98076\\
4080	0.98082\\
4090	0.9809\\
4100	0.98098\\
4110	0.98104\\
4120	0.98113\\
4130	0.98121\\
4140	0.98131\\
4150	0.98134\\
4160	0.98139\\
4170	0.98145\\
4180	0.98155\\
4190	0.98168\\
4200	0.98173\\
4210	0.98182\\
4220	0.98184\\
4230	0.98189\\
4240	0.98196\\
4250	0.98206\\
4260	0.9821\\
4270	0.98218\\
4280	0.98224\\
4290	0.98227\\
4300	0.98233\\
4310	0.98244\\
4320	0.98247\\
4330	0.98256\\
4340	0.98262\\
4350	0.98269\\
4360	0.98276\\
4370	0.98281\\
4380	0.98282\\
4390	0.98288\\
4400	0.98291\\
4410	0.98299\\
4420	0.98307\\
4430	0.9832\\
4440	0.98326\\
4450	0.98335\\
4460	0.98336\\
4470	0.9834\\
4480	0.98347\\
4490	0.9835\\
4500	0.98353\\
4510	0.98359\\
4520	0.98366\\
4530	0.98368\\
4540	0.98373\\
4550	0.98383\\
4560	0.98393\\
4570	0.98398\\
4580	0.98404\\
4590	0.98416\\
4600	0.98419\\
4610	0.98425\\
4620	0.98429\\
4630	0.9843\\
4640	0.98436\\
4650	0.9844\\
4660	0.98444\\
4670	0.9845\\
4680	0.98457\\
4690	0.98468\\
4700	0.98473\\
4710	0.98474\\
4720	0.98476\\
4730	0.98481\\
4740	0.98481\\
4750	0.98486\\
4760	0.98488\\
4770	0.98489\\
4780	0.98497\\
4790	0.98504\\
4800	0.98504\\
4810	0.9851\\
4820	0.98514\\
4830	0.9852\\
4840	0.98527\\
4850	0.9853\\
4860	0.98532\\
4870	0.98536\\
4880	0.98541\\
4890	0.98545\\
4900	0.98548\\
4910	0.98555\\
4920	0.98563\\
4930	0.98569\\
4940	0.98576\\
4950	0.98579\\
4960	0.98583\\
4970	0.98585\\
4980	0.98591\\
4990	0.98594\\
5000	0.98597\\
5010	0.98603\\
5020	0.98608\\
5030	0.98611\\
5040	0.98612\\
5050	0.98615\\
5060	0.98619\\
5070	0.98629\\
5080	0.98632\\
5090	0.9864\\
5100	0.98644\\
5110	0.98647\\
5120	0.98651\\
5130	0.98657\\
5140	0.98664\\
5150	0.98669\\
5160	0.9867\\
5170	0.98673\\
5180	0.98677\\
5190	0.98682\\
5200	1\\
};

\addplot [color=varcol,semithick,mark=diamond,mark repeat=25,mark phase=11]
  table[row sep=crcr]{%
0	0\\
10	0.26042\\
20	0.3608\\
30	0.43119\\
40	0.48473\\
50	0.529\\
60	0.56504\\
70	0.59521\\
80	0.62195\\
90	0.64484\\
100	0.66522\\
110	0.6831\\
120	0.69873\\
130	0.71378\\
140	0.72721\\
150	0.73997\\
160	0.75075\\
170	0.76093\\
180	0.77124\\
190	0.7802\\
200	0.78917\\
210	0.7966\\
220	0.80395\\
230	0.81089\\
240	0.81777\\
250	0.82403\\
260	0.83053\\
270	0.83601\\
280	0.84144\\
290	0.84632\\
300	0.85133\\
310	0.85619\\
320	0.86081\\
330	0.86545\\
340	0.86946\\
350	0.87307\\
360	0.87677\\
370	0.88049\\
380	0.88416\\
390	0.88732\\
400	0.89029\\
410	0.89337\\
420	0.89614\\
430	0.89906\\
440	0.90156\\
450	0.90413\\
460	0.90676\\
470	0.90912\\
480	0.91135\\
490	0.9135\\
500	0.91546\\
510	0.9174\\
520	0.91939\\
530	0.92136\\
540	0.92332\\
550	0.92531\\
560	0.92717\\
570	0.92877\\
580	0.9305\\
590	0.93211\\
600	0.93357\\
610	0.93501\\
620	0.93654\\
630	0.93792\\
640	0.9392\\
650	0.94042\\
660	0.94176\\
670	0.94293\\
680	0.94411\\
690	0.94521\\
700	0.94633\\
710	0.94735\\
720	0.94845\\
730	0.94942\\
740	0.95054\\
750	0.95141\\
760	0.95238\\
770	0.95343\\
780	0.95439\\
790	0.9552\\
800	0.95599\\
810	0.95674\\
820	0.95752\\
830	0.95827\\
840	0.95907\\
850	0.95988\\
860	0.96064\\
870	0.96148\\
880	0.96225\\
890	0.96299\\
900	0.96371\\
910	0.96443\\
920	0.96516\\
930	0.96587\\
940	0.96652\\
950	0.96707\\
960	0.9676\\
970	0.96814\\
980	0.96863\\
990	0.96928\\
1000	0.96972\\
1010	0.97025\\
1020	0.9706\\
1030	0.97116\\
1040	0.97151\\
1050	0.97193\\
1060	0.9723\\
1070	0.97279\\
1080	0.97317\\
1090	0.97357\\
1100	0.9739\\
1110	0.97435\\
1120	0.97484\\
1130	0.97519\\
1140	0.97556\\
1150	0.97597\\
1160	0.97631\\
1170	0.97664\\
1180	0.97697\\
1190	0.97734\\
1200	0.97772\\
1210	0.97806\\
1220	0.97843\\
1230	0.97879\\
1240	0.97907\\
1250	0.97938\\
1260	0.97974\\
1270	0.97999\\
1280	0.98031\\
1290	0.98061\\
1300	0.98085\\
1310	0.98111\\
1320	0.98139\\
1330	0.98164\\
1340	0.982\\
1350	0.98221\\
1360	0.98256\\
1370	0.98281\\
1380	0.98305\\
1390	0.98328\\
1400	0.98352\\
1410	0.98374\\
1420	0.98392\\
1430	0.98417\\
1440	0.98438\\
1450	0.98455\\
1460	0.98478\\
1470	0.985\\
1480	0.98515\\
1490	0.98532\\
1500	0.98561\\
1510	0.98576\\
1520	0.98595\\
1530	0.98618\\
1540	0.98635\\
1550	0.98657\\
1560	0.9867\\
1570	0.98688\\
1580	0.98708\\
1590	0.98723\\
1600	0.98742\\
1610	0.98755\\
1620	0.98769\\
1630	0.98783\\
1640	0.98799\\
1650	0.98813\\
1660	0.98831\\
1670	0.98842\\
1680	0.98859\\
1690	0.98875\\
1700	0.98884\\
1710	0.98898\\
1720	0.98912\\
1730	0.98924\\
1740	0.98939\\
1750	0.98954\\
1760	0.98965\\
1770	0.98983\\
1780	0.98995\\
1790	0.99008\\
1800	0.99025\\
1810	0.99037\\
1820	0.9905\\
1830	0.99058\\
1840	0.99066\\
1850	0.99078\\
1860	0.99086\\
1870	0.99099\\
1880	0.99108\\
1890	0.99119\\
1900	0.99128\\
1910	0.99138\\
1920	0.99147\\
1930	0.99155\\
1940	0.99162\\
1950	0.99174\\
1960	0.99187\\
1970	0.99197\\
1980	0.99212\\
1990	0.99224\\
2000	0.99235\\
2010	0.99245\\
2020	0.99255\\
2030	0.99262\\
2040	0.9927\\
2050	0.99274\\
2060	0.99284\\
2070	0.99295\\
2080	0.99302\\
2090	0.99311\\
2100	0.99324\\
2110	0.99329\\
2120	0.99334\\
2130	0.99342\\
2140	0.99353\\
2150	0.99359\\
2160	0.99365\\
2170	0.99369\\
2180	0.99373\\
2190	0.99378\\
2200	0.99385\\
2210	0.99395\\
2220	0.99401\\
2230	0.99408\\
2240	0.99419\\
2250	0.99433\\
2260	0.99438\\
2270	0.99442\\
2280	0.99446\\
2290	0.99452\\
2300	0.99456\\
2310	0.9946\\
2320	0.99465\\
2330	0.99468\\
2340	0.9948\\
2350	0.99489\\
2360	0.99489\\
2370	0.99498\\
2380	0.99506\\
2390	0.9951\\
2400	0.99515\\
2410	0.99524\\
2420	0.99527\\
2430	0.9953\\
2440	0.99534\\
2450	0.99541\\
2460	0.99545\\
2470	0.99548\\
2480	0.9955\\
2490	0.99556\\
2500	0.99559\\
2510	0.99563\\
2520	0.99568\\
2530	0.99571\\
2540	0.99578\\
2550	0.99584\\
2560	0.99586\\
2570	0.99591\\
2580	0.99596\\
2590	0.99599\\
2600	0.99603\\
2610	0.99609\\
2620	0.99613\\
2630	0.99623\\
2640	0.99628\\
2650	0.9963\\
2660	0.99635\\
2670	0.99638\\
2680	0.99641\\
2690	0.99647\\
2700	0.99649\\
2710	0.99657\\
2720	0.99664\\
2730	0.99669\\
2740	0.99672\\
2750	0.99679\\
2760	0.99684\\
2770	0.99687\\
2780	0.99688\\
2790	0.99691\\
2800	0.99692\\
2810	0.99693\\
2820	0.99696\\
2830	0.99699\\
2840	0.99704\\
2850	0.99706\\
2860	0.99706\\
2870	0.99707\\
2880	0.99711\\
2890	0.99715\\
2900	0.99718\\
2910	0.9972\\
2920	0.9972\\
2930	0.99722\\
2940	0.99725\\
2950	0.9973\\
2960	0.99733\\
2970	0.99735\\
2980	0.99738\\
2990	0.99741\\
3000	0.99741\\
3010	0.99744\\
3020	0.99745\\
3030	0.99748\\
3040	0.99748\\
3050	0.99753\\
3060	0.99757\\
3070	0.99759\\
3080	0.99766\\
3090	0.99769\\
3100	0.99773\\
3110	0.99775\\
3120	0.99778\\
3130	0.99782\\
3140	0.99784\\
3150	0.99785\\
3160	0.99787\\
3170	0.99789\\
3180	0.9979\\
3190	0.99793\\
3200	0.99795\\
3210	0.998\\
3220	0.99802\\
3230	0.99804\\
3240	0.99804\\
3250	0.99811\\
3260	0.99811\\
3270	0.99815\\
3280	0.99815\\
3290	0.99815\\
3300	0.99818\\
3310	0.9982\\
3320	0.99822\\
3330	0.99823\\
3340	0.99824\\
3350	0.99826\\
3360	0.9983\\
3370	0.99831\\
3380	0.99835\\
3390	0.99836\\
3400	0.99837\\
3410	0.99839\\
3420	0.99843\\
3430	0.99846\\
3440	0.99846\\
3450	0.99848\\
3460	0.9985\\
3470	0.99854\\
3480	0.99856\\
3490	0.99857\\
3500	0.99859\\
3510	0.99861\\
3520	0.99862\\
3530	0.99863\\
3540	0.99865\\
3550	0.99865\\
3560	0.99866\\
3570	0.99869\\
3580	0.99869\\
3590	0.99871\\
3600	0.99872\\
3610	0.99874\\
3620	0.99876\\
3630	0.99877\\
3640	0.99877\\
3650	0.99878\\
3660	0.99878\\
3670	0.9988\\
3680	0.99882\\
3690	0.99882\\
3700	0.99884\\
3710	0.99887\\
3720	0.99887\\
3730	0.99889\\
3740	0.99889\\
3750	0.9989\\
3760	0.9989\\
3770	0.99891\\
3780	0.99892\\
3790	0.99894\\
3800	0.99894\\
3810	0.99894\\
3820	0.99895\\
3830	0.99896\\
3840	0.99896\\
3850	0.999\\
3860	0.999\\
3870	0.999\\
3880	0.999\\
3890	0.999\\
3900	0.99901\\
3910	0.99901\\
3920	0.99901\\
3930	0.99901\\
3940	0.99902\\
3950	0.99903\\
3960	0.99903\\
3970	0.99903\\
3980	0.99905\\
3990	0.99905\\
4000	0.99905\\
4010	0.99907\\
4020	0.99907\\
4030	0.9991\\
4040	0.99911\\
4050	0.99911\\
4060	0.99912\\
4070	0.99912\\
4080	0.99914\\
4090	0.99914\\
4100	0.99914\\
4110	0.99915\\
4120	0.99916\\
4130	0.99918\\
4140	0.99919\\
4150	0.99919\\
4160	0.99921\\
4170	0.99923\\
4180	0.99924\\
4190	0.99925\\
4200	0.99926\\
4210	0.99927\\
4220	0.99927\\
4230	0.99928\\
4240	0.9993\\
4250	0.9993\\
4260	0.99931\\
4270	0.99931\\
4280	0.99931\\
4290	0.99931\\
4300	0.99932\\
4310	0.99932\\
4320	0.99932\\
4330	0.99932\\
4340	0.99933\\
4350	0.99934\\
4360	0.99935\\
4370	0.99936\\
4380	0.99936\\
4390	0.99936\\
4400	0.99937\\
4410	0.99937\\
4420	0.99937\\
4430	0.99937\\
4440	0.99937\\
4450	0.99938\\
4460	0.9994\\
4470	0.99941\\
4480	0.99941\\
4490	0.99941\\
4500	0.99941\\
4510	0.99941\\
4520	0.99941\\
4530	0.99941\\
4540	0.99942\\
4550	0.99944\\
4560	0.99945\\
4570	0.99946\\
4580	0.99947\\
4590	0.99947\\
4600	0.99947\\
4610	0.99949\\
4620	0.9995\\
4630	0.9995\\
4640	0.99951\\
4650	0.99954\\
4660	0.99954\\
4670	0.99955\\
4680	0.99956\\
4690	0.99956\\
4700	0.99957\\
4710	0.99957\\
4720	0.99958\\
4730	0.99958\\
4740	0.99958\\
4750	0.99958\\
4760	0.99958\\
4770	0.99959\\
4780	0.99959\\
4790	0.9996\\
4800	0.9996\\
4810	0.99961\\
4820	0.99961\\
4830	0.99961\\
4840	0.99961\\
4850	0.99961\\
4860	0.99961\\
4870	0.99961\\
4880	0.99961\\
4890	0.99961\\
4900	0.99961\\
4910	0.99963\\
4920	0.99964\\
4930	0.99964\\
4940	0.99964\\
4950	0.99964\\
4960	0.99964\\
4970	0.99964\\
4980	0.99965\\
4990	0.99965\\
5000	0.99965\\
5010	0.99965\\
5020	0.99965\\
5030	0.99965\\
5040	0.99965\\
5050	0.99966\\
5060	0.99966\\
5070	0.99966\\
5080	0.99966\\
5090	0.99966\\
5100	0.99966\\
5110	0.99966\\
5120	0.99967\\
5130	0.99967\\
5140	0.99967\\
5150	0.99968\\
5160	0.99969\\
5170	0.99969\\
5180	0.99969\\
5190	0.99969\\
5200	1\\
};

\addplot [color=maxcol,semithick,mark=square,mark options={solid},mark repeat=25,mark phase=16]
  table[row sep=crcr]{%
0	0\\
10	0.23073\\
20	0.3206\\
30	0.38355\\
40	0.43416\\
50	0.47443\\
60	0.50818\\
70	0.53877\\
80	0.56473\\
90	0.58702\\
100	0.60711\\
110	0.62505\\
120	0.64181\\
130	0.65602\\
140	0.66921\\
150	0.68167\\
160	0.69348\\
170	0.70457\\
180	0.71436\\
190	0.72387\\
200	0.73274\\
210	0.74128\\
220	0.74897\\
230	0.75676\\
240	0.76356\\
250	0.76979\\
260	0.77627\\
270	0.78237\\
280	0.78808\\
290	0.79383\\
300	0.79929\\
310	0.80406\\
320	0.80904\\
330	0.81382\\
340	0.81829\\
350	0.82245\\
360	0.82655\\
370	0.83077\\
380	0.83463\\
390	0.83814\\
400	0.842\\
410	0.84535\\
420	0.84843\\
430	0.85136\\
440	0.85453\\
450	0.8577\\
460	0.86087\\
470	0.8637\\
480	0.86632\\
490	0.8687\\
500	0.87119\\
510	0.87373\\
520	0.8759\\
530	0.87832\\
540	0.88036\\
550	0.88237\\
560	0.88453\\
570	0.88675\\
580	0.88879\\
590	0.89083\\
600	0.89276\\
610	0.89453\\
620	0.89632\\
630	0.89803\\
640	0.89957\\
650	0.90137\\
660	0.90306\\
670	0.90456\\
680	0.90629\\
690	0.90786\\
700	0.90927\\
710	0.91058\\
720	0.91203\\
730	0.91336\\
740	0.91443\\
750	0.91571\\
760	0.91695\\
770	0.91804\\
780	0.91921\\
790	0.92044\\
800	0.92173\\
810	0.92273\\
820	0.92375\\
830	0.92471\\
840	0.92589\\
850	0.92685\\
860	0.92772\\
870	0.92873\\
880	0.92967\\
890	0.93065\\
900	0.93151\\
910	0.93243\\
920	0.93326\\
930	0.9342\\
940	0.93504\\
950	0.93593\\
960	0.93662\\
970	0.93737\\
980	0.93814\\
990	0.93885\\
1000	0.93967\\
1010	0.94041\\
1020	0.94121\\
1030	0.94186\\
1040	0.94237\\
1050	0.943\\
1060	0.94375\\
1070	0.94443\\
1080	0.9449\\
1090	0.94567\\
1100	0.94628\\
1110	0.94689\\
1120	0.94748\\
1130	0.94808\\
1140	0.94865\\
1150	0.94932\\
1160	0.94989\\
1170	0.95045\\
1180	0.95106\\
1190	0.95165\\
1200	0.9523\\
1210	0.9529\\
1220	0.95337\\
1230	0.95393\\
1240	0.95435\\
1250	0.95486\\
1260	0.95537\\
1270	0.95583\\
1280	0.95629\\
1290	0.95668\\
1300	0.95716\\
1310	0.95755\\
1320	0.95814\\
1330	0.95848\\
1340	0.95895\\
1350	0.95935\\
1360	0.95979\\
1370	0.96022\\
1380	0.96059\\
1390	0.96099\\
1400	0.96135\\
1410	0.96164\\
1420	0.96206\\
1430	0.9625\\
1440	0.96286\\
1450	0.9632\\
1460	0.96355\\
1470	0.96391\\
1480	0.96424\\
1490	0.96452\\
1500	0.96487\\
1510	0.9653\\
1520	0.96569\\
1530	0.96601\\
1540	0.96623\\
1550	0.9665\\
1560	0.9669\\
1570	0.96724\\
1580	0.96752\\
1590	0.96775\\
1600	0.96801\\
1610	0.96823\\
1620	0.96851\\
1630	0.96872\\
1640	0.96891\\
1650	0.96914\\
1660	0.96944\\
1670	0.96964\\
1680	0.96995\\
1690	0.97023\\
1700	0.9705\\
1710	0.97072\\
1720	0.97098\\
1730	0.97119\\
1740	0.97145\\
1750	0.97165\\
1760	0.97183\\
1770	0.97208\\
1780	0.97235\\
1790	0.97256\\
1800	0.97276\\
1810	0.97292\\
1820	0.97313\\
1830	0.97339\\
1840	0.97363\\
1850	0.97386\\
1860	0.97402\\
1870	0.97418\\
1880	0.97444\\
1890	0.97463\\
1900	0.97491\\
1910	0.97515\\
1920	0.97529\\
1930	0.97553\\
1940	0.97573\\
1950	0.9759\\
1960	0.9761\\
1970	0.97629\\
1980	0.97649\\
1990	0.97664\\
2000	0.97678\\
2010	0.97701\\
2020	0.97723\\
2030	0.97744\\
2040	0.97759\\
2050	0.97772\\
2060	0.97779\\
2070	0.97797\\
2080	0.97816\\
2090	0.97833\\
2100	0.97853\\
2110	0.97863\\
2120	0.97869\\
2130	0.97878\\
2140	0.97891\\
2150	0.97906\\
2160	0.9792\\
2170	0.97938\\
2180	0.97953\\
2190	0.97967\\
2200	0.97985\\
2210	0.97998\\
2220	0.98015\\
2230	0.98024\\
2240	0.98046\\
2250	0.98063\\
2260	0.98085\\
2270	0.98095\\
2280	0.98106\\
2290	0.98119\\
2300	0.98127\\
2310	0.98137\\
2320	0.98151\\
2330	0.98156\\
2340	0.98169\\
2350	0.98181\\
2360	0.98192\\
2370	0.982\\
2380	0.98215\\
2390	0.98228\\
2400	0.98239\\
2410	0.98255\\
2420	0.98268\\
2430	0.98282\\
2440	0.98292\\
2450	0.98308\\
2460	0.98315\\
2470	0.98328\\
2480	0.98341\\
2490	0.98346\\
2500	0.98361\\
2510	0.98374\\
2520	0.98384\\
2530	0.98396\\
2540	0.98401\\
2550	0.9841\\
2560	0.98419\\
2570	0.98431\\
2580	0.98438\\
2590	0.98444\\
2600	0.98454\\
2610	0.98464\\
2620	0.98471\\
2630	0.98478\\
2640	0.98484\\
2650	0.98496\\
2660	0.98508\\
2670	0.98517\\
2680	0.98523\\
2690	0.98536\\
2700	0.98548\\
2710	0.98555\\
2720	0.98561\\
2730	0.98565\\
2740	0.98576\\
2750	0.98581\\
2760	0.98591\\
2770	0.98593\\
2780	0.98602\\
2790	0.98614\\
2800	0.9862\\
2810	0.98624\\
2820	0.98633\\
2830	0.98641\\
2840	0.98645\\
2850	0.98649\\
2860	0.98655\\
2870	0.98661\\
2880	0.98667\\
2890	0.98674\\
2900	0.98683\\
2910	0.98694\\
2920	0.98709\\
2930	0.98716\\
2940	0.98726\\
2950	0.98734\\
2960	0.9874\\
2970	0.98747\\
2980	0.98756\\
2990	0.98768\\
3000	0.98775\\
3010	0.98777\\
3020	0.98781\\
3030	0.98787\\
3040	0.98794\\
3050	0.98802\\
3060	0.98808\\
3070	0.98813\\
3080	0.98817\\
3090	0.98824\\
3100	0.98829\\
3110	0.98833\\
3120	0.98839\\
3130	0.9885\\
3140	0.98853\\
3150	0.98862\\
3160	0.98873\\
3170	0.98877\\
3180	0.98885\\
3190	0.98893\\
3200	0.98899\\
3210	0.989\\
3220	0.98906\\
3230	0.9891\\
3240	0.98916\\
3250	0.98922\\
3260	0.98926\\
3270	0.98933\\
3280	0.98934\\
3290	0.98939\\
3300	0.98949\\
3310	0.98958\\
3320	0.98961\\
3330	0.98966\\
3340	0.9897\\
3350	0.98977\\
3360	0.98987\\
3370	0.98993\\
3380	0.98997\\
3390	0.98999\\
3400	0.99007\\
3410	0.99012\\
3420	0.99015\\
3430	0.9902\\
3440	0.99023\\
3450	0.99028\\
3460	0.99035\\
3470	0.99041\\
3480	0.99045\\
3490	0.9905\\
3500	0.99057\\
3510	0.99061\\
3520	0.99062\\
3530	0.99067\\
3540	0.99069\\
3550	0.99074\\
3560	0.99077\\
3570	0.99079\\
3580	0.99086\\
3590	0.99091\\
3600	0.99097\\
3610	0.99102\\
3620	0.99106\\
3630	0.99113\\
3640	0.99116\\
3650	0.99118\\
3660	0.99123\\
3670	0.99123\\
3680	0.99128\\
3690	0.99133\\
3700	0.99136\\
3710	0.99144\\
3720	0.99147\\
3730	0.99154\\
3740	0.99158\\
3750	0.99161\\
3760	0.99167\\
3770	0.99169\\
3780	0.99172\\
3790	0.99177\\
3800	0.99181\\
3810	0.99189\\
3820	0.99192\\
3830	0.99198\\
3840	0.99202\\
3850	0.99208\\
3860	0.99209\\
3870	0.99213\\
3880	0.99217\\
3890	0.9922\\
3900	0.99222\\
3910	0.99224\\
3920	0.99226\\
3930	0.99232\\
3940	0.99232\\
3950	0.99234\\
3960	0.99238\\
3970	0.9924\\
3980	0.99242\\
3990	0.99248\\
4000	0.9925\\
4010	0.99257\\
4020	0.99263\\
4030	0.99267\\
4040	0.99269\\
4050	0.99273\\
4060	0.99278\\
4070	0.99281\\
4080	0.99281\\
4090	0.99283\\
4100	0.99283\\
4110	0.99288\\
4120	0.99294\\
4130	0.99295\\
4140	0.99296\\
4150	0.99301\\
4160	0.99304\\
4170	0.99307\\
4180	0.99309\\
4190	0.99314\\
4200	0.99315\\
4210	0.99316\\
4220	0.99317\\
4230	0.99319\\
4240	0.99323\\
4250	0.99329\\
4260	0.99331\\
4270	0.99335\\
4280	0.99337\\
4290	0.99337\\
4300	0.99342\\
4310	0.99344\\
4320	0.99345\\
4330	0.99347\\
4340	0.9935\\
4350	0.9935\\
4360	0.99352\\
4370	0.99357\\
4380	0.9936\\
4390	0.99364\\
4400	0.99365\\
4410	0.99368\\
4420	0.99372\\
4430	0.99375\\
4440	0.99376\\
4450	0.99377\\
4460	0.99378\\
4470	0.99381\\
4480	0.99382\\
4490	0.99385\\
4500	0.99386\\
4510	0.99388\\
4520	0.99391\\
4530	0.99394\\
4540	0.99397\\
4550	0.99398\\
4560	0.994\\
4570	0.99401\\
4580	0.99404\\
4590	0.99404\\
4600	0.99407\\
4610	0.99409\\
4620	0.99412\\
4630	0.99416\\
4640	0.99417\\
4650	0.9942\\
4660	0.99424\\
4670	0.99424\\
4680	0.99428\\
4690	0.99431\\
4700	0.99431\\
4710	0.99434\\
4720	0.99437\\
4730	0.99438\\
4740	0.9944\\
4750	0.99442\\
4760	0.99446\\
4770	0.99447\\
4780	0.99449\\
4790	0.99451\\
4800	0.99452\\
4810	0.99454\\
4820	0.99456\\
4830	0.99461\\
4840	0.99465\\
4850	0.99468\\
4860	0.99472\\
4870	0.99476\\
4880	0.99477\\
4890	0.99479\\
4900	0.99481\\
4910	0.99486\\
4920	0.99488\\
4930	0.99495\\
4940	0.99496\\
4950	0.99498\\
4960	0.995\\
4970	0.99503\\
4980	0.99507\\
4990	0.99509\\
5000	0.99512\\
5010	0.99517\\
5020	0.99519\\
5030	0.99523\\
5040	0.99525\\
5050	0.99526\\
5060	0.99528\\
5070	0.99528\\
5080	0.99528\\
5090	0.9953\\
5100	0.99533\\
5110	0.99535\\
5120	0.99535\\
5130	0.99535\\
5140	0.99536\\
5150	0.99537\\
5160	0.99537\\
5170	0.9954\\
5180	0.99542\\
5190	0.99543\\
5200	1\\
};

\addplot [color=cntcol,semithick,mark=triangle,mark options={solid,rotate=180},mark repeat=25,mark phase=20]
  table[row sep=crcr]{%
0	0\\
10	0.25371\\
20	0.35138\\
30	0.41893\\
40	0.47099\\
50	0.51264\\
60	0.54806\\
70	0.57732\\
80	0.60202\\
90	0.62305\\
100	0.64292\\
110	0.66104\\
120	0.67808\\
130	0.69313\\
140	0.70668\\
150	0.71941\\
160	0.73112\\
170	0.74169\\
180	0.75233\\
190	0.76179\\
200	0.77033\\
210	0.77887\\
220	0.78665\\
230	0.79393\\
240	0.80119\\
250	0.80785\\
260	0.81437\\
270	0.82003\\
280	0.82567\\
290	0.83093\\
300	0.83605\\
310	0.84082\\
320	0.8455\\
330	0.84991\\
340	0.85424\\
350	0.85851\\
360	0.86241\\
370	0.86618\\
380	0.86967\\
390	0.87297\\
400	0.87634\\
410	0.87933\\
420	0.88258\\
430	0.88557\\
440	0.88806\\
450	0.89098\\
460	0.89366\\
470	0.89608\\
480	0.89842\\
490	0.9008\\
500	0.90334\\
510	0.9056\\
520	0.90759\\
530	0.90988\\
540	0.91168\\
550	0.91368\\
560	0.91559\\
570	0.91759\\
580	0.9194\\
590	0.92098\\
600	0.92262\\
610	0.92442\\
620	0.92597\\
630	0.92746\\
640	0.92899\\
650	0.93041\\
660	0.93172\\
670	0.93303\\
680	0.93447\\
690	0.93569\\
700	0.93701\\
710	0.93802\\
720	0.93907\\
730	0.94024\\
740	0.94145\\
750	0.94251\\
760	0.94355\\
770	0.94454\\
780	0.94567\\
790	0.94664\\
800	0.94748\\
810	0.94824\\
820	0.94909\\
830	0.95001\\
840	0.95085\\
850	0.95168\\
860	0.95245\\
870	0.95348\\
880	0.95432\\
890	0.95519\\
900	0.95606\\
910	0.9568\\
920	0.95762\\
930	0.95837\\
940	0.95904\\
950	0.95965\\
960	0.96038\\
970	0.96106\\
980	0.96154\\
990	0.96219\\
1000	0.96278\\
1010	0.96331\\
1020	0.96378\\
1030	0.96447\\
1040	0.96498\\
1050	0.9655\\
1060	0.96609\\
1070	0.96664\\
1080	0.96713\\
1090	0.96774\\
1100	0.96824\\
1110	0.96874\\
1120	0.96918\\
1130	0.96957\\
1140	0.97\\
1150	0.97046\\
1160	0.97088\\
1170	0.97135\\
1180	0.97171\\
1190	0.9722\\
1200	0.97268\\
1210	0.97306\\
1220	0.97341\\
1230	0.97389\\
1240	0.97448\\
1250	0.97495\\
1260	0.97523\\
1270	0.97566\\
1280	0.97597\\
1290	0.97632\\
1300	0.97663\\
1310	0.977\\
1320	0.97742\\
1330	0.97776\\
1340	0.97801\\
1350	0.97831\\
1360	0.9786\\
1370	0.9789\\
1380	0.97922\\
1390	0.97953\\
1400	0.97982\\
1410	0.98014\\
1420	0.98038\\
1430	0.98059\\
1440	0.98081\\
1450	0.98106\\
1460	0.98134\\
1470	0.98163\\
1480	0.98188\\
1490	0.98215\\
1500	0.98236\\
1510	0.98253\\
1520	0.98281\\
1530	0.98294\\
1540	0.98317\\
1550	0.98341\\
1560	0.98367\\
1570	0.98384\\
1580	0.9841\\
1590	0.98428\\
1600	0.98451\\
1610	0.98473\\
1620	0.98486\\
1630	0.98503\\
1640	0.98523\\
1650	0.98541\\
1660	0.98554\\
1670	0.98572\\
1680	0.98587\\
1690	0.98612\\
1700	0.98634\\
1710	0.98648\\
1720	0.98665\\
1730	0.98683\\
1740	0.98698\\
1750	0.98712\\
1760	0.98727\\
1770	0.98737\\
1780	0.9875\\
1790	0.98764\\
1800	0.98786\\
1810	0.98799\\
1820	0.98815\\
1830	0.98825\\
1840	0.98832\\
1850	0.98852\\
1860	0.98866\\
1870	0.98882\\
1880	0.98895\\
1890	0.98906\\
1900	0.98912\\
1910	0.98923\\
1920	0.98932\\
1930	0.98945\\
1940	0.98959\\
1950	0.98972\\
1960	0.98981\\
1970	0.98992\\
1980	0.99002\\
1990	0.99016\\
2000	0.99029\\
2010	0.99037\\
2020	0.99043\\
2030	0.99055\\
2040	0.99066\\
2050	0.99076\\
2060	0.99084\\
2070	0.9909\\
2080	0.99099\\
2090	0.99109\\
2100	0.99117\\
2110	0.99128\\
2120	0.99136\\
2130	0.99146\\
2140	0.99157\\
2150	0.99165\\
2160	0.99172\\
2170	0.9918\\
2180	0.99187\\
2190	0.99201\\
2200	0.99207\\
2210	0.99216\\
2220	0.99224\\
2230	0.99228\\
2240	0.99238\\
2250	0.99251\\
2260	0.99256\\
2270	0.99264\\
2280	0.99272\\
2290	0.99278\\
2300	0.99282\\
2310	0.99286\\
2320	0.99293\\
2330	0.99296\\
2340	0.99304\\
2350	0.99309\\
2360	0.9931\\
2370	0.99312\\
2380	0.99321\\
2390	0.99325\\
2400	0.99332\\
2410	0.99343\\
2420	0.99345\\
2430	0.99352\\
2440	0.99361\\
2450	0.99371\\
2460	0.99376\\
2470	0.99378\\
2480	0.99384\\
2490	0.9939\\
2500	0.99394\\
2510	0.99406\\
2520	0.99407\\
2530	0.99409\\
2540	0.99414\\
2550	0.99419\\
2560	0.99423\\
2570	0.99427\\
2580	0.99436\\
2590	0.9944\\
2600	0.99444\\
2610	0.99448\\
2620	0.9945\\
2630	0.99453\\
2640	0.99459\\
2650	0.99463\\
2660	0.99467\\
2670	0.99476\\
2680	0.99479\\
2690	0.99479\\
2700	0.99482\\
2710	0.99487\\
2720	0.99492\\
2730	0.99496\\
2740	0.99499\\
2750	0.99505\\
2760	0.9951\\
2770	0.99515\\
2780	0.99518\\
2790	0.9952\\
2800	0.99527\\
2810	0.99529\\
2820	0.99536\\
2830	0.99539\\
2840	0.99544\\
2850	0.99549\\
2860	0.99551\\
2870	0.99558\\
2880	0.9956\\
2890	0.99563\\
2900	0.99566\\
2910	0.99568\\
2920	0.99569\\
2930	0.99574\\
2940	0.99579\\
2950	0.99583\\
2960	0.9959\\
2970	0.99593\\
2980	0.99597\\
2990	0.99602\\
3000	0.99605\\
3010	0.99611\\
3020	0.99613\\
3030	0.99617\\
3040	0.99618\\
3050	0.99621\\
3060	0.99626\\
3070	0.99628\\
3080	0.99633\\
3090	0.99636\\
3100	0.99637\\
3110	0.99639\\
3120	0.99645\\
3130	0.99647\\
3140	0.99652\\
3150	0.99654\\
3160	0.99655\\
3170	0.99657\\
3180	0.99661\\
3190	0.99664\\
3200	0.99666\\
3210	0.99667\\
3220	0.99669\\
3230	0.99671\\
3240	0.99676\\
3250	0.99678\\
3260	0.9968\\
3270	0.99683\\
3280	0.99685\\
3290	0.99689\\
3300	0.99693\\
3310	0.99697\\
3320	0.99702\\
3330	0.99703\\
3340	0.99705\\
3350	0.99708\\
3360	0.99712\\
3370	0.99715\\
3380	0.99715\\
3390	0.99715\\
3400	0.99715\\
3410	0.99718\\
3420	0.99719\\
3430	0.9972\\
3440	0.99722\\
3450	0.99722\\
3460	0.99723\\
3470	0.99724\\
3480	0.99725\\
3490	0.99729\\
3500	0.9973\\
3510	0.99731\\
3520	0.99732\\
3530	0.99736\\
3540	0.99737\\
3550	0.99737\\
3560	0.99739\\
3570	0.99741\\
3580	0.99742\\
3590	0.99743\\
3600	0.99744\\
3610	0.99746\\
3620	0.99747\\
3630	0.99748\\
3640	0.99751\\
3650	0.99752\\
3660	0.99755\\
3670	0.99757\\
3680	0.9976\\
3690	0.99764\\
3700	0.99766\\
3710	0.9977\\
3720	0.99772\\
3730	0.99776\\
3740	0.99778\\
3750	0.99781\\
3760	0.99783\\
3770	0.99785\\
3780	0.99787\\
3790	0.99789\\
3800	0.99789\\
3810	0.9979\\
3820	0.99791\\
3830	0.99793\\
3840	0.99794\\
3850	0.99794\\
3860	0.99796\\
3870	0.99798\\
3880	0.998\\
3890	0.99801\\
3900	0.99802\\
3910	0.99803\\
3920	0.99803\\
3930	0.99804\\
3940	0.99808\\
3950	0.9981\\
3960	0.9981\\
3970	0.99813\\
3980	0.99814\\
3990	0.99816\\
4000	0.99818\\
4010	0.99818\\
4020	0.99819\\
4030	0.99819\\
4040	0.9982\\
4050	0.9982\\
4060	0.99823\\
4070	0.99825\\
4080	0.99827\\
4090	0.99829\\
4100	0.99829\\
4110	0.99829\\
4120	0.9983\\
4130	0.99832\\
4140	0.99835\\
4150	0.99836\\
4160	0.99837\\
4170	0.99837\\
4180	0.99837\\
4190	0.9984\\
4200	0.99841\\
4210	0.99841\\
4220	0.99842\\
4230	0.99842\\
4240	0.99842\\
4250	0.99843\\
4260	0.99843\\
4270	0.99843\\
4280	0.99846\\
4290	0.99847\\
4300	0.99848\\
4310	0.9985\\
4320	0.9985\\
4330	0.99851\\
4340	0.99853\\
4350	0.99854\\
4360	0.99854\\
4370	0.99855\\
4380	0.99855\\
4390	0.99855\\
4400	0.99855\\
4410	0.99855\\
4420	0.99856\\
4430	0.99857\\
4440	0.99859\\
4450	0.99859\\
4460	0.99859\\
4470	0.9986\\
4480	0.9986\\
4490	0.99861\\
4500	0.99863\\
4510	0.99866\\
4520	0.99866\\
4530	0.99866\\
4540	0.99866\\
4550	0.99867\\
4560	0.99867\\
4570	0.99867\\
4580	0.9987\\
4590	0.9987\\
4600	0.9987\\
4610	0.9987\\
4620	0.9987\\
4630	0.99871\\
4640	0.99871\\
4650	0.99873\\
4660	0.99874\\
4670	0.99875\\
4680	0.99877\\
4690	0.99877\\
4700	0.99878\\
4710	0.99879\\
4720	0.99881\\
4730	0.99881\\
4740	0.99881\\
4750	0.99881\\
4760	0.99881\\
4770	0.99882\\
4780	0.99883\\
4790	0.99885\\
4800	0.99885\\
4810	0.99886\\
4820	0.99888\\
4830	0.99889\\
4840	0.99889\\
4850	0.9989\\
4860	0.9989\\
4870	0.99891\\
4880	0.99891\\
4890	0.99892\\
4900	0.99894\\
4910	0.99895\\
4920	0.99896\\
4930	0.99896\\
4940	0.99897\\
4950	0.99897\\
4960	0.99897\\
4970	0.99898\\
4980	0.99898\\
4990	0.99899\\
5000	0.99899\\
5010	0.999\\
5020	0.999\\
5030	0.999\\
5040	0.99901\\
5050	0.99902\\
5060	0.99902\\
5070	0.99903\\
5080	0.99904\\
5090	0.99904\\
5100	0.99904\\
5110	0.99904\\
5120	0.99905\\
5130	0.99905\\
5140	0.99905\\
5150	0.99906\\
5160	0.99907\\
5170	0.99907\\
5180	0.99908\\
5190	0.99908\\
5200	1\\
};

\addplot [color=mafcol, semithick, mark=triangle,mark repeat=25,mark phase=24]
  table[row sep=crcr]{%
0	0\\
10	0.22685\\
20	0.3169\\
30	0.38089\\
40	0.42982\\
50	0.47174\\
60	0.50752\\
70	0.53697\\
80	0.56316\\
90	0.58685\\
100	0.60728\\
110	0.62542\\
120	0.64128\\
130	0.65673\\
140	0.67134\\
150	0.6842\\
160	0.69633\\
170	0.70732\\
180	0.71758\\
190	0.72688\\
200	0.73567\\
210	0.74451\\
220	0.75243\\
230	0.75981\\
240	0.76724\\
250	0.774\\
260	0.78074\\
270	0.78712\\
280	0.79281\\
290	0.79823\\
300	0.80352\\
310	0.80872\\
320	0.81323\\
330	0.8182\\
340	0.82271\\
350	0.82707\\
360	0.83124\\
370	0.83537\\
380	0.83943\\
390	0.84306\\
400	0.84678\\
410	0.85016\\
420	0.85334\\
430	0.85662\\
440	0.86008\\
450	0.86292\\
460	0.86583\\
470	0.86886\\
480	0.87163\\
490	0.87422\\
500	0.87679\\
510	0.87925\\
520	0.88175\\
530	0.88399\\
540	0.88638\\
550	0.88867\\
560	0.89093\\
570	0.89324\\
580	0.89528\\
590	0.89707\\
600	0.89906\\
610	0.90102\\
620	0.90291\\
630	0.90455\\
640	0.90623\\
650	0.90795\\
660	0.90959\\
670	0.91109\\
680	0.9127\\
690	0.91441\\
700	0.91573\\
710	0.91738\\
720	0.9187\\
730	0.91993\\
740	0.92141\\
750	0.92251\\
760	0.92385\\
770	0.9249\\
780	0.9261\\
790	0.92731\\
800	0.92861\\
810	0.92969\\
820	0.93075\\
830	0.93198\\
840	0.93296\\
850	0.9341\\
860	0.93512\\
870	0.93638\\
880	0.93727\\
890	0.9382\\
900	0.93914\\
910	0.93984\\
920	0.94089\\
930	0.94189\\
940	0.94274\\
950	0.94344\\
960	0.94419\\
970	0.94494\\
980	0.94572\\
990	0.94654\\
1000	0.94737\\
1010	0.94804\\
1020	0.94876\\
1030	0.94941\\
1040	0.95009\\
1050	0.9508\\
1060	0.95139\\
1070	0.95205\\
1080	0.95281\\
1090	0.95353\\
1100	0.95405\\
1110	0.95459\\
1120	0.95535\\
1130	0.95593\\
1140	0.95649\\
1150	0.95703\\
1160	0.95757\\
1170	0.95799\\
1180	0.9585\\
1190	0.959\\
1200	0.95952\\
1210	0.95996\\
1220	0.96049\\
1230	0.96098\\
1240	0.96141\\
1250	0.962\\
1260	0.96239\\
1270	0.96291\\
1280	0.96335\\
1290	0.96372\\
1300	0.96413\\
1310	0.96452\\
1320	0.96495\\
1330	0.96538\\
1340	0.96579\\
1350	0.96611\\
1360	0.96642\\
1370	0.96679\\
1380	0.96711\\
1390	0.96747\\
1400	0.96782\\
1410	0.9682\\
1420	0.9685\\
1430	0.96882\\
1440	0.96921\\
1450	0.96961\\
1460	0.96998\\
1470	0.97033\\
1480	0.97068\\
1490	0.97107\\
1500	0.97141\\
1510	0.97167\\
1520	0.97199\\
1530	0.97236\\
1540	0.97259\\
1550	0.97292\\
1560	0.97324\\
1570	0.97356\\
1580	0.97386\\
1590	0.97411\\
1600	0.97451\\
1610	0.97474\\
1620	0.97514\\
1630	0.97542\\
1640	0.97573\\
1650	0.97586\\
1660	0.97606\\
1670	0.97637\\
1680	0.97664\\
1690	0.97683\\
1700	0.97705\\
1710	0.97729\\
1720	0.97751\\
1730	0.97773\\
1740	0.97804\\
1750	0.97832\\
1760	0.97843\\
1770	0.97871\\
1780	0.97896\\
1790	0.97913\\
1800	0.97934\\
1810	0.97954\\
1820	0.97975\\
1830	0.97996\\
1840	0.98012\\
1850	0.98024\\
1860	0.9805\\
1870	0.9807\\
1880	0.98083\\
1890	0.98103\\
1900	0.98126\\
1910	0.98148\\
1920	0.98177\\
1930	0.98199\\
1940	0.98218\\
1950	0.9823\\
1960	0.98243\\
1970	0.98263\\
1980	0.98275\\
1990	0.98293\\
2000	0.98306\\
2010	0.98328\\
2020	0.9834\\
2030	0.98363\\
2040	0.98374\\
2050	0.98392\\
2060	0.98406\\
2070	0.98417\\
2080	0.98439\\
2090	0.98458\\
2100	0.98485\\
2110	0.98499\\
2120	0.98516\\
2130	0.98526\\
2140	0.98537\\
2150	0.98558\\
2160	0.9857\\
2170	0.98587\\
2180	0.98597\\
2190	0.98613\\
2200	0.98628\\
2210	0.98634\\
2220	0.98641\\
2230	0.98654\\
2240	0.98659\\
2250	0.98669\\
2260	0.98674\\
2270	0.98685\\
2280	0.98693\\
2290	0.98709\\
2300	0.9872\\
2310	0.98723\\
2320	0.98733\\
2330	0.9874\\
2340	0.98749\\
2350	0.98759\\
2360	0.98776\\
2370	0.98788\\
2380	0.98793\\
2390	0.98803\\
2400	0.9881\\
2410	0.98819\\
2420	0.98837\\
2430	0.98846\\
2440	0.98857\\
2450	0.98868\\
2460	0.98878\\
2470	0.98893\\
2480	0.98901\\
2490	0.98909\\
2500	0.98918\\
2510	0.98924\\
2520	0.98938\\
2530	0.98946\\
2540	0.98951\\
2550	0.98959\\
2560	0.98968\\
2570	0.9898\\
2580	0.98986\\
2590	0.98995\\
2600	0.98999\\
2610	0.99003\\
2620	0.99013\\
2630	0.9902\\
2640	0.99027\\
2650	0.9904\\
2660	0.99048\\
2670	0.99058\\
2680	0.99065\\
2690	0.99072\\
2700	0.99081\\
2710	0.99087\\
2720	0.99093\\
2730	0.99101\\
2740	0.99108\\
2750	0.99114\\
2760	0.9912\\
2770	0.99123\\
2780	0.99128\\
2790	0.99132\\
2800	0.99137\\
2810	0.99146\\
2820	0.99151\\
2830	0.99159\\
2840	0.99165\\
2850	0.99177\\
2860	0.99185\\
2870	0.99188\\
2880	0.99194\\
2890	0.99203\\
2900	0.99205\\
2910	0.99211\\
2920	0.99218\\
2930	0.99225\\
2940	0.99235\\
2950	0.99241\\
2960	0.99244\\
2970	0.99248\\
2980	0.99253\\
2990	0.99261\\
3000	0.99266\\
3010	0.9927\\
3020	0.99273\\
3030	0.99277\\
3040	0.99287\\
3050	0.9929\\
3060	0.99297\\
3070	0.99306\\
3080	0.99312\\
3090	0.99318\\
3100	0.99325\\
3110	0.99328\\
3120	0.99332\\
3130	0.99334\\
3140	0.99336\\
3150	0.99339\\
3160	0.99343\\
3170	0.99351\\
3180	0.99354\\
3190	0.99359\\
3200	0.99367\\
3210	0.99371\\
3220	0.99374\\
3230	0.99381\\
3240	0.99387\\
3250	0.9939\\
3260	0.99398\\
3270	0.99402\\
3280	0.99404\\
3290	0.99405\\
3300	0.99405\\
3310	0.99408\\
3320	0.99409\\
3330	0.99413\\
3340	0.99417\\
3350	0.99418\\
3360	0.99425\\
3370	0.99428\\
3380	0.99434\\
3390	0.99435\\
3400	0.99439\\
3410	0.99444\\
3420	0.99446\\
3430	0.99451\\
3440	0.99455\\
3450	0.99462\\
3460	0.99463\\
3470	0.99468\\
3480	0.9947\\
3490	0.99471\\
3500	0.99476\\
3510	0.9948\\
3520	0.99483\\
3530	0.99485\\
3540	0.99485\\
3550	0.9949\\
3560	0.99493\\
3570	0.99493\\
3580	0.99498\\
3590	0.99501\\
3600	0.99503\\
3610	0.99505\\
3620	0.99507\\
3630	0.99509\\
3640	0.9951\\
3650	0.99512\\
3660	0.99515\\
3670	0.99517\\
3680	0.9952\\
3690	0.99525\\
3700	0.99529\\
3710	0.99533\\
3720	0.99536\\
3730	0.99544\\
3740	0.99546\\
3750	0.99552\\
3760	0.99554\\
3770	0.99559\\
3780	0.99564\\
3790	0.99565\\
3800	0.99568\\
3810	0.99572\\
3820	0.99575\\
3830	0.99577\\
3840	0.99579\\
3850	0.99579\\
3860	0.99584\\
3870	0.99585\\
3880	0.99589\\
3890	0.99592\\
3900	0.99594\\
3910	0.99595\\
3920	0.99597\\
3930	0.99601\\
3940	0.99602\\
3950	0.99603\\
3960	0.99607\\
3970	0.9961\\
3980	0.99613\\
3990	0.99615\\
4000	0.99617\\
4010	0.99619\\
4020	0.9962\\
4030	0.9962\\
4040	0.99622\\
4050	0.99622\\
4060	0.99622\\
4070	0.99623\\
4080	0.99625\\
4090	0.99626\\
4100	0.99628\\
4110	0.99628\\
4120	0.9963\\
4130	0.99631\\
4140	0.99633\\
4150	0.99634\\
4160	0.99635\\
4170	0.99638\\
4180	0.9964\\
4190	0.99643\\
4200	0.99645\\
4210	0.99645\\
4220	0.99648\\
4230	0.99651\\
4240	0.99652\\
4250	0.99653\\
4260	0.99655\\
4270	0.99658\\
4280	0.99658\\
4290	0.99663\\
4300	0.99663\\
4310	0.99666\\
4320	0.99666\\
4330	0.99668\\
4340	0.99669\\
4350	0.9967\\
4360	0.99672\\
4370	0.99675\\
4380	0.99677\\
4390	0.99677\\
4400	0.99678\\
4410	0.99679\\
4420	0.9968\\
4430	0.99682\\
4440	0.99684\\
4450	0.99686\\
4460	0.99689\\
4470	0.9969\\
4480	0.99693\\
4490	0.99696\\
4500	0.99696\\
4510	0.99697\\
4520	0.99698\\
4530	0.99699\\
4540	0.997\\
4550	0.99702\\
4560	0.99703\\
4570	0.99703\\
4580	0.99706\\
4590	0.9971\\
4600	0.9971\\
4610	0.9971\\
4620	0.99714\\
4630	0.99714\\
4640	0.99714\\
4650	0.99715\\
4660	0.99715\\
4670	0.99718\\
4680	0.9972\\
4690	0.99722\\
4700	0.99722\\
4710	0.99722\\
4720	0.99723\\
4730	0.99724\\
4740	0.99724\\
4750	0.99725\\
4760	0.99727\\
4770	0.9973\\
4780	0.99732\\
4790	0.99734\\
4800	0.99736\\
4810	0.99737\\
4820	0.9974\\
4830	0.9974\\
4840	0.99745\\
4850	0.99748\\
4860	0.99749\\
4870	0.99751\\
4880	0.99752\\
4890	0.99754\\
4900	0.99756\\
4910	0.99758\\
4920	0.99758\\
4930	0.99758\\
4940	0.99759\\
4950	0.99762\\
4960	0.99762\\
4970	0.99765\\
4980	0.99765\\
4990	0.99767\\
5000	0.99767\\
5010	0.99767\\
5020	0.99767\\
5030	0.99768\\
5040	0.99772\\
5050	0.99772\\
5060	0.99774\\
5070	0.99774\\
5080	0.99774\\
5090	0.99775\\
5100	0.99777\\
5110	0.99778\\
5120	0.99781\\
5130	0.99781\\
5140	0.99784\\
5150	0.99785\\
5160	0.99785\\
5170	0.99785\\
5180	0.99786\\
5190	0.99788\\
5200	1\\
};

\end{axis}
\end{tikzpicture}%

%% file: fig/max_err2.tex
\begin{tikzpicture}
\pgfplotsset{every tick label/.append style={font=\tiny}}
\tikzstyle{dotted}= [dash pattern=on \pgflinewidth off 0.5mm] 
\tikzstyle{dashed}= [dash pattern=on 7.5*0.8*0.8pt off 7.5*0.4*0.8pt]
\tikzstyle{dashdotted} = [dash pattern=on 7.5*0.8*0.6pt off 7.5*0.8*0.3pt on \the\pgflinewidth off 7.5*0.8*0.3pt]
\tikzstyle{dotted2} = [dash pattern=on 7.5*0.8*0.3pt off 7.5*0.8*0.2pt]

\begin{axis}[%
width=\fwidth,
height=\fheight,
at={(0,0)},
xmin=0,
xmax=50,
xlabel near ticks,
xlabel style={font=\scriptsize\color{white!15!black}},
xlabel={$\nu_{\text{max}}$},
ylabel near ticks,
ymin=0.5,
ymax=1,
ylabel style={font=\scriptsize\color{white!15!black}},
ylabel={Empirical CDF},axis background/.style={fill=white},
xmajorgrids,
ymajorgrids,
legend style={font=\tiny, at={(0.992,0.02)}, anchor=south east, legend cell align=left, align=left, fill opacity=0.8, draw opacity=1, text opacity=1, draw=white!80!black}
]
\addplot [color=msecol, semithick,mark=x,mark repeat=25,mark phase=1]
  table[row sep=crcr]{%
0	0\\
0.1	0.10901\\
0.2	0.15315\\
0.3	0.18684\\
0.4	0.21453\\
0.5	0.23928\\
0.6	0.26162\\
0.7	0.28159\\
0.8	0.29932\\
0.9	0.31593\\
1	0.33207\\
1.1	0.3468\\
1.2	0.36038\\
1.3	0.37417\\
1.4	0.38678\\
1.5	0.39865\\
1.6	0.40999\\
1.7	0.42089\\
1.8	0.4317\\
1.9	0.44156\\
2	0.45168\\
2.1	0.46066\\
2.2	0.46999\\
2.3	0.4791\\
2.4	0.4879\\
2.5	0.4966\\
2.6	0.50407\\
2.7	0.51188\\
2.8	0.51956\\
2.9	0.52715\\
3	0.5345\\
3.1	0.54112\\
3.2	0.54737\\
3.3	0.55413\\
3.4	0.56102\\
3.5	0.56737\\
3.6	0.57338\\
3.7	0.57928\\
3.8	0.58518\\
3.9	0.59077\\
4	0.59649\\
4.1	0.60189\\
4.2	0.60748\\
4.3	0.61241\\
4.4	0.61765\\
4.5	0.62282\\
4.6	0.6277\\
4.7	0.63301\\
4.8	0.63762\\
4.9	0.64233\\
5	0.64683\\
5.1	0.65106\\
5.2	0.6556\\
5.3	0.65991\\
5.4	0.66432\\
5.5	0.66875\\
5.6	0.67266\\
5.7	0.67638\\
5.8	0.68041\\
5.9	0.68434\\
6	0.68785\\
6.1	0.69177\\
6.2	0.6957\\
6.3	0.699\\
6.4	0.70243\\
6.5	0.70583\\
6.6	0.70908\\
6.7	0.7124\\
6.8	0.71536\\
6.9	0.71853\\
7	0.72141\\
7.1	0.7245\\
7.2	0.72757\\
7.3	0.73096\\
7.4	0.73389\\
7.5	0.73681\\
7.6	0.73996\\
7.7	0.7429\\
7.8	0.74567\\
7.9	0.74859\\
8	0.75149\\
8.1	0.75369\\
8.2	0.7565\\
8.3	0.75919\\
8.4	0.7621\\
8.5	0.76449\\
8.6	0.76652\\
8.7	0.76908\\
8.8	0.77158\\
8.9	0.77404\\
9	0.77637\\
9.1	0.77856\\
9.2	0.78058\\
9.3	0.78278\\
9.4	0.78496\\
9.5	0.7873\\
9.6	0.78938\\
9.7	0.79163\\
9.8	0.79369\\
9.9	0.79612\\
10	0.79812\\
10.1	0.79998\\
10.2	0.80175\\
10.3	0.80359\\
10.4	0.80552\\
10.5	0.80749\\
10.6	0.80956\\
10.7	0.81155\\
10.8	0.81334\\
10.9	0.8151\\
11	0.81691\\
11.1	0.81857\\
11.2	0.82016\\
11.3	0.82178\\
11.4	0.82322\\
11.5	0.82495\\
11.6	0.82664\\
11.7	0.82836\\
11.8	0.82992\\
11.9	0.83151\\
12	0.83316\\
12.1	0.83455\\
12.2	0.83605\\
12.3	0.83784\\
12.4	0.83929\\
12.5	0.84083\\
12.6	0.84231\\
12.7	0.84379\\
12.8	0.84512\\
12.9	0.8465\\
13	0.84788\\
13.1	0.84937\\
13.2	0.85087\\
13.3	0.85216\\
13.4	0.85356\\
13.5	0.85493\\
13.6	0.85623\\
13.7	0.85749\\
13.8	0.85876\\
13.9	0.85995\\
14	0.86108\\
14.1	0.86242\\
14.2	0.86365\\
14.3	0.86494\\
14.4	0.86608\\
14.5	0.86732\\
14.6	0.86851\\
14.7	0.86958\\
14.8	0.87072\\
14.9	0.87165\\
15	0.87275\\
15.1	0.8737\\
15.2	0.87477\\
15.3	0.87575\\
15.4	0.8768\\
15.5	0.87763\\
15.6	0.87845\\
15.7	0.87958\\
15.8	0.8806\\
15.9	0.88164\\
16	0.88256\\
16.1	0.88362\\
16.2	0.88466\\
16.3	0.88569\\
16.4	0.88676\\
16.5	0.88752\\
16.6	0.88844\\
16.7	0.88932\\
16.8	0.89014\\
16.9	0.89111\\
17	0.89198\\
17.1	0.89271\\
17.2	0.89361\\
17.3	0.89442\\
17.4	0.89527\\
17.5	0.8962\\
17.6	0.89708\\
17.7	0.89787\\
17.8	0.8986\\
17.9	0.89942\\
18	0.90002\\
18.1	0.90072\\
18.2	0.90133\\
18.3	0.90233\\
18.4	0.90299\\
18.5	0.90379\\
18.6	0.90467\\
18.7	0.9054\\
18.8	0.90623\\
18.9	0.90701\\
19	0.90773\\
19.1	0.90821\\
19.2	0.90901\\
19.3	0.90975\\
19.4	0.91045\\
19.5	0.91116\\
19.6	0.91191\\
19.7	0.91254\\
19.8	0.91325\\
19.9	0.91397\\
20	0.91471\\
20.1	0.91542\\
20.2	0.91606\\
20.3	0.91679\\
20.4	0.91732\\
20.5	0.91805\\
20.6	0.91863\\
20.7	0.91914\\
20.8	0.91976\\
20.9	0.92051\\
21	0.92108\\
21.1	0.92171\\
21.2	0.92234\\
21.3	0.92302\\
21.4	0.92351\\
21.5	0.92411\\
21.6	0.92453\\
21.7	0.92507\\
21.8	0.92558\\
21.9	0.92605\\
22	0.92657\\
22.1	0.92707\\
22.2	0.92758\\
22.3	0.92809\\
22.4	0.92861\\
22.5	0.92917\\
22.6	0.92973\\
22.7	0.9303\\
22.8	0.93085\\
22.9	0.93139\\
23	0.93196\\
23.1	0.93237\\
23.2	0.93278\\
23.3	0.93332\\
23.4	0.93369\\
23.5	0.93408\\
23.6	0.93445\\
23.7	0.93481\\
23.8	0.93537\\
23.9	0.93571\\
24	0.93619\\
24.1	0.9365\\
24.2	0.9369\\
24.3	0.93735\\
24.4	0.93758\\
24.5	0.938\\
24.6	0.93839\\
24.7	0.93877\\
24.8	0.93928\\
24.9	0.93978\\
25	0.94011\\
25.1	0.94055\\
25.2	0.94097\\
25.3	0.94136\\
25.4	0.94169\\
25.5	0.94212\\
25.6	0.9425\\
25.7	0.94294\\
25.8	0.9434\\
25.9	0.94375\\
26	0.94412\\
26.1	0.94457\\
26.2	0.94488\\
26.3	0.9452\\
26.4	0.94554\\
26.5	0.94589\\
26.6	0.9462\\
26.7	0.94652\\
26.8	0.9468\\
26.9	0.94715\\
27	0.9475\\
27.1	0.94789\\
27.2	0.94824\\
27.3	0.94863\\
27.4	0.94887\\
27.5	0.94923\\
27.6	0.94948\\
27.7	0.94966\\
27.8	0.95\\
27.9	0.95021\\
28	0.95056\\
28.1	0.95077\\
28.2	0.95099\\
28.3	0.95131\\
28.4	0.95161\\
28.5	0.95188\\
28.6	0.95226\\
28.7	0.95254\\
28.8	0.95277\\
28.9	0.95307\\
29	0.95331\\
29.1	0.9536\\
29.2	0.95387\\
29.3	0.95409\\
29.4	0.95438\\
29.5	0.95465\\
29.6	0.95495\\
29.7	0.95524\\
29.8	0.95551\\
29.9	0.95569\\
30	0.95597\\
30.1	0.95629\\
30.2	0.95651\\
30.3	0.9568\\
30.4	0.95708\\
30.5	0.95739\\
30.6	0.95773\\
30.7	0.95791\\
30.8	0.95827\\
30.9	0.95851\\
31	0.95872\\
31.1	0.95901\\
31.2	0.95932\\
31.3	0.95956\\
31.4	0.95975\\
31.5	0.96\\
31.6	0.96023\\
31.7	0.96041\\
31.8	0.96074\\
31.9	0.961\\
32	0.96126\\
32.1	0.96152\\
32.2	0.96173\\
32.3	0.96194\\
32.4	0.96221\\
32.5	0.9624\\
32.6	0.96268\\
32.7	0.96287\\
32.8	0.96308\\
32.9	0.96324\\
33	0.96335\\
33.1	0.96361\\
33.2	0.9638\\
33.3	0.96405\\
33.4	0.96425\\
33.5	0.96452\\
33.6	0.96467\\
33.7	0.9648\\
33.8	0.96512\\
33.9	0.96533\\
34	0.96558\\
34.1	0.96579\\
34.2	0.966\\
34.3	0.96626\\
34.4	0.96641\\
34.5	0.96672\\
34.6	0.96684\\
34.7	0.96698\\
34.8	0.96728\\
34.9	0.96744\\
35	0.96761\\
35.1	0.96788\\
35.2	0.9681\\
35.3	0.96823\\
35.4	0.96838\\
35.5	0.96849\\
35.6	0.96865\\
35.7	0.96883\\
35.8	0.96898\\
35.9	0.96913\\
36	0.96931\\
36.1	0.96946\\
36.2	0.96963\\
36.3	0.96986\\
36.4	0.97008\\
36.5	0.97026\\
36.6	0.97044\\
36.7	0.97057\\
36.8	0.97079\\
36.9	0.97096\\
37	0.97113\\
37.1	0.97129\\
37.2	0.97152\\
37.3	0.97174\\
37.4	0.97193\\
37.5	0.97212\\
37.6	0.97231\\
37.7	0.97255\\
37.8	0.97268\\
37.9	0.97292\\
38	0.97308\\
38.1	0.97325\\
38.2	0.97338\\
38.3	0.97344\\
38.4	0.97358\\
38.5	0.97378\\
38.6	0.974\\
38.7	0.97416\\
38.8	0.97435\\
38.9	0.97448\\
39	0.97458\\
39.1	0.97473\\
39.2	0.97489\\
39.3	0.97502\\
39.4	0.97521\\
39.5	0.9753\\
39.6	0.9754\\
39.7	0.97547\\
39.8	0.97562\\
39.9	0.9758\\
40	0.97593\\
40.1	0.97608\\
40.2	0.97622\\
40.3	0.97641\\
40.4	0.97645\\
40.5	0.97658\\
40.6	0.9768\\
40.7	0.97692\\
40.8	0.97709\\
40.9	0.97724\\
41	0.97737\\
41.1	0.97748\\
41.2	0.9776\\
41.3	0.9777\\
41.4	0.97782\\
41.5	0.97791\\
41.6	0.97808\\
41.7	0.97815\\
41.8	0.97828\\
41.9	0.97836\\
42	0.97855\\
42.1	0.97873\\
42.2	0.97885\\
42.3	0.97894\\
42.4	0.97907\\
42.5	0.97917\\
42.6	0.97926\\
42.7	0.97937\\
42.8	0.97952\\
42.9	0.97959\\
43	0.97963\\
43.1	0.97977\\
43.2	0.97986\\
43.3	0.97992\\
43.4	0.98008\\
43.5	0.98023\\
43.6	0.98036\\
43.7	0.98048\\
43.8	0.98063\\
43.9	0.98072\\
44	0.98082\\
44.1	0.98096\\
44.2	0.98102\\
44.3	0.98115\\
44.4	0.98127\\
44.5	0.98137\\
44.6	0.98148\\
44.7	0.9816\\
44.8	0.98166\\
44.9	0.98178\\
45	0.98187\\
45.1	0.98198\\
45.2	0.98208\\
45.3	0.9822\\
45.4	0.98228\\
45.5	0.98238\\
45.6	0.98251\\
45.7	0.98259\\
45.8	0.98266\\
45.9	0.98273\\
46	0.98278\\
46.1	0.98286\\
46.2	0.98296\\
46.3	0.98306\\
46.4	0.98313\\
46.5	0.98325\\
46.6	0.98335\\
46.7	0.98341\\
46.8	0.9835\\
46.9	0.9836\\
47	0.98364\\
47.1	0.98378\\
47.2	0.98388\\
47.3	0.984\\
47.4	0.98405\\
47.5	0.98415\\
47.6	0.98417\\
47.7	0.98422\\
47.8	0.98428\\
47.9	0.98436\\
48	0.98445\\
48.1	0.98454\\
48.2	0.98461\\
48.3	0.9847\\
48.4	0.9848\\
48.5	0.98483\\
48.6	0.98488\\
48.7	0.98494\\
48.8	0.98504\\
48.9	0.98512\\
49	0.98523\\
49.1	0.98531\\
49.2	0.98541\\
49.3	0.98543\\
49.4	0.98548\\
49.5	0.98557\\
49.6	0.98565\\
49.7	0.98575\\
49.8	0.98581\\
49.9	0.98587\\
50	0.98591\\
50.1	0.98598\\
50.2	0.98606\\
50.3	0.98611\\
50.4	0.98616\\
50.5	0.98621\\
50.6	0.98627\\
50.7	0.98633\\
50.8	0.98641\\
50.9	0.98647\\
51	1\\
};

\addplot [color=avgcol,semithick,mark=o,mark repeat=25,mark phase=6]
  table[row sep=crcr]{%
0	0\\
0.1	0.08943\\
0.2	0.1256\\
0.3	0.1533\\
0.4	0.17598\\
0.5	0.19556\\
0.6	0.21335\\
0.7	0.23042\\
0.8	0.24552\\
0.9	0.25953\\
1	0.27282\\
1.1	0.28518\\
1.2	0.29739\\
1.3	0.30858\\
1.4	0.31931\\
1.5	0.32983\\
1.6	0.3399\\
1.7	0.34944\\
1.8	0.35878\\
1.9	0.3676\\
2	0.37615\\
2.1	0.38484\\
2.2	0.39289\\
2.3	0.40083\\
2.4	0.40816\\
2.5	0.41504\\
2.6	0.42187\\
2.7	0.42898\\
2.8	0.4354\\
2.9	0.44175\\
3	0.44803\\
3.1	0.45414\\
3.2	0.45979\\
3.3	0.46545\\
3.4	0.47164\\
3.5	0.47755\\
3.6	0.48307\\
3.7	0.48856\\
3.8	0.49381\\
3.9	0.4987\\
4	0.50367\\
4.1	0.50904\\
4.2	0.51387\\
4.3	0.51877\\
4.4	0.52406\\
4.5	0.52885\\
4.6	0.53369\\
4.7	0.53806\\
4.8	0.54243\\
4.9	0.54703\\
5	0.5515\\
5.1	0.55587\\
5.2	0.56031\\
5.3	0.56442\\
5.4	0.56857\\
5.5	0.57236\\
5.6	0.57635\\
5.7	0.58039\\
5.8	0.58416\\
5.9	0.58757\\
6	0.5912\\
6.1	0.59494\\
6.2	0.5989\\
6.3	0.6029\\
6.4	0.6062\\
6.5	0.60928\\
6.6	0.61251\\
6.7	0.61593\\
6.8	0.61908\\
6.9	0.6222\\
7	0.62539\\
7.1	0.62865\\
7.2	0.63189\\
7.3	0.63493\\
7.4	0.63809\\
7.5	0.64129\\
7.6	0.64448\\
7.7	0.64728\\
7.8	0.6503\\
7.9	0.65313\\
8	0.65575\\
8.1	0.65859\\
8.2	0.66116\\
8.3	0.66392\\
8.4	0.66661\\
8.5	0.66934\\
8.6	0.67207\\
8.7	0.67465\\
8.8	0.67729\\
8.9	0.67969\\
9	0.68203\\
9.1	0.68446\\
9.2	0.68692\\
9.3	0.68915\\
9.4	0.69156\\
9.5	0.69368\\
9.6	0.696\\
9.7	0.69845\\
9.8	0.70072\\
9.9	0.70296\\
10	0.70501\\
10.1	0.70707\\
10.2	0.70925\\
10.3	0.71128\\
10.4	0.71338\\
10.5	0.71561\\
10.6	0.71756\\
10.7	0.71945\\
10.8	0.72153\\
10.9	0.72347\\
11	0.72541\\
11.1	0.72713\\
11.2	0.72883\\
11.3	0.73087\\
11.4	0.7327\\
11.5	0.7347\\
11.6	0.73679\\
11.7	0.73855\\
11.8	0.74025\\
11.9	0.74204\\
12	0.74383\\
12.1	0.74576\\
12.2	0.74757\\
12.3	0.74928\\
12.4	0.75101\\
12.5	0.75264\\
12.6	0.7544\\
12.7	0.756\\
12.8	0.7578\\
12.9	0.7596\\
13	0.76108\\
13.1	0.76258\\
13.2	0.76414\\
13.3	0.76585\\
13.4	0.7674\\
13.5	0.76886\\
13.6	0.77023\\
13.7	0.77169\\
13.8	0.77322\\
13.9	0.77467\\
14	0.77635\\
14.1	0.7776\\
14.2	0.77898\\
14.3	0.78039\\
14.4	0.78195\\
14.5	0.7831\\
14.6	0.78439\\
14.7	0.78573\\
14.8	0.78698\\
14.9	0.7884\\
15	0.78981\\
15.1	0.79111\\
15.2	0.7923\\
15.3	0.7935\\
15.4	0.79483\\
15.5	0.79612\\
15.6	0.79745\\
15.7	0.79856\\
15.8	0.79981\\
15.9	0.80083\\
16	0.80193\\
16.1	0.80287\\
16.2	0.80421\\
16.3	0.80542\\
16.4	0.80662\\
16.5	0.80792\\
16.6	0.80879\\
16.7	0.81\\
16.8	0.81109\\
16.9	0.81216\\
17	0.81329\\
17.1	0.81441\\
17.2	0.81546\\
17.3	0.81648\\
17.4	0.81757\\
17.5	0.81862\\
17.6	0.81965\\
17.7	0.82068\\
17.8	0.82154\\
17.9	0.82244\\
18	0.82336\\
18.1	0.82441\\
18.2	0.82545\\
18.3	0.82659\\
18.4	0.82763\\
18.5	0.82865\\
18.6	0.82957\\
18.7	0.83043\\
18.8	0.83114\\
18.9	0.8319\\
19	0.83288\\
19.1	0.83373\\
19.2	0.83467\\
19.3	0.8357\\
19.4	0.83657\\
19.5	0.83738\\
19.6	0.83838\\
19.7	0.83919\\
19.8	0.84001\\
19.9	0.84105\\
20	0.84202\\
20.1	0.84272\\
20.2	0.84362\\
20.3	0.84439\\
20.4	0.8454\\
20.5	0.84619\\
20.6	0.84704\\
20.7	0.84772\\
20.8	0.84852\\
20.9	0.84944\\
21	0.85016\\
21.1	0.85087\\
21.2	0.85185\\
21.3	0.85267\\
21.4	0.85349\\
21.5	0.85413\\
21.6	0.85498\\
21.7	0.85575\\
21.8	0.85664\\
21.9	0.85741\\
22	0.85823\\
22.1	0.85912\\
22.2	0.85977\\
22.3	0.86066\\
22.4	0.86139\\
22.5	0.86221\\
22.6	0.86281\\
22.7	0.86333\\
22.8	0.86411\\
22.9	0.8649\\
23	0.86554\\
23.1	0.86615\\
23.2	0.86671\\
23.3	0.86748\\
23.4	0.86836\\
23.5	0.86895\\
23.6	0.86962\\
23.7	0.87023\\
23.8	0.87085\\
23.9	0.87143\\
24	0.872\\
24.1	0.87269\\
24.2	0.87331\\
24.3	0.87405\\
24.4	0.8747\\
24.5	0.87536\\
24.6	0.8761\\
24.7	0.87671\\
24.8	0.87732\\
24.9	0.87797\\
25	0.87844\\
25.1	0.87904\\
25.2	0.87955\\
25.3	0.88022\\
25.4	0.88091\\
25.5	0.88156\\
25.6	0.88216\\
25.7	0.88269\\
25.8	0.88332\\
25.9	0.88388\\
26	0.88436\\
26.1	0.8848\\
26.2	0.88515\\
26.3	0.8857\\
26.4	0.88626\\
26.5	0.88688\\
26.6	0.88739\\
26.7	0.88793\\
26.8	0.88852\\
26.9	0.88906\\
27	0.8895\\
27.1	0.89007\\
27.2	0.89064\\
27.3	0.89118\\
27.4	0.89165\\
27.5	0.89227\\
27.6	0.89277\\
27.7	0.8932\\
27.8	0.89378\\
27.9	0.8942\\
28	0.8948\\
28.1	0.89527\\
28.2	0.8956\\
28.3	0.89605\\
28.4	0.89643\\
28.5	0.89695\\
28.6	0.89742\\
28.7	0.89786\\
28.8	0.89836\\
28.9	0.8989\\
29	0.89934\\
29.1	0.89984\\
29.2	0.90022\\
29.3	0.90069\\
29.4	0.90122\\
29.5	0.90162\\
29.6	0.90201\\
29.7	0.90253\\
29.8	0.90292\\
29.9	0.90338\\
30	0.90378\\
30.1	0.90424\\
30.2	0.90466\\
30.3	0.90513\\
30.4	0.90555\\
30.5	0.90589\\
30.6	0.90627\\
30.7	0.90658\\
30.8	0.90703\\
30.9	0.90745\\
31	0.90789\\
31.1	0.90824\\
31.2	0.90859\\
31.3	0.90892\\
31.4	0.90935\\
31.5	0.90972\\
31.6	0.91004\\
31.7	0.91048\\
31.8	0.91084\\
31.9	0.91114\\
32	0.91143\\
32.1	0.91181\\
32.2	0.9121\\
32.3	0.91255\\
32.4	0.91298\\
32.5	0.91328\\
32.6	0.91358\\
32.7	0.91399\\
32.8	0.91442\\
32.9	0.91481\\
33	0.91516\\
33.1	0.91543\\
33.2	0.91582\\
33.3	0.91617\\
33.4	0.91643\\
33.5	0.91678\\
33.6	0.9172\\
33.7	0.91753\\
33.8	0.91791\\
33.9	0.91819\\
34	0.91863\\
34.1	0.91904\\
34.2	0.91932\\
34.3	0.91975\\
34.4	0.92006\\
34.5	0.92031\\
34.6	0.92064\\
34.7	0.92094\\
34.8	0.92132\\
34.9	0.92162\\
35	0.92194\\
35.1	0.92219\\
35.2	0.92256\\
35.3	0.92288\\
35.4	0.92323\\
35.5	0.9236\\
35.6	0.92389\\
35.7	0.92415\\
35.8	0.92449\\
35.9	0.92488\\
36	0.92523\\
36.1	0.92558\\
36.2	0.92593\\
36.3	0.92623\\
36.4	0.92652\\
36.5	0.92689\\
36.6	0.92713\\
36.7	0.9274\\
36.8	0.92769\\
36.9	0.92795\\
37	0.92833\\
37.1	0.92868\\
37.2	0.92903\\
37.3	0.92929\\
37.4	0.92969\\
37.5	0.93004\\
37.6	0.93026\\
37.7	0.9305\\
37.8	0.93074\\
37.9	0.93099\\
38	0.93136\\
38.1	0.93167\\
38.2	0.93188\\
38.3	0.93211\\
38.4	0.9324\\
38.5	0.93268\\
38.6	0.93293\\
38.7	0.93317\\
38.8	0.93347\\
38.9	0.93379\\
39	0.93395\\
39.1	0.93419\\
39.2	0.9345\\
39.3	0.93473\\
39.4	0.93493\\
39.5	0.93515\\
39.6	0.9355\\
39.7	0.93575\\
39.8	0.93598\\
39.9	0.93621\\
40	0.93652\\
40.1	0.93675\\
40.2	0.937\\
40.3	0.93728\\
40.4	0.93757\\
40.5	0.93783\\
40.6	0.93799\\
40.7	0.93817\\
40.8	0.93848\\
40.9	0.93866\\
41	0.93887\\
41.1	0.93912\\
41.2	0.93938\\
41.3	0.93961\\
41.4	0.93987\\
41.5	0.94016\\
41.6	0.9404\\
41.7	0.9406\\
41.8	0.9408\\
41.9	0.94097\\
42	0.94123\\
42.1	0.9415\\
42.2	0.9417\\
42.3	0.94194\\
42.4	0.9421\\
42.5	0.94233\\
42.6	0.9425\\
42.7	0.94273\\
42.8	0.94298\\
42.9	0.9432\\
43	0.94344\\
43.1	0.94361\\
43.2	0.94387\\
43.3	0.94404\\
43.4	0.94424\\
43.5	0.9444\\
43.6	0.94468\\
43.7	0.94486\\
43.8	0.94504\\
43.9	0.94522\\
44	0.94542\\
44.1	0.94554\\
44.2	0.94577\\
44.3	0.94592\\
44.4	0.94616\\
44.5	0.94637\\
44.6	0.94655\\
44.7	0.9467\\
44.8	0.94699\\
44.9	0.94713\\
45	0.94726\\
45.1	0.94752\\
45.2	0.94775\\
45.3	0.94793\\
45.4	0.9481\\
45.5	0.94827\\
45.6	0.94841\\
45.7	0.94861\\
45.8	0.94877\\
45.9	0.94896\\
46	0.94914\\
46.1	0.94939\\
46.2	0.94962\\
46.3	0.94983\\
46.4	0.94996\\
46.5	0.95013\\
46.6	0.95034\\
46.7	0.95052\\
46.8	0.95072\\
46.9	0.95094\\
47	0.95113\\
47.1	0.95134\\
47.2	0.95158\\
47.3	0.95173\\
47.4	0.95191\\
47.5	0.95211\\
47.6	0.95236\\
47.7	0.95248\\
47.8	0.95263\\
47.9	0.95275\\
48	0.95292\\
48.1	0.95309\\
48.2	0.95332\\
48.3	0.95357\\
48.4	0.95372\\
48.5	0.95395\\
48.6	0.95415\\
48.7	0.95434\\
48.8	0.95453\\
48.9	0.95468\\
49	0.95489\\
49.1	0.95507\\
49.2	0.95521\\
49.3	0.95537\\
49.4	0.95552\\
49.5	0.95566\\
49.6	0.95581\\
49.7	0.95601\\
49.8	0.95611\\
49.9	0.95628\\
50	0.95654\\
50.1	0.95677\\
50.2	0.95696\\
50.3	0.95719\\
50.4	0.95744\\
50.5	0.95754\\
50.6	0.95763\\
50.7	0.9578\\
50.8	0.95795\\
50.9	0.95814\\
51	1\\
};

\addplot [color=varcol,semithick,mark=diamond,mark repeat=25,mark phase=11]
  table[row sep=crcr]{%
0	0\\
0.1	0.11594\\
0.2	0.16357\\
0.3	0.19913\\
0.4	0.22868\\
0.5	0.25457\\
0.6	0.2774\\
0.7	0.29768\\
0.8	0.3169\\
0.9	0.33437\\
1	0.35006\\
1.1	0.36555\\
1.2	0.37999\\
1.3	0.39356\\
1.4	0.40613\\
1.5	0.41867\\
1.6	0.43057\\
1.7	0.44205\\
1.8	0.45264\\
1.9	0.46351\\
2	0.47327\\
2.1	0.48282\\
2.2	0.49164\\
2.3	0.50071\\
2.4	0.50955\\
2.5	0.51789\\
2.6	0.52586\\
2.7	0.53378\\
2.8	0.54096\\
2.9	0.5483\\
3	0.55513\\
3.1	0.56236\\
3.2	0.56924\\
3.3	0.57586\\
3.4	0.58259\\
3.5	0.58845\\
3.6	0.59431\\
3.7	0.60011\\
3.8	0.60581\\
3.9	0.61156\\
4	0.61698\\
4.1	0.62264\\
4.2	0.62798\\
4.3	0.63315\\
4.4	0.63825\\
4.5	0.64314\\
4.6	0.64782\\
4.7	0.65261\\
4.8	0.65687\\
4.9	0.66127\\
5	0.66574\\
5.1	0.6703\\
5.2	0.67461\\
5.3	0.67896\\
5.4	0.68277\\
5.5	0.68703\\
5.6	0.69115\\
5.7	0.69512\\
5.8	0.69881\\
5.9	0.7027\\
6	0.70648\\
6.1	0.71021\\
6.2	0.71388\\
6.3	0.71724\\
6.4	0.72105\\
6.5	0.7244\\
6.6	0.72792\\
6.7	0.73091\\
6.8	0.73416\\
6.9	0.73707\\
7	0.73992\\
7.1	0.74268\\
7.2	0.7457\\
7.3	0.74866\\
7.4	0.75161\\
7.5	0.75451\\
7.6	0.75738\\
7.7	0.76005\\
7.8	0.76281\\
7.9	0.76549\\
8	0.76821\\
8.1	0.77074\\
8.2	0.77316\\
8.3	0.77597\\
8.4	0.7785\\
8.5	0.78085\\
8.6	0.78337\\
8.7	0.78562\\
8.8	0.78791\\
8.9	0.79007\\
9	0.79227\\
9.1	0.79425\\
9.2	0.79639\\
9.3	0.79837\\
9.4	0.80007\\
9.5	0.80207\\
9.6	0.80416\\
9.7	0.80609\\
9.8	0.80814\\
9.9	0.80993\\
10	0.81205\\
10.1	0.81413\\
10.2	0.81605\\
10.3	0.81777\\
10.4	0.81975\\
10.5	0.82164\\
10.6	0.82347\\
10.7	0.82504\\
10.8	0.8269\\
10.9	0.82851\\
11	0.83012\\
11.1	0.83193\\
11.2	0.83351\\
11.3	0.83536\\
11.4	0.83703\\
11.5	0.83842\\
11.6	0.84025\\
11.7	0.84168\\
11.8	0.84311\\
11.9	0.84467\\
12	0.84631\\
12.1	0.84759\\
12.2	0.84891\\
12.3	0.85039\\
12.4	0.85188\\
12.5	0.8533\\
12.6	0.8548\\
12.7	0.85611\\
12.8	0.85769\\
12.9	0.85912\\
13	0.86045\\
13.1	0.86187\\
13.2	0.8632\\
13.3	0.86439\\
13.4	0.86563\\
13.5	0.86691\\
13.6	0.86793\\
13.7	0.86916\\
13.8	0.87017\\
13.9	0.87143\\
14	0.87255\\
14.1	0.87392\\
14.2	0.87499\\
14.3	0.87596\\
14.4	0.87699\\
14.5	0.87822\\
14.6	0.87923\\
14.7	0.88036\\
14.8	0.88134\\
14.9	0.88253\\
15	0.8835\\
15.1	0.88455\\
15.2	0.88565\\
15.3	0.88666\\
15.4	0.88758\\
15.5	0.88858\\
15.6	0.88955\\
15.7	0.89043\\
15.8	0.89125\\
15.9	0.8922\\
16	0.89312\\
16.1	0.89403\\
16.2	0.89488\\
16.3	0.89564\\
16.4	0.89644\\
16.5	0.89727\\
16.6	0.89804\\
16.7	0.8989\\
16.8	0.89975\\
16.9	0.90059\\
17	0.90132\\
17.1	0.90215\\
17.2	0.90303\\
17.3	0.90382\\
17.4	0.90443\\
17.5	0.90508\\
17.6	0.90588\\
17.7	0.90679\\
17.8	0.90754\\
17.9	0.90823\\
18	0.90892\\
18.1	0.9098\\
18.2	0.91065\\
18.3	0.91132\\
18.4	0.91185\\
18.5	0.91261\\
18.6	0.91335\\
18.7	0.91409\\
18.8	0.91475\\
18.9	0.91532\\
19	0.91597\\
19.1	0.91654\\
19.2	0.91707\\
19.3	0.91757\\
19.4	0.91827\\
19.5	0.91892\\
19.6	0.91954\\
19.7	0.9202\\
19.8	0.92076\\
19.9	0.92147\\
20	0.92209\\
20.1	0.92274\\
20.2	0.92337\\
20.3	0.92396\\
20.4	0.92454\\
20.5	0.92507\\
20.6	0.92585\\
20.7	0.92655\\
20.8	0.92714\\
20.9	0.9277\\
21	0.92821\\
21.1	0.9288\\
21.2	0.92929\\
21.3	0.92973\\
21.4	0.93036\\
21.5	0.93081\\
21.6	0.93126\\
21.7	0.93163\\
21.8	0.93217\\
21.9	0.93261\\
22	0.93304\\
22.1	0.93351\\
22.2	0.93397\\
22.3	0.93443\\
22.4	0.93489\\
22.5	0.93529\\
22.6	0.93583\\
22.7	0.93635\\
22.8	0.93682\\
22.9	0.93734\\
23	0.93778\\
23.1	0.93819\\
23.2	0.93864\\
23.3	0.93914\\
23.4	0.93957\\
23.5	0.93999\\
23.6	0.94043\\
23.7	0.94091\\
23.8	0.94136\\
23.9	0.94187\\
24	0.94234\\
24.1	0.94284\\
24.2	0.9432\\
24.3	0.9436\\
24.4	0.94402\\
24.5	0.94432\\
24.6	0.94474\\
24.7	0.94522\\
24.8	0.94553\\
24.9	0.94592\\
25	0.94627\\
25.1	0.94664\\
25.2	0.94696\\
25.3	0.94724\\
25.4	0.94753\\
25.5	0.94789\\
25.6	0.94827\\
25.7	0.9486\\
25.8	0.94895\\
25.9	0.94934\\
26	0.94981\\
26.1	0.95015\\
26.2	0.95059\\
26.3	0.95085\\
26.4	0.95115\\
26.5	0.95152\\
26.6	0.95186\\
26.7	0.95225\\
26.8	0.95249\\
26.9	0.95275\\
27	0.95309\\
27.1	0.95346\\
27.2	0.95387\\
27.3	0.95414\\
27.4	0.95448\\
27.5	0.95469\\
27.6	0.95493\\
27.7	0.95522\\
27.8	0.95552\\
27.9	0.95575\\
28	0.95603\\
28.1	0.95639\\
28.2	0.95665\\
28.3	0.95698\\
28.4	0.95722\\
28.5	0.95749\\
28.6	0.95773\\
28.7	0.95804\\
28.8	0.95833\\
28.9	0.95868\\
29	0.9589\\
29.1	0.95909\\
29.2	0.95937\\
29.3	0.95966\\
29.4	0.95984\\
29.5	0.96008\\
29.6	0.9603\\
29.7	0.96054\\
29.8	0.96076\\
29.9	0.96093\\
30	0.9612\\
30.1	0.96143\\
30.2	0.96172\\
30.3	0.96195\\
30.4	0.96221\\
30.5	0.96254\\
30.6	0.96277\\
30.7	0.96294\\
30.8	0.9632\\
30.9	0.96352\\
31	0.96378\\
31.1	0.964\\
31.2	0.96428\\
31.3	0.96452\\
31.4	0.96473\\
31.5	0.96494\\
31.6	0.96525\\
31.7	0.96543\\
31.8	0.96564\\
31.9	0.96587\\
32	0.96606\\
32.1	0.96626\\
32.2	0.96646\\
32.3	0.96667\\
32.4	0.96687\\
32.5	0.96698\\
32.6	0.96713\\
32.7	0.96731\\
32.8	0.96752\\
32.9	0.96773\\
33	0.96789\\
33.1	0.96812\\
33.2	0.9684\\
33.3	0.96857\\
33.4	0.96878\\
33.5	0.96901\\
33.6	0.96922\\
33.7	0.96936\\
33.8	0.96952\\
33.9	0.96964\\
34	0.96984\\
34.1	0.97003\\
34.2	0.9702\\
34.3	0.97042\\
34.4	0.97061\\
34.5	0.97074\\
34.6	0.97087\\
34.7	0.971\\
34.8	0.9711\\
34.9	0.97125\\
35	0.97143\\
35.1	0.9716\\
35.2	0.97178\\
35.3	0.97199\\
35.4	0.97211\\
35.5	0.97232\\
35.6	0.97243\\
35.7	0.97255\\
35.8	0.97273\\
35.9	0.97291\\
36	0.97304\\
36.1	0.97316\\
36.2	0.97336\\
36.3	0.97347\\
36.4	0.97357\\
36.5	0.97367\\
36.6	0.97386\\
36.7	0.97394\\
36.8	0.97412\\
36.9	0.9743\\
37	0.97448\\
37.1	0.97462\\
37.2	0.97481\\
37.3	0.97492\\
37.4	0.97509\\
37.5	0.97523\\
37.6	0.97542\\
37.7	0.97557\\
37.8	0.97573\\
37.9	0.97593\\
38	0.97604\\
38.1	0.97622\\
38.2	0.97634\\
38.3	0.97652\\
38.4	0.97671\\
38.5	0.97684\\
38.6	0.97691\\
38.7	0.97703\\
38.8	0.97717\\
38.9	0.97731\\
39	0.97741\\
39.1	0.97751\\
39.2	0.97768\\
39.3	0.97781\\
39.4	0.97798\\
39.5	0.97814\\
39.6	0.9783\\
39.7	0.97843\\
39.8	0.97858\\
39.9	0.97874\\
40	0.97887\\
40.1	0.97899\\
40.2	0.97909\\
40.3	0.97921\\
40.4	0.97929\\
40.5	0.97946\\
40.6	0.9796\\
40.7	0.97969\\
40.8	0.9798\\
40.9	0.97988\\
41	0.97999\\
41.1	0.98009\\
41.2	0.98018\\
41.3	0.9803\\
41.4	0.98045\\
41.5	0.98054\\
41.6	0.98062\\
41.7	0.98081\\
41.8	0.98085\\
41.9	0.9809\\
42	0.98099\\
42.1	0.98109\\
42.2	0.98118\\
42.3	0.98125\\
42.4	0.98139\\
42.5	0.98147\\
42.6	0.98157\\
42.7	0.98174\\
42.8	0.98184\\
42.9	0.9819\\
43	0.98201\\
43.1	0.98207\\
43.2	0.98218\\
43.3	0.98232\\
43.4	0.98241\\
43.5	0.98248\\
43.6	0.98261\\
43.7	0.9827\\
43.8	0.98281\\
43.9	0.98289\\
44	0.98299\\
44.1	0.98308\\
44.2	0.98317\\
44.3	0.98327\\
44.4	0.98335\\
44.5	0.98343\\
44.6	0.98353\\
44.7	0.98369\\
44.8	0.98381\\
44.9	0.98389\\
45	0.98399\\
45.1	0.98406\\
45.2	0.98416\\
45.3	0.98427\\
45.4	0.98435\\
45.5	0.98442\\
45.6	0.98447\\
45.7	0.98455\\
45.8	0.98463\\
45.9	0.98468\\
46	0.98482\\
46.1	0.9849\\
46.2	0.98502\\
46.3	0.98514\\
46.4	0.98521\\
46.5	0.9853\\
46.6	0.98535\\
46.7	0.98543\\
46.8	0.98551\\
46.9	0.98558\\
47	0.98566\\
47.1	0.98578\\
47.2	0.98582\\
47.3	0.98592\\
47.4	0.98595\\
47.5	0.98608\\
47.6	0.98613\\
47.7	0.98627\\
47.8	0.98637\\
47.9	0.98641\\
48	0.98652\\
48.1	0.98656\\
48.2	0.98664\\
48.3	0.98679\\
48.4	0.98688\\
48.5	0.98701\\
48.6	0.98707\\
48.7	0.98719\\
48.8	0.98726\\
48.9	0.98737\\
49	0.98745\\
49.1	0.98747\\
49.2	0.98754\\
49.3	0.98762\\
49.4	0.98768\\
49.5	0.98776\\
49.6	0.98782\\
49.7	0.98788\\
49.8	0.98793\\
49.9	0.98797\\
50	0.98801\\
50.1	0.98804\\
50.2	0.98809\\
50.3	0.98817\\
50.4	0.98823\\
50.5	0.98833\\
50.6	0.98835\\
50.7	0.98841\\
50.8	0.9885\\
50.9	0.98858\\
51	1\\
};

\addplot [color=maxcol,semithick,mark=square,mark options={solid},mark repeat=25,mark phase=16]
  table[row sep=crcr]{%
0	0\\
0.1	0.14478\\
0.2	0.20258\\
0.3	0.24556\\
0.4	0.28082\\
0.5	0.3112\\
0.6	0.3383\\
0.7	0.36229\\
0.8	0.38433\\
0.9	0.40412\\
1	0.42319\\
1.1	0.44067\\
1.2	0.45712\\
1.3	0.47207\\
1.4	0.48695\\
1.5	0.50105\\
1.6	0.5141\\
1.7	0.52603\\
1.8	0.53785\\
1.9	0.54914\\
2	0.55964\\
2.1	0.56988\\
2.2	0.57995\\
2.3	0.58956\\
2.4	0.59824\\
2.5	0.60617\\
2.6	0.61427\\
2.7	0.622\\
2.8	0.62988\\
2.9	0.63715\\
3	0.64451\\
3.1	0.65127\\
3.2	0.65774\\
3.3	0.66424\\
3.4	0.67035\\
3.5	0.67573\\
3.6	0.68121\\
3.7	0.68658\\
3.8	0.69166\\
3.9	0.69691\\
4	0.7017\\
4.1	0.70687\\
4.2	0.71121\\
4.3	0.71558\\
4.4	0.72038\\
4.5	0.72502\\
4.6	0.72963\\
4.7	0.73365\\
4.8	0.7377\\
4.9	0.74152\\
5	0.74495\\
5.1	0.74886\\
5.2	0.75272\\
5.3	0.75618\\
5.4	0.75938\\
5.5	0.76315\\
5.6	0.76666\\
5.7	0.76977\\
5.8	0.77301\\
5.9	0.77611\\
6	0.77949\\
6.1	0.78258\\
6.2	0.78544\\
6.3	0.78844\\
6.4	0.79134\\
6.5	0.79411\\
6.6	0.79699\\
6.7	0.79945\\
6.8	0.80212\\
6.9	0.80465\\
7	0.80715\\
7.1	0.80968\\
7.2	0.81198\\
7.3	0.81427\\
7.4	0.81663\\
7.5	0.81891\\
7.6	0.82091\\
7.7	0.82308\\
7.8	0.8251\\
7.9	0.82705\\
8	0.82908\\
8.1	0.83135\\
8.2	0.83364\\
8.3	0.83558\\
8.4	0.83728\\
8.5	0.83916\\
8.6	0.84103\\
8.7	0.84269\\
8.8	0.84428\\
8.9	0.84609\\
9	0.84783\\
9.1	0.84943\\
9.2	0.85119\\
9.3	0.8527\\
9.4	0.85445\\
9.5	0.85606\\
9.6	0.85761\\
9.7	0.85884\\
9.8	0.86029\\
9.9	0.86162\\
10	0.86309\\
10.1	0.86432\\
10.2	0.86546\\
10.3	0.86699\\
10.4	0.86842\\
10.5	0.86961\\
10.6	0.87109\\
10.7	0.87239\\
10.8	0.87363\\
10.9	0.87514\\
11	0.87632\\
11.1	0.87764\\
11.2	0.87868\\
11.3	0.87978\\
11.4	0.88095\\
11.5	0.88217\\
11.6	0.88329\\
11.7	0.8846\\
11.8	0.88581\\
11.9	0.88692\\
12	0.88804\\
12.1	0.88911\\
12.2	0.89018\\
12.3	0.89138\\
12.4	0.89253\\
12.5	0.89357\\
12.6	0.89444\\
12.7	0.89554\\
12.8	0.89646\\
12.9	0.89751\\
13	0.89835\\
13.1	0.89929\\
13.2	0.90037\\
13.3	0.90119\\
13.4	0.90204\\
13.5	0.90295\\
13.6	0.90378\\
13.7	0.9046\\
13.8	0.9055\\
13.9	0.9064\\
14	0.90716\\
14.1	0.90798\\
14.2	0.90906\\
14.3	0.90989\\
14.4	0.91072\\
14.5	0.91171\\
14.6	0.9126\\
14.7	0.91339\\
14.8	0.91411\\
14.9	0.91463\\
15	0.91542\\
15.1	0.91606\\
15.2	0.91678\\
15.3	0.91748\\
15.4	0.9182\\
15.5	0.91895\\
15.6	0.91964\\
15.7	0.92039\\
15.8	0.92085\\
15.9	0.92151\\
16	0.9221\\
16.1	0.92276\\
16.2	0.9234\\
16.3	0.92388\\
16.4	0.92471\\
16.5	0.92548\\
16.6	0.92603\\
16.7	0.92674\\
16.8	0.92743\\
16.9	0.92804\\
17	0.92862\\
17.1	0.92927\\
17.2	0.92979\\
17.3	0.93029\\
17.4	0.93077\\
17.5	0.93129\\
17.6	0.93181\\
17.7	0.93243\\
17.8	0.93293\\
17.9	0.93362\\
18	0.93414\\
18.1	0.93456\\
18.2	0.935\\
18.3	0.93555\\
18.4	0.93589\\
18.5	0.93627\\
18.6	0.93674\\
18.7	0.93726\\
18.8	0.93768\\
18.9	0.93825\\
19	0.93876\\
19.1	0.9393\\
19.2	0.93977\\
19.3	0.94025\\
19.4	0.94074\\
19.5	0.94108\\
19.6	0.94151\\
19.7	0.94195\\
19.8	0.94236\\
19.9	0.94281\\
20	0.94321\\
20.1	0.94365\\
20.2	0.94411\\
20.3	0.94451\\
20.4	0.94498\\
20.5	0.94535\\
20.6	0.94567\\
20.7	0.94606\\
20.8	0.94651\\
20.9	0.94687\\
21	0.94722\\
21.1	0.94755\\
21.2	0.94799\\
21.3	0.9483\\
21.4	0.94872\\
21.5	0.94903\\
21.6	0.94938\\
21.7	0.94966\\
21.8	0.95002\\
21.9	0.95032\\
22	0.95074\\
22.1	0.95111\\
22.2	0.95134\\
22.3	0.9516\\
22.4	0.95202\\
22.5	0.95235\\
22.6	0.95272\\
22.7	0.95308\\
22.8	0.95343\\
22.9	0.95392\\
23	0.95421\\
23.1	0.95443\\
23.2	0.95469\\
23.3	0.95501\\
23.4	0.95529\\
23.5	0.95568\\
23.6	0.95585\\
23.7	0.95616\\
23.8	0.9564\\
23.9	0.9568\\
24	0.95709\\
24.1	0.95735\\
24.2	0.95767\\
24.3	0.95801\\
24.4	0.95831\\
24.5	0.95862\\
24.6	0.95898\\
24.7	0.95923\\
24.8	0.95944\\
24.9	0.95975\\
25	0.96009\\
25.1	0.96033\\
25.2	0.96049\\
25.3	0.96081\\
25.4	0.96107\\
25.5	0.9613\\
25.6	0.96151\\
25.7	0.96178\\
25.8	0.96209\\
25.9	0.96229\\
26	0.9626\\
26.1	0.96279\\
26.2	0.963\\
26.3	0.96322\\
26.4	0.96345\\
26.5	0.96357\\
26.6	0.9638\\
26.7	0.964\\
26.8	0.96422\\
26.9	0.96438\\
27	0.9646\\
27.1	0.96482\\
27.2	0.96501\\
27.3	0.96525\\
27.4	0.96554\\
27.5	0.96568\\
27.6	0.96593\\
27.7	0.96615\\
27.8	0.96639\\
27.9	0.96661\\
28	0.96681\\
28.1	0.96715\\
28.2	0.96739\\
28.3	0.96767\\
28.4	0.96785\\
28.5	0.96808\\
28.6	0.9683\\
28.7	0.96847\\
28.8	0.9687\\
28.9	0.96891\\
29	0.96915\\
29.1	0.96938\\
29.2	0.96959\\
29.3	0.96978\\
29.4	0.96986\\
29.5	0.97\\
29.6	0.97021\\
29.7	0.97054\\
29.8	0.9708\\
29.9	0.97098\\
30	0.9711\\
30.1	0.97127\\
30.2	0.97148\\
30.3	0.97168\\
30.4	0.97188\\
30.5	0.97199\\
30.6	0.97211\\
30.7	0.97228\\
30.8	0.97244\\
30.9	0.97256\\
31	0.97278\\
31.1	0.97291\\
31.2	0.97304\\
31.3	0.9732\\
31.4	0.9733\\
31.5	0.97344\\
31.6	0.97359\\
31.7	0.97373\\
31.8	0.97382\\
31.9	0.97398\\
32	0.97416\\
32.1	0.97434\\
32.2	0.97449\\
32.3	0.97466\\
32.4	0.97478\\
32.5	0.97491\\
32.6	0.97503\\
32.7	0.97515\\
32.8	0.97531\\
32.9	0.97548\\
33	0.97559\\
33.1	0.97574\\
33.2	0.97588\\
33.3	0.97606\\
33.4	0.97617\\
33.5	0.97626\\
33.6	0.9764\\
33.7	0.97651\\
33.8	0.97666\\
33.9	0.97675\\
34	0.97683\\
34.1	0.977\\
34.2	0.9772\\
34.3	0.97733\\
34.4	0.97747\\
34.5	0.97765\\
34.6	0.97776\\
34.7	0.97783\\
34.8	0.97797\\
34.9	0.97803\\
35	0.97813\\
35.1	0.97827\\
35.2	0.97832\\
35.3	0.97842\\
35.4	0.9785\\
35.5	0.97861\\
35.6	0.97872\\
35.7	0.97882\\
35.8	0.97892\\
35.9	0.97905\\
36	0.97916\\
36.1	0.97926\\
36.2	0.97936\\
36.3	0.97946\\
36.4	0.97959\\
36.5	0.97967\\
36.6	0.97978\\
36.7	0.97986\\
36.8	0.97994\\
36.9	0.98008\\
37	0.98016\\
37.1	0.98024\\
37.2	0.9804\\
37.3	0.98053\\
37.4	0.98059\\
37.5	0.98069\\
37.6	0.98078\\
37.7	0.98084\\
37.8	0.98094\\
37.9	0.98107\\
38	0.98114\\
38.1	0.98127\\
38.2	0.98135\\
38.3	0.98143\\
38.4	0.98152\\
38.5	0.98166\\
38.6	0.98181\\
38.7	0.98192\\
38.8	0.98204\\
38.9	0.98218\\
39	0.98231\\
39.1	0.98241\\
39.2	0.98249\\
39.3	0.98257\\
39.4	0.98273\\
39.5	0.98289\\
39.6	0.98298\\
39.7	0.98307\\
39.8	0.98311\\
39.9	0.9832\\
40	0.98334\\
40.1	0.98347\\
40.2	0.98354\\
40.3	0.98362\\
40.4	0.98374\\
40.5	0.98385\\
40.6	0.98394\\
40.7	0.984\\
40.8	0.98407\\
40.9	0.98411\\
41	0.98417\\
41.1	0.98422\\
41.2	0.98432\\
41.3	0.98436\\
41.4	0.98448\\
41.5	0.98459\\
41.6	0.98468\\
41.7	0.98473\\
41.8	0.98482\\
41.9	0.9849\\
42	0.98495\\
42.1	0.98499\\
42.2	0.98507\\
42.3	0.98513\\
42.4	0.98516\\
42.5	0.98526\\
42.6	0.98531\\
42.7	0.98539\\
42.8	0.98546\\
42.9	0.98558\\
43	0.98564\\
43.1	0.98569\\
43.2	0.98577\\
43.3	0.98581\\
43.4	0.9859\\
43.5	0.98596\\
43.6	0.98601\\
43.7	0.98604\\
43.8	0.98613\\
43.9	0.98619\\
44	0.98623\\
44.1	0.98629\\
44.2	0.98636\\
44.3	0.9864\\
44.4	0.98653\\
44.5	0.98659\\
44.6	0.98663\\
44.7	0.98669\\
44.8	0.98674\\
44.9	0.98684\\
45	0.98694\\
45.1	0.98701\\
45.2	0.98711\\
45.3	0.98715\\
45.4	0.98723\\
45.5	0.98733\\
45.6	0.98744\\
45.7	0.9875\\
45.8	0.98757\\
45.9	0.98762\\
46	0.98768\\
46.1	0.98775\\
46.2	0.9878\\
46.3	0.98786\\
46.4	0.98792\\
46.5	0.98796\\
46.6	0.988\\
46.7	0.98812\\
46.8	0.98819\\
46.9	0.98826\\
47	0.98833\\
47.1	0.98841\\
47.2	0.98851\\
47.3	0.98859\\
47.4	0.98869\\
47.5	0.98873\\
47.6	0.98875\\
47.7	0.98879\\
47.8	0.98887\\
47.9	0.98895\\
48	0.98899\\
48.1	0.98907\\
48.2	0.98913\\
48.3	0.98922\\
48.4	0.98928\\
48.5	0.98932\\
48.6	0.98936\\
48.7	0.98941\\
48.8	0.98952\\
48.9	0.98955\\
49	0.98959\\
49.1	0.98968\\
49.2	0.98974\\
49.3	0.98977\\
49.4	0.98985\\
49.5	0.98987\\
49.6	0.98997\\
49.7	0.99004\\
49.8	0.99008\\
49.9	0.99014\\
50	0.99018\\
50.1	0.99027\\
50.2	0.99036\\
50.3	0.99043\\
50.4	0.99048\\
50.5	0.99051\\
50.6	0.99053\\
50.7	0.99057\\
50.8	0.99062\\
50.9	0.99066\\
51	1\\
};

\addplot [color=cntcol,semithick,mark=triangle,mark options={solid,rotate=180},mark repeat=25,mark phase=20]
  table[row sep=crcr]{%
0	0\\
0.1	0.10448\\
0.2	0.14669\\
0.3	0.17936\\
0.4	0.20683\\
0.5	0.23028\\
0.6	0.25086\\
0.7	0.27084\\
0.8	0.28845\\
0.9	0.30447\\
1	0.31991\\
1.1	0.33391\\
1.2	0.34734\\
1.3	0.36018\\
1.4	0.37252\\
1.5	0.38427\\
1.6	0.39556\\
1.7	0.40671\\
1.8	0.41754\\
1.9	0.42706\\
2	0.43655\\
2.1	0.44564\\
2.2	0.45527\\
2.3	0.46405\\
2.4	0.47225\\
2.5	0.48087\\
2.6	0.48857\\
2.7	0.49658\\
2.8	0.50414\\
2.9	0.51087\\
3	0.51746\\
3.1	0.52425\\
3.2	0.53107\\
3.3	0.53744\\
3.4	0.54377\\
3.5	0.54992\\
3.6	0.55619\\
3.7	0.56224\\
3.8	0.56764\\
3.9	0.5732\\
4	0.57857\\
4.1	0.58436\\
4.2	0.58971\\
4.3	0.59496\\
4.4	0.60003\\
4.5	0.60491\\
4.6	0.60978\\
4.7	0.61463\\
4.8	0.61932\\
4.9	0.62341\\
5	0.62774\\
5.1	0.63213\\
5.2	0.63635\\
5.3	0.64071\\
5.4	0.64527\\
5.5	0.64917\\
5.6	0.6532\\
5.7	0.65766\\
5.8	0.66179\\
5.9	0.66546\\
6	0.66934\\
6.1	0.67292\\
6.2	0.67631\\
6.3	0.68016\\
6.4	0.68362\\
6.5	0.68696\\
6.6	0.69001\\
6.7	0.69334\\
6.8	0.6967\\
6.9	0.69999\\
7	0.70332\\
7.1	0.7063\\
7.2	0.70967\\
7.3	0.7128\\
7.4	0.71583\\
7.5	0.71907\\
7.6	0.72208\\
7.7	0.72471\\
7.8	0.72746\\
7.9	0.7303\\
8	0.7331\\
8.1	0.73587\\
8.2	0.73848\\
8.3	0.74131\\
8.4	0.74378\\
8.5	0.74632\\
8.6	0.74905\\
8.7	0.7515\\
8.8	0.75399\\
8.9	0.75655\\
9	0.75906\\
9.1	0.76116\\
9.2	0.76343\\
9.3	0.76567\\
9.4	0.76797\\
9.5	0.77018\\
9.6	0.77241\\
9.7	0.77455\\
9.8	0.77669\\
9.9	0.77888\\
10	0.78119\\
10.1	0.78347\\
10.2	0.78528\\
10.3	0.78737\\
10.4	0.78918\\
10.5	0.79121\\
10.6	0.79321\\
10.7	0.79542\\
10.8	0.79741\\
10.9	0.79911\\
11	0.80121\\
11.1	0.80279\\
11.2	0.80463\\
11.3	0.80644\\
11.4	0.80802\\
11.5	0.80992\\
11.6	0.81176\\
11.7	0.81372\\
11.8	0.81541\\
11.9	0.81719\\
12	0.81876\\
12.1	0.82032\\
12.2	0.82165\\
12.3	0.82328\\
12.4	0.82472\\
12.5	0.82636\\
12.6	0.8281\\
12.7	0.82969\\
12.8	0.83127\\
12.9	0.83265\\
13	0.8339\\
13.1	0.83541\\
13.2	0.83661\\
13.3	0.838\\
13.4	0.8393\\
13.5	0.8407\\
13.6	0.84212\\
13.7	0.8436\\
13.8	0.84491\\
13.9	0.84622\\
14	0.84747\\
14.1	0.84858\\
14.2	0.8497\\
14.3	0.85099\\
14.4	0.85224\\
14.5	0.8535\\
14.6	0.85489\\
14.7	0.85613\\
14.8	0.85738\\
14.9	0.85842\\
15	0.85951\\
15.1	0.86075\\
15.2	0.86173\\
15.3	0.86301\\
15.4	0.86407\\
15.5	0.8652\\
15.6	0.86624\\
15.7	0.86739\\
15.8	0.8684\\
15.9	0.86933\\
16	0.87053\\
16.1	0.87158\\
16.2	0.87267\\
16.3	0.8734\\
16.4	0.87431\\
16.5	0.87545\\
16.6	0.8764\\
16.7	0.87742\\
16.8	0.8784\\
16.9	0.87932\\
17	0.88015\\
17.1	0.88121\\
17.2	0.8822\\
17.3	0.88294\\
17.4	0.88396\\
17.5	0.88466\\
17.6	0.88564\\
17.7	0.88641\\
17.8	0.88727\\
17.9	0.88805\\
18	0.88891\\
18.1	0.88975\\
18.2	0.89067\\
18.3	0.89156\\
18.4	0.89239\\
18.5	0.89331\\
18.6	0.89403\\
18.7	0.89478\\
18.8	0.89553\\
18.9	0.89609\\
19	0.89687\\
19.1	0.89767\\
19.2	0.89831\\
19.3	0.89923\\
19.4	0.90012\\
19.5	0.90091\\
19.6	0.90147\\
19.7	0.90226\\
19.8	0.90299\\
19.9	0.90366\\
20	0.9044\\
20.1	0.90504\\
20.2	0.90565\\
20.3	0.90637\\
20.4	0.90713\\
20.5	0.90785\\
20.6	0.90855\\
20.7	0.90912\\
20.8	0.90974\\
20.9	0.91041\\
21	0.91103\\
21.1	0.91157\\
21.2	0.91207\\
21.3	0.91264\\
21.4	0.91314\\
21.5	0.91384\\
21.6	0.91453\\
21.7	0.91523\\
21.8	0.91592\\
21.9	0.91653\\
22	0.9173\\
22.1	0.91785\\
22.2	0.91845\\
22.3	0.91895\\
22.4	0.91953\\
22.5	0.92015\\
22.6	0.92069\\
22.7	0.92137\\
22.8	0.92181\\
22.9	0.92229\\
23	0.92286\\
23.1	0.92344\\
23.2	0.92388\\
23.3	0.9245\\
23.4	0.92491\\
23.5	0.92546\\
23.6	0.92604\\
23.7	0.9266\\
23.8	0.92702\\
23.9	0.92744\\
24	0.92801\\
24.1	0.92841\\
24.2	0.92877\\
24.3	0.92926\\
24.4	0.92979\\
24.5	0.93033\\
24.6	0.9308\\
24.7	0.93128\\
24.8	0.93173\\
24.9	0.93209\\
25	0.93256\\
25.1	0.933\\
25.2	0.93348\\
25.3	0.93397\\
25.4	0.93431\\
25.5	0.93469\\
25.6	0.93507\\
25.7	0.9355\\
25.8	0.93595\\
25.9	0.93626\\
26	0.93669\\
26.1	0.93703\\
26.2	0.93735\\
26.3	0.93769\\
26.4	0.93811\\
26.5	0.93842\\
26.6	0.93884\\
26.7	0.93914\\
26.8	0.9395\\
26.9	0.93978\\
27	0.94016\\
27.1	0.94058\\
27.2	0.941\\
27.3	0.94137\\
27.4	0.94181\\
27.5	0.9421\\
27.6	0.94234\\
27.7	0.94271\\
27.8	0.94306\\
27.9	0.94336\\
28	0.94372\\
28.1	0.944\\
28.2	0.9444\\
28.3	0.94474\\
28.4	0.94508\\
28.5	0.94544\\
28.6	0.9458\\
28.7	0.94611\\
28.8	0.94642\\
28.9	0.94669\\
29	0.94693\\
29.1	0.94716\\
29.2	0.94741\\
29.3	0.94767\\
29.4	0.94794\\
29.5	0.94826\\
29.6	0.94854\\
29.7	0.94882\\
29.8	0.94921\\
29.9	0.94943\\
30	0.94979\\
30.1	0.95012\\
30.2	0.95048\\
30.3	0.95084\\
30.4	0.95106\\
30.5	0.95146\\
30.6	0.95174\\
30.7	0.95197\\
30.8	0.95226\\
30.9	0.95244\\
31	0.95267\\
31.1	0.95285\\
31.2	0.95315\\
31.3	0.95345\\
31.4	0.95359\\
31.5	0.95381\\
31.6	0.9541\\
31.7	0.95431\\
31.8	0.95453\\
31.9	0.95477\\
32	0.95499\\
32.1	0.95521\\
32.2	0.95544\\
32.3	0.95579\\
32.4	0.9561\\
32.5	0.95632\\
32.6	0.95657\\
32.7	0.95681\\
32.8	0.95708\\
32.9	0.95733\\
33	0.95756\\
33.1	0.95776\\
33.2	0.95792\\
33.3	0.95816\\
33.4	0.95839\\
33.5	0.95856\\
33.6	0.95878\\
33.7	0.95898\\
33.8	0.95922\\
33.9	0.95944\\
34	0.95975\\
34.1	0.95997\\
34.2	0.96021\\
34.3	0.96036\\
34.4	0.96061\\
34.5	0.96082\\
34.6	0.96101\\
34.7	0.96131\\
34.8	0.96155\\
34.9	0.96181\\
35	0.96199\\
35.1	0.96221\\
35.2	0.96246\\
35.3	0.96265\\
35.4	0.96288\\
35.5	0.963\\
35.6	0.9632\\
35.7	0.96336\\
35.8	0.9635\\
35.9	0.96363\\
36	0.96375\\
36.1	0.9639\\
36.2	0.96406\\
36.3	0.96431\\
36.4	0.96445\\
36.5	0.9647\\
36.6	0.9649\\
36.7	0.96505\\
36.8	0.96521\\
36.9	0.96537\\
37	0.9655\\
37.1	0.96573\\
37.2	0.96591\\
37.3	0.96607\\
37.4	0.96628\\
37.5	0.96644\\
37.6	0.96664\\
37.7	0.9669\\
37.8	0.96702\\
37.9	0.96718\\
38	0.96737\\
38.1	0.96754\\
38.2	0.9677\\
38.3	0.96779\\
38.4	0.96796\\
38.5	0.96812\\
38.6	0.96829\\
38.7	0.9684\\
38.8	0.96859\\
38.9	0.96873\\
39	0.96889\\
39.1	0.96907\\
39.2	0.96927\\
39.3	0.96943\\
39.4	0.96961\\
39.5	0.96973\\
39.6	0.96988\\
39.7	0.97003\\
39.8	0.97019\\
39.9	0.97038\\
40	0.97051\\
40.1	0.97077\\
40.2	0.97094\\
40.3	0.97105\\
40.4	0.97122\\
40.5	0.97139\\
40.6	0.97154\\
40.7	0.97169\\
40.8	0.9718\\
40.9	0.97194\\
41	0.97215\\
41.1	0.97224\\
41.2	0.97238\\
41.3	0.97263\\
41.4	0.97277\\
41.5	0.9729\\
41.6	0.973\\
41.7	0.97314\\
41.8	0.97332\\
41.9	0.97346\\
42	0.97364\\
42.1	0.97376\\
42.2	0.97387\\
42.3	0.97394\\
42.4	0.97414\\
42.5	0.97426\\
42.6	0.97436\\
42.7	0.97443\\
42.8	0.97454\\
42.9	0.97466\\
43	0.97474\\
43.1	0.97487\\
43.2	0.975\\
43.3	0.97512\\
43.4	0.97522\\
43.5	0.97538\\
43.6	0.97556\\
43.7	0.97571\\
43.8	0.97578\\
43.9	0.9759\\
44	0.97601\\
44.1	0.9761\\
44.2	0.97619\\
44.3	0.97627\\
44.4	0.97639\\
44.5	0.97646\\
44.6	0.97652\\
44.7	0.97664\\
44.8	0.97679\\
44.9	0.97692\\
45	0.977\\
45.1	0.97715\\
45.2	0.97729\\
45.3	0.97738\\
45.4	0.9775\\
45.5	0.97756\\
45.6	0.97762\\
45.7	0.9778\\
45.8	0.9779\\
45.9	0.97797\\
46	0.97799\\
46.1	0.97813\\
46.2	0.97826\\
46.3	0.97833\\
46.4	0.97843\\
46.5	0.97848\\
46.6	0.9786\\
46.7	0.97879\\
46.8	0.97898\\
46.9	0.9791\\
47	0.97923\\
47.1	0.97936\\
47.2	0.97945\\
47.3	0.97954\\
47.4	0.9796\\
47.5	0.97967\\
47.6	0.97975\\
47.7	0.97981\\
47.8	0.97995\\
47.9	0.98006\\
48	0.98013\\
48.1	0.98022\\
48.2	0.98029\\
48.3	0.98035\\
48.4	0.98042\\
48.5	0.98053\\
48.6	0.98067\\
48.7	0.98076\\
48.8	0.98092\\
48.9	0.98101\\
49	0.98111\\
49.1	0.98121\\
49.2	0.98129\\
49.3	0.9814\\
49.4	0.98149\\
49.5	0.98157\\
49.6	0.98165\\
49.7	0.98169\\
49.8	0.98178\\
49.9	0.98184\\
50	0.98192\\
50.1	0.98205\\
50.2	0.98211\\
50.3	0.98216\\
50.4	0.9823\\
50.5	0.9824\\
50.6	0.98249\\
50.7	0.98258\\
50.8	0.98267\\
50.9	0.98277\\
51	1\\
};

\addplot [color=mafcol, semithick, mark=triangle,mark repeat=25,mark phase=24]
  table[row sep=crcr]{%
0	0\\
0.1	0.09951\\
0.2	0.14012\\
0.3	0.17083\\
0.4	0.19654\\
0.5	0.21937\\
0.6	0.24022\\
0.7	0.25863\\
0.8	0.2762\\
0.9	0.29157\\
1	0.3065\\
1.1	0.32049\\
1.2	0.33389\\
1.3	0.34697\\
1.4	0.35885\\
1.5	0.36965\\
1.6	0.38069\\
1.7	0.39104\\
1.8	0.40126\\
1.9	0.41111\\
2	0.42084\\
2.1	0.42974\\
2.2	0.4389\\
2.3	0.44739\\
2.4	0.45569\\
2.5	0.46395\\
2.6	0.4719\\
2.7	0.47962\\
2.8	0.48709\\
2.9	0.49382\\
3	0.50092\\
3.1	0.5076\\
3.2	0.51435\\
3.3	0.52068\\
3.4	0.52665\\
3.5	0.53291\\
3.6	0.53883\\
3.7	0.5445\\
3.8	0.55047\\
3.9	0.55617\\
4	0.56158\\
4.1	0.56682\\
4.2	0.57251\\
4.3	0.5773\\
4.4	0.58222\\
4.5	0.58749\\
4.6	0.59185\\
4.7	0.59654\\
4.8	0.60129\\
4.9	0.60617\\
5	0.61089\\
5.1	0.61519\\
5.2	0.61935\\
5.3	0.62375\\
5.4	0.62763\\
5.5	0.6318\\
5.6	0.63581\\
5.7	0.6402\\
5.8	0.64431\\
5.9	0.6481\\
6	0.65165\\
6.1	0.6556\\
6.2	0.65906\\
6.3	0.66244\\
6.4	0.66641\\
6.5	0.67034\\
6.6	0.67379\\
6.7	0.67704\\
6.8	0.68018\\
6.9	0.6833\\
7	0.68681\\
7.1	0.68983\\
7.2	0.69273\\
7.3	0.69569\\
7.4	0.69869\\
7.5	0.7021\\
7.6	0.70484\\
7.7	0.70782\\
7.8	0.71085\\
7.9	0.71382\\
8	0.71667\\
8.1	0.71932\\
8.2	0.72229\\
8.3	0.72497\\
8.4	0.72773\\
8.5	0.73061\\
8.6	0.73309\\
8.7	0.73538\\
8.8	0.73788\\
8.9	0.74007\\
9	0.7425\\
9.1	0.74509\\
9.2	0.74747\\
9.3	0.75004\\
9.4	0.75228\\
9.5	0.7546\\
9.6	0.75707\\
9.7	0.75931\\
9.8	0.76159\\
9.9	0.76368\\
10	0.76584\\
10.1	0.76793\\
10.2	0.77004\\
10.3	0.77229\\
10.4	0.77399\\
10.5	0.77607\\
10.6	0.77794\\
10.7	0.78003\\
10.8	0.78224\\
10.9	0.78425\\
11	0.78614\\
11.1	0.78817\\
11.2	0.78998\\
11.3	0.79171\\
11.4	0.79364\\
11.5	0.79536\\
11.6	0.79697\\
11.7	0.79885\\
11.8	0.80052\\
11.9	0.8021\\
12	0.80379\\
12.1	0.80541\\
12.2	0.80721\\
12.3	0.80894\\
12.4	0.81054\\
12.5	0.81187\\
12.6	0.81348\\
12.7	0.81494\\
12.8	0.81652\\
12.9	0.81794\\
13	0.81932\\
13.1	0.82097\\
13.2	0.82258\\
13.3	0.82402\\
13.4	0.8254\\
13.5	0.827\\
13.6	0.82812\\
13.7	0.82966\\
13.8	0.83087\\
13.9	0.83216\\
14	0.83346\\
14.1	0.83471\\
14.2	0.83611\\
14.3	0.8373\\
14.4	0.83851\\
14.5	0.83976\\
14.6	0.84106\\
14.7	0.84235\\
14.8	0.84379\\
14.9	0.845\\
15	0.8461\\
15.1	0.84759\\
15.2	0.84874\\
15.3	0.84984\\
15.4	0.85107\\
15.5	0.85212\\
15.6	0.85324\\
15.7	0.85429\\
15.8	0.85557\\
15.9	0.85667\\
16	0.85783\\
16.1	0.85882\\
16.2	0.85993\\
16.3	0.86101\\
16.4	0.86225\\
16.5	0.86341\\
16.6	0.86434\\
16.7	0.86538\\
16.8	0.86635\\
16.9	0.86744\\
17	0.86866\\
17.1	0.86984\\
17.2	0.87104\\
17.3	0.87195\\
17.4	0.87293\\
17.5	0.87389\\
17.6	0.87493\\
17.7	0.876\\
17.8	0.87704\\
17.9	0.87784\\
18	0.87866\\
18.1	0.87954\\
18.2	0.88032\\
18.3	0.88116\\
18.4	0.88207\\
18.5	0.883\\
18.6	0.88366\\
18.7	0.88467\\
18.8	0.88547\\
18.9	0.88636\\
19	0.88714\\
19.1	0.88801\\
19.2	0.88883\\
19.3	0.88962\\
19.4	0.89022\\
19.5	0.891\\
19.6	0.89177\\
19.7	0.89256\\
19.8	0.89333\\
19.9	0.894\\
20	0.89487\\
20.1	0.89567\\
20.2	0.89646\\
20.3	0.89726\\
20.4	0.89807\\
20.5	0.89881\\
20.6	0.89952\\
20.7	0.90019\\
20.8	0.90092\\
20.9	0.90165\\
21	0.90216\\
21.1	0.90275\\
21.2	0.90344\\
21.3	0.904\\
21.4	0.9047\\
21.5	0.90535\\
21.6	0.90612\\
21.7	0.90682\\
21.8	0.90733\\
21.9	0.90781\\
22	0.90828\\
22.1	0.90884\\
22.2	0.90943\\
22.3	0.90997\\
22.4	0.91051\\
22.5	0.91109\\
22.6	0.91173\\
22.7	0.91225\\
22.8	0.91287\\
22.9	0.91341\\
23	0.91399\\
23.1	0.91451\\
23.2	0.91508\\
23.3	0.91573\\
23.4	0.91639\\
23.5	0.91678\\
23.6	0.91732\\
23.7	0.91782\\
23.8	0.91838\\
23.9	0.91888\\
24	0.91929\\
24.1	0.91977\\
24.2	0.92024\\
24.3	0.92076\\
24.4	0.92128\\
24.5	0.92187\\
24.6	0.92237\\
24.7	0.9229\\
24.8	0.92349\\
24.9	0.92406\\
25	0.92452\\
25.1	0.92523\\
25.2	0.9256\\
25.3	0.92609\\
25.4	0.92652\\
25.5	0.92695\\
25.6	0.92741\\
25.7	0.92793\\
25.8	0.92839\\
25.9	0.92879\\
26	0.92922\\
26.1	0.92964\\
26.2	0.93015\\
26.3	0.93064\\
26.4	0.93119\\
26.5	0.93157\\
26.6	0.93196\\
26.7	0.93244\\
26.8	0.93287\\
26.9	0.93327\\
27	0.93367\\
27.1	0.934\\
27.2	0.93441\\
27.3	0.93472\\
27.4	0.9352\\
27.5	0.93562\\
27.6	0.93596\\
27.7	0.93626\\
27.8	0.93666\\
27.9	0.937\\
28	0.93746\\
28.1	0.93788\\
28.2	0.93817\\
28.3	0.93849\\
28.4	0.93883\\
28.5	0.93922\\
28.6	0.93957\\
28.7	0.93997\\
28.8	0.94037\\
28.9	0.94072\\
29	0.94107\\
29.1	0.94141\\
29.2	0.94175\\
29.3	0.942\\
29.4	0.94238\\
29.5	0.94267\\
29.6	0.94297\\
29.7	0.94329\\
29.8	0.94363\\
29.9	0.9439\\
30	0.94417\\
30.1	0.94447\\
30.2	0.9446\\
30.3	0.94491\\
30.4	0.94526\\
30.5	0.9456\\
30.6	0.94593\\
30.7	0.94619\\
30.8	0.94656\\
30.9	0.94686\\
31	0.94718\\
31.1	0.9474\\
31.2	0.94773\\
31.3	0.94807\\
31.4	0.9483\\
31.5	0.94861\\
31.6	0.94892\\
31.7	0.94916\\
31.8	0.94941\\
31.9	0.94967\\
32	0.95\\
32.1	0.95025\\
32.2	0.95055\\
32.3	0.95076\\
32.4	0.95098\\
32.5	0.95132\\
32.6	0.95164\\
32.7	0.95195\\
32.8	0.95216\\
32.9	0.95233\\
33	0.95253\\
33.1	0.9528\\
33.2	0.95315\\
33.3	0.9534\\
33.4	0.95372\\
33.5	0.95398\\
33.6	0.95414\\
33.7	0.95454\\
33.8	0.95472\\
33.9	0.95487\\
34	0.95516\\
34.1	0.95545\\
34.2	0.95561\\
34.3	0.95579\\
34.4	0.95601\\
34.5	0.9563\\
34.6	0.95656\\
34.7	0.95685\\
34.8	0.95702\\
34.9	0.95731\\
35	0.9575\\
35.1	0.95779\\
35.2	0.95805\\
35.3	0.95836\\
35.4	0.9586\\
35.5	0.95884\\
35.6	0.95903\\
35.7	0.95925\\
35.8	0.95939\\
35.9	0.95963\\
36	0.95994\\
36.1	0.96016\\
36.2	0.96032\\
36.3	0.96052\\
36.4	0.96071\\
36.5	0.96093\\
36.6	0.9612\\
36.7	0.96143\\
36.8	0.96166\\
36.9	0.9618\\
37	0.96203\\
37.1	0.9622\\
37.2	0.96239\\
37.3	0.96263\\
37.4	0.96283\\
37.5	0.96304\\
37.6	0.96317\\
37.7	0.96335\\
37.8	0.96355\\
37.9	0.96366\\
38	0.96387\\
38.1	0.96395\\
38.2	0.96416\\
38.3	0.96438\\
38.4	0.9646\\
38.5	0.96474\\
38.6	0.96493\\
38.7	0.96511\\
38.8	0.96529\\
38.9	0.96551\\
39	0.96573\\
39.1	0.96591\\
39.2	0.96611\\
39.3	0.9663\\
39.4	0.96644\\
39.5	0.9666\\
39.6	0.96672\\
39.7	0.9669\\
39.8	0.96706\\
39.9	0.96725\\
40	0.96745\\
40.1	0.96755\\
40.2	0.96772\\
40.3	0.96792\\
40.4	0.9682\\
40.5	0.9683\\
40.6	0.96851\\
40.7	0.96865\\
40.8	0.96881\\
40.9	0.96892\\
41	0.96904\\
41.1	0.96921\\
41.2	0.96936\\
41.3	0.96959\\
41.4	0.96965\\
41.5	0.96977\\
41.6	0.96993\\
41.7	0.97003\\
41.8	0.97016\\
41.9	0.97029\\
42	0.97051\\
42.1	0.97068\\
42.2	0.97085\\
42.3	0.97093\\
42.4	0.97107\\
42.5	0.9712\\
42.6	0.97134\\
42.7	0.97149\\
42.8	0.97162\\
42.9	0.97186\\
43	0.972\\
43.1	0.9721\\
43.2	0.97227\\
43.3	0.97238\\
43.4	0.97252\\
43.5	0.97261\\
43.6	0.97272\\
43.7	0.97283\\
43.8	0.97292\\
43.9	0.97305\\
44	0.97325\\
44.1	0.97339\\
44.2	0.97349\\
44.3	0.97361\\
44.4	0.97377\\
44.5	0.97387\\
44.6	0.97395\\
44.7	0.97405\\
44.8	0.97418\\
44.9	0.97427\\
45	0.97438\\
45.1	0.97458\\
45.2	0.97472\\
45.3	0.97482\\
45.4	0.975\\
45.5	0.97513\\
45.6	0.97528\\
45.7	0.97537\\
45.8	0.97542\\
45.9	0.97559\\
46	0.9757\\
46.1	0.97578\\
46.2	0.97587\\
46.3	0.97591\\
46.4	0.97604\\
46.5	0.97626\\
46.6	0.97636\\
46.7	0.97646\\
46.8	0.97656\\
46.9	0.97663\\
47	0.97668\\
47.1	0.97678\\
47.2	0.97689\\
47.3	0.97704\\
47.4	0.97715\\
47.5	0.9773\\
47.6	0.97742\\
47.7	0.97755\\
47.8	0.97766\\
47.9	0.97775\\
48	0.97784\\
48.1	0.97793\\
48.2	0.97799\\
48.3	0.9781\\
48.4	0.97823\\
48.5	0.97835\\
48.6	0.97846\\
48.7	0.97852\\
48.8	0.97865\\
48.9	0.97873\\
49	0.97878\\
49.1	0.97892\\
49.2	0.979\\
49.3	0.97912\\
49.4	0.97919\\
49.5	0.97922\\
49.6	0.97928\\
49.7	0.97941\\
49.8	0.97946\\
49.9	0.97954\\
50	0.97966\\
50.1	0.9797\\
50.2	0.97979\\
50.3	0.97987\\
50.4	0.97997\\
50.5	0.98007\\
50.6	0.98017\\
50.7	0.98025\\
50.8	0.98034\\
50.9	0.98042\\
51	1\\
};

\end{axis}
\end{tikzpicture}%